\documentclass[a4paper,11pt]{article}

\usepackage[utf8]{inputenc}
\usepackage[margin=0.8in]{geometry}
\usepackage{siunitx}
\usepackage{amsmath}
\usepackage{amssymb}
\usepackage{bm}
\usepackage{graphicx}
\usepackage{caption} 
\usepackage{subcaption}
\graphicspath{{figures/}}
\usepackage{sectsty}
\usepackage{color}
\usepackage{titlesec}
\usepackage[standard]{ntheorem}
\usepackage{enumitem}

\titleformat*{\section}{\large\bfseries}
\titleformat*{\subsection}{\normalsize\bfseries}
\titleformat*{\subsubsection}{\normalsize\bfseries}
\titleformat*{\paragraph}{\normalsize\bfseries}
\titleformat*{\subparagraph}{\normalsize\bfseries}

\newcommand{\q}[1]{\lvert #1 \rangle}
\newcommand{\qd}[1]{\langle #1 \rvert}

\title{Stabilization of approximate GHZ state with quasi-local couplings}
\author{Vincent Martin$^1$ and Alain Sarlette$^{1,2}$\\ $^1$ LPENS, Département de physique, Ecole normale supérieure,\\ Centre Automatique et Systèmes (CAS) MINES ParisTech, Université PSL,\\ Sorbonne Université, CNRS, Inria, 75005 Paris \\ $^2$ Department of Electronics and Information Systems, Ghent University} 

\begin{document}
\maketitle

\begin{abstract}
We propose a reservoir design, composed of fixed dissipation operators acting each on few local subsystems, to stabilize an approximate GHZ state on $n$ qubits. The main idea is to work out how a previously proposed sequence of two stabilization steps can be applied instead in appropriate (probabilistic) superposition. We examine alternatives to synchronize the superposition using local couplings only, thanks to a chain of ``clock'' ancillas or to additional levels on the data subsystems. The practical value of these alternatives depends on experimental constraints. They all feature a design tradeoff between approximate stabilization fidelity and protection against perturbations. These proposals illustrate how simple autonomous automata can be implemented in quantum reservoir engineering to replace sequential state preparation procedures. Encoding automaton actions via additional data levels only, appears particularly efficient in this context. Our analysis method, reducing the Lindblad master equation to a Markov chain on virtual output signals, may be of independent interest.
\end{abstract}


\section{Introduction}\label{sec:intro}

The entanglement of multiple subsystems is a major feature of quantum technology, with applications including metrology \cite{escher2011general}, secure communication \cite{ursin2007entanglement}, (especially measurement-based) computation schemes \cite{briegel2009measurement} and quantum error correction \cite{bennett1996mixed}. In line with the power of such states, the physical resources required for robustly generating or stabilizing them are not trivial. High-fidelity production of multi-partite entangled states can be approached from several angles. A most direct way is to start from an easy, typically separable state, and apply a sequence of gates to neighboring qubits \cite{nielsen2010quantum}. For instance, starting from state $\q{+}\q{00...0}$, with $\q{+} = (\q{0}+\q{1})/\sqrt{2}$ on the first qubit, and applying a chain of CNOT gates acting on qubits (1,2), then (2,3),... generates the Greenberger-Horne-Zeilinger state (GHZ, \cite{greenberger1990bell}) on all qubits: 
\begin{equation}\label{eq:GHZstate}
\q{GHZ_+} = (\q{00...0} + \q{11...1}) \, / \, \sqrt{2}\; .
\end{equation}
Along similar lines, applying Control-phase gates generates ``cluster states'' as used in measurement-based quantum computing for instance \cite{yu2011deterministic}. These approaches require fast and robust gates, to dominate decoherence effects without inducing imprecisions during state preparation, and after perfect initialization of the individual qubits. Another approach is to use an ancilla, which interacts consecutively with each subsystem or which could be emitted by any of several subsystems \cite{feng2019chip,zheng1997generation}, such that the system is projected onto the target multi-partite entangled state if the ancilla measurement gives a particular outcome. Such ``heralded preparation'' can be complemented with feedback actions. It could be used as an additional error correction procedure after preparation with a gate sequence, but most interestingly it allows entanglement between distant (i.e.~not directly interacting) subsystems. Entanglement distillation enables to further improve entanglement fidelity, by combining several lower-quality copies of the target state into one higher-quality instance \cite{bennett1996purification}. All these approaches involve sequential procedures for generating an entangled state at a given time.

In contrast, so-called ``reservoir engineering'' aims to stabilize the target state with well-designed fixed dissipative dynamics \cite{poyatos1996quantum}. The system thus corrects for perturbations autonomously without assuming any particular starting point or explicit measurement. The state stabilized by such reservoir engineering might possibly serve as high-quality input for further entanglement purification approaches. This fully passive setting is the topic of the present paper. The authors of \cite{ticozzi2014steady}, while building a more general toolset, have initiated the investigation of GHZ state stabilization through reservoir engineering and a method to stabilize a GHZ state over $n \leq 3$ qubits. In more recent work, \cite{cervera2022experimental} describe another way to stabilize the state $\q{GHZ_{3+}} = (\q{000} + \q{111}+\q{222}) \, / \, \sqrt{3}\;$ on three \emph{qutrits}. However, beyond $n=3$, the first message is an impossibility result. In particular, \cite{ticozzi2014steady} has established the impossibility to stabilize a GHZ state on $n$ qubits, if no quasi-local dissipation operator spans more than $n/2$ qubits. In follow-up work, they propose a particular construction to approximately stabilize the $n=4$ case \cite{johnson2016general}. Importantly for the present work, \cite{ticozzi2014steady} also proposes a conditional stabilization scheme, where the GHZ state is stabilized provided each qubit is initialized in the particular state $\q{+}$; this approach is thus (slightly) sequential again, and if phase-flip perturbations $\q{\pm} \leftrightarrow \q{\mp}$ act on any qubits any time after initialization in $\q{++...+}$, the system will not recover. Other engineered reservoirs have been proposed for stabilizing a submanifold of highly \emph{correlated} states like span$\{ \q{00...0}$, $\q{11...1}\}$, while the nonlocal quantum phase ``+'' in $(\q{00...0} + \q{11...1}) \, / \, \sqrt{2}$ remains unprotected \cite{zapletal2022stabilization}; this is essentially a more concrete proposal of what the conditional stabilizer of \cite{ticozzi2014steady} is doing.
Given the impossibility result of \cite{ticozzi2014steady}, the best global stabilization we can hope to achieve with fixed, quasi-local reservoir engineering is of \emph{approximately} a GHZ state. 

The present paper proposes concrete ways to globally stabilize an approximate GHZ state with reservoir engineering using fixed, quasi-local operators only. The main idea is to apply the two steps of the sequential stabilization proposed in \cite{ticozzi2014steady} ``alternatively'': most of the time, the qubits apply the conditional GHZ stabilizers of \cite{ticozzi2014steady}; but occasionally, they all reset back to $\q{+}$. Applying these resets at random times, the reservoir ``average'' dynamics, which is all that matters for quantum predictions, involves a mixture between recently reset states (thus, far from target) and those which have evolved with the GHZ stabilizers for some time. If resets are scarce enough, then the average state should be close enough to the target $\q{GHZ_+}$. The tradeoff is of course that, with fewer resets, the system will recover more slowly from general perturbations. One main point of the paper is thus to analyze this tradeoff and its scaling with the number of qubits. Another main point is to design a mechanism for \emph{synchronized} reset of all the qubits, without which the scheme would not work. We propose two types of architecture to achieve this synchronization with \emph{local interactions} only, by which we mean each subsystem interacting with a few neighbors only, and each interaction operator involving few (ideally two) subsystems.

Our first architecture, described in Section \ref{sec:ancillas}, builds on a modified version of the clock ancilla used in the dissipative computing proposal of \cite{Verstraete2009} and related papers. Their clock ancilla is directly coupled to all the data qubits and their target outcome is heralded by a particular state of the unitarily evolving clock. We propose a more local and non-heralded scheme. Regarding locality, we associate one ancilla `clock' subsystem to each data qubit, letting the `clock' subsystems interact with each other and with their respective data qubit, all with local operators only. Regarding non-heralding, we modify the Hamiltonian-based symmetric clock evolution of \cite{Verstraete2009}, into a cyclic clock evolution under jump operators, and adjust the jump rates to have a larger population on the target state. The clock synchronization only requires classical correlation of ancillas. Its efficiency depends on the time-scale separations that are achievable among various components of the dynamics. We present a few alternatives in this direction, including a wave propagating through the qubit chain. Our second architecture, presented in Section \ref{sec:3level}, proposes to replace the data qubits by qutrits, using the additional level to operate the resets' synchronization. The target state is still specified by \eqref{eq:GHZstate}, with timescale separation ensuring that only a small fraction of the state has leaked to levels $\q{2}$ of the qutrits. These constructions are meant to revive interest in developing local Lindbladian automata whose dynamics protect valuable quantum information states, see also \cite{herold2015cellular}.

In Section \ref{sec:analysis1} and Section \ref{sec:analysis2}, we characterize the performance of those schemes, both in simulations and through approximate analysis methods. The aim is to compute the steady-state fidelity of the reservoir in presence of general perturbations. Our approximate analysis is based on timescale separations and classical Markov chain models. For this we develop an original method to translate our Lindbladian dissipative system into classical Markov chains over hypothetical output signals. This method may be of independent interest, see e.g.~Section \ref{ssec:MC1definition}.


\section{Problem description}\label{sec:description}

The original setting involves a chain of $n$ (data) qubits which have to be stabilized in the entangled superposition \eqref{eq:GHZstate}. Our proposals rely on enlarging the full Hilbert space, either by using additional levels on the data subsystems, or by adding ancillary subsystems to the setting. We introduce these novelties from the start in our system description. We keep the spatial arrangement of a chain and require local interactions along this chain.

Consider thus a chain of $n$ (data) subsystems, each of finite dimension $Q$, to which we adjoin a chain of $m$ ancillas, each of dimension $D$. The full Hilbert space $\mathcal{H}$ is thus of dimension $Q^n D^m$ and we want a procedure that works for arbitrary $n$.
\begin{itemize}
\item Our first type of proposal relies on data qubits ($Q=2$) and associates typically one ancilla per data qubit ($m=n$) or per pair of neighboring data qubits ($m=n-1$). Ancillas have dimension $D \in \{3,4\}$.
\item Our second type of proposal relies on data qutrits ($Q=3$) and requires no ancillas ($m=0$).
\end{itemize}
We denote by $\q{0},\q{1},\q{2}$ the canonical states of the data subsystems, while we use letters $\q{g},\q{e},\q{m},\q{f},...$ for canonical states of the ancillas. We use an index when appropriate to label which ancilla and/or data subsystem is meant.

We let this system evolve according to a time-independent Lindblad equation of the form:
\begin{align} \nonumber
\tfrac{d}{dt} \rho_t =& -i[ H,\; \rho_t] \; + \; \sum_{k=1}^{K_1} \; L_k \rho_t L_k^\dagger - \tfrac{1}{2}\, (\; L_k^\dagger L_k \rho_t + \rho_t L_k^\dagger L_k  \;) \; + \; \sum_{k=1}^{K_2} \; M_k \rho_t M_k^\dagger - \tfrac{1}{2}\, (\; M_k^\dagger M_k \rho_t + \rho_t M_k^\dagger M_k  \;)  \; \\ \label{eq:L0}
& \; + \sum_{k=1}^{K_3} \; N_k \rho_t N_k^\dagger - \tfrac{1}{2}\, (\; N_k^\dagger N_k \rho_t + \rho_t N_k^\dagger N_k  \;)    \; .
\end{align}
The distinction of operators $L_k$, $M_k$ and $N_k$ for the dissipation channels is to facilitate later discussion. The objective is to design constant operators $H$ and $L_k,M_k,N_k$ such that any initial state on $\mathcal{H}$ converges towards the so-called Greenberger-Horne-Zeilinger (GHZ) state \eqref{eq:GHZstate}. It is important to insist on the objective followed throughout this paper: the dynamics follows a \emph{time-independent} Lindblad equation, with constant operators, and the state of interest is the \emph{steady state} $\bar{\rho}$ reached with this Lindbladian when tracing over the ancilla degrees of freedom (global and unconditional asymptotic stabilization). In particular, if the steady state is not unique and there exist initial states from which stabilization of $\q{GHZ_+}$ fails, then we consider that the scheme is not working. On the other hand, when an engineered reservoir does stabilize a unique steady state, we use fidelity $\mathcal{F}(\bar{\rho}) = \qd{GHZ_+}\, \bar{\rho} \,\q{GHZ_+}$ as a measure of closeness to our objective. To evaluate the protective power of the engineered reservoir, we will consider the case where on top of \eqref{eq:L0} each subsystem is subject to perturbation channels.

The essential constraint for reservoir design is that each term in $H$ and each of the operators  $L_k$, $M_k$ and $N_k$ must be quasi-local, namely~each one of them must act like the identity on all of the Hilbert space except a few subsystems; furthermore, we require that these subsystems are neighbors according to the physical layout of a chain. Thus each data or ancilla subsystem can only be coupled to a small and fixed number of neighbors, independent of $n$. The catch is that \cite{ticozzi2014steady} has proved the impossibility to globally asymptotically stabilize a GHZ state with such fixed Lindbladian reservoir on $n>3$ qubits, when $m=0$ and $Q=2$. In fact, their general theorem also covers the case with $Q>2$, and the impossibility result  remains true in presence of additional ancilla subsystems: see our Appendix. Therefore, our objective is to stabilize a state $\bar{\rho}$ which is a good approximation of  $\q{GHZ_+} \qd{GHZ_+}$.

In addition to the impossibility result, the authors in \cite{ticozzi2014steady} also notice a particular procedure to generate the GHZ state: first stabilize each qubit individually towards $\q{+} = (\q{0}+\q{1})/\sqrt{2}$, then apply the dissipation channels
\begin{equation}\label{eq:LTV}
L_k = \sqrt{\kappa_c}(\q{11}\qd{10} + \q{00}\qd{01})  \text{ on qubits } (k,k+1) \; , \quad   \text{ for }  k=1,2,...,n-1 \; .
\end{equation}
This $L_k$ asymptotically sets qubit $k+1$ to the same bit-value as qubit $k$, coherently for both possible bit-values and at a rate $\kappa_c$. The aim of the present paper is to investigate how to combine these two steps into a single time-invariant Lindbladian, still with local interactions only, and whose steady state $\bar{\rho}$ would be close to the target GHZ state. Concretely, we propose several stabilization procedures based on the same simple idea: at each time step $t$ the data subsystems are all reset to $\q{+}$ with a small probability, and from there they have a high probability to keep applying just the dissipation operators $L_k$ of \eqref{eq:LTV}, hence approaching $\q{GHZ_+}$. When resets are scarcer, the resulting Lindbladian's steady state $\bar{\rho}$ would get closer to $\q{GHZ_+}$, but recovery from a phase-flip error gets slower; at the limit of infinitesimal reset rate, the system would feature fast convergence towards a subspace infinitesimally close to $\text{span}\{\q{00..0},\,\q{11..1} \}$, and infinitesimally slow convergence towards $\q{GHZ_+}$ within this subspace.

It is essential though for this idea, that the reset to $\q{+}$ takes place synchronously, and only synchronously, on all the data subsystems. The first reason is that a reset pulls the state away from $\q{GHZ_+}$, so we want to minimize the fraction of time doing resets. In this sense, it is more efficient to correct potential phase errors on \emph{all} qubits at every reset round. The second and more important reason is that even when starting on $\q{GHZ_+}$, when a single data subsystem undergoes a reset to $\q{+}$ and we let the system converge back with $\eqref{eq:LTV}$, the state will \emph{not} converge to $\q{GHZ_+}$. In other words, every reset round involving some but not all data subsystems, would not only be useless but even deteriorate the fidelity \emph{until the next all-data-subsystems reset round}. To make synchronous resets (significantly) more probable than partial ones, a dedicated synchronization procedure is needed. In our proposals, the enlarged Hilbert space serves the essential role of implementing this \emph{synchronous} reset of all qubits to $\q{+}$. In the following sections, we describe and analyze different ways to obtain engineered reservoirs from this principle, first with $Q=2$ and a chain of ancillas (Section \ref{sec:ancillas}), then without ancillas but exploiting a third level on each data subsystem  (Section \ref{sec:3level}).\\

\noindent \emph{Remark 1:} The reader could notice that compared to \cite{ticozzi2014steady}, we target \emph{approximate} stabilization \emph{and} we enlarge the Hilbert space. In \cite{johnson2016general} the authors propose a way to approximately stabilize a GHZ state on $n=4$ qubits. The present work provides no particular ideas for efficiently stabilizing an \emph{approximate} GHZ state of arbitrary $n$ using the Hilbert space of the \emph{data qubits only}; this possibility remains open.


\begin{figure}
\begin{center}
\includegraphics[width=150mm, trim=55 160 80 125, clip]{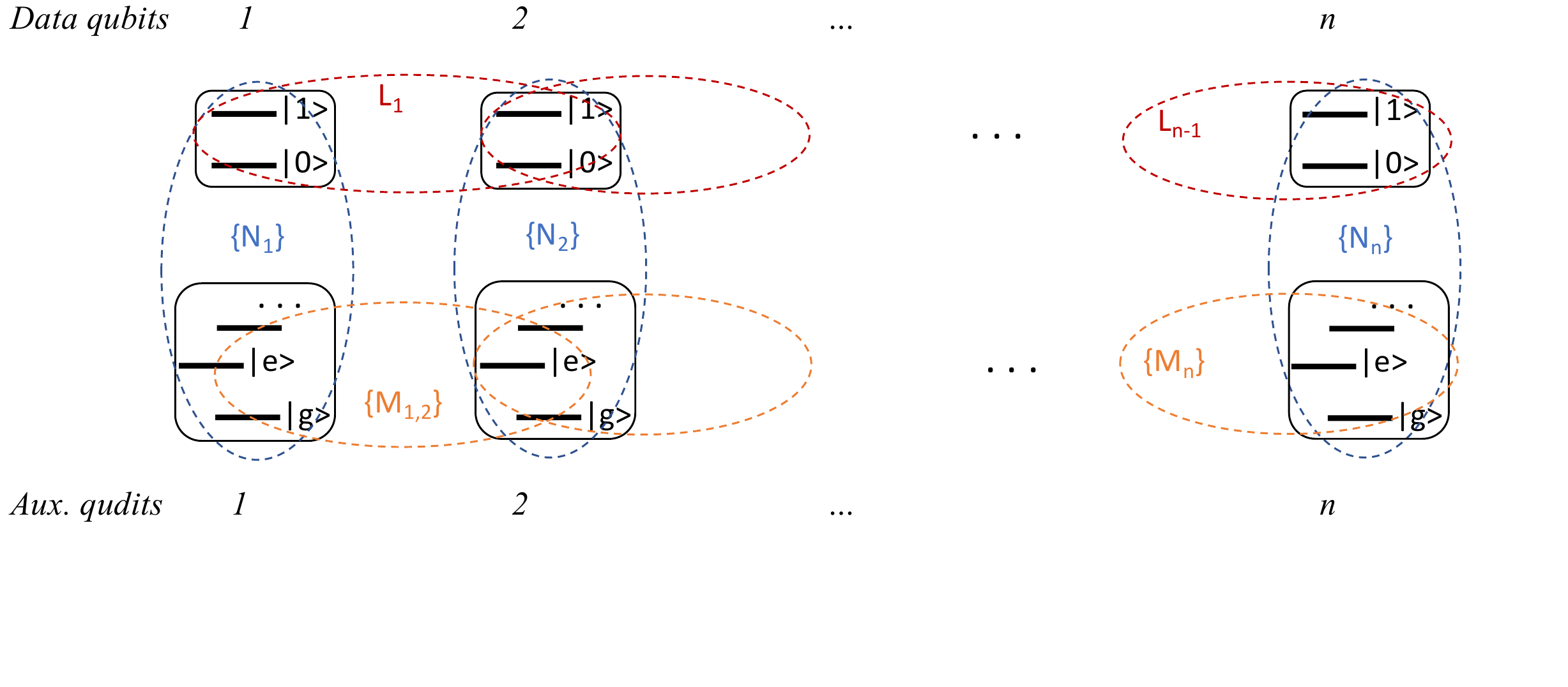}
\includegraphics[width=150mm, trim=70 160 80 130, clip]{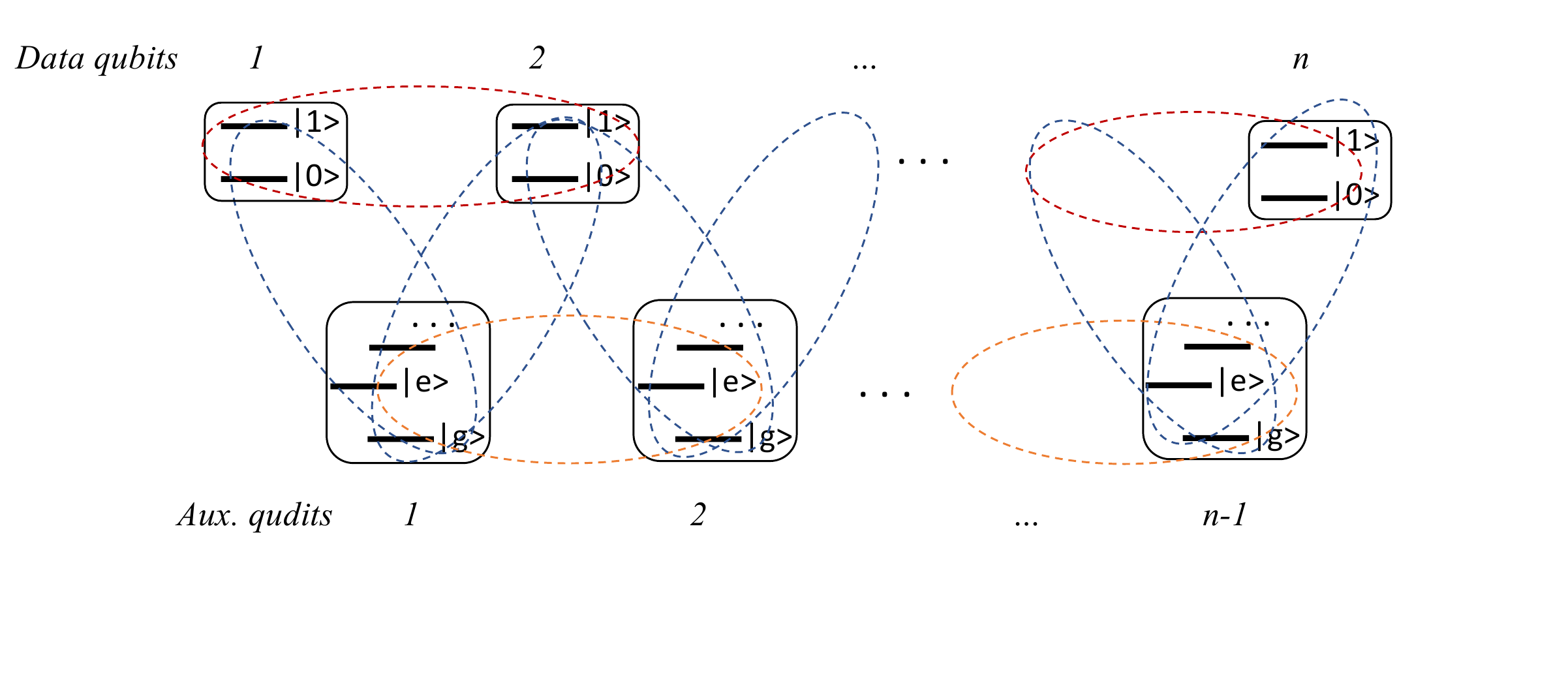}
\end{center}
\caption{General architecture of our approximate GHZ reservoir with ``clock-ancillas''. An auxiliary ``ancilla'' quDit is associated to each data qubit (top), or to each pair of adjacent data qubits (bottom). Only neighboring qubits or quDits can interact through dissipation operators, according to a double-chain topology (dotted ellipses, labeled with corresponding operators). The data qubits interact according to dissipation operators \eqref{eq:LTV} proposed by \cite{ticozzi2014steady}; taken alone, this would stabilize them in the manifold $\mathcal{H}_s := \text{span}\{\;\q{00...0},\,\q{11...1}\;\}$. The mechanism with auxiliary quDits is meant to softly reset the data into an initial state from which they converge towards the $+$ superposition of the basis states in $\mathcal{H}_s$ ; see Figure \ref{fig:clockwork} for more details.}\label{fig:chains}
\end{figure}

\section{Architectures with a ``clock'' of ancilla quDits}\label{sec:ancillas}

In this type of architecture, the system consists of the chain of $n$ data qubits (Q=2), to which we adjoin a chain of ancilla quDits, see Figure \ref{fig:chains}. In \eqref{eq:L0}, the dissipation operators
\begin{itemize}
\item $L_k$ will always be related to stabilizing the subspace $\text{span}\{\q{00...0},\q{11...1}\}$  like in \eqref{eq:LTV},
\item $M_k$ will govern the evolution of the ancillas, and  
\item $N_k$ are used to reset data qubits to $\q{+}$ conditional on ancillas.
\end{itemize}
Our proposals take Hamiltonian $H=0$. One may revise this choice, e.g.~to combine some dissipation operators into a single coherent superposition while killing its dark states with $H$. In the most constraining setting, we only allow bipartite interactions: each $L_k$ can act nontrivially only on two neighboring data qubits $k$ and $k+1$; each $M_k$ on two neighboring ancilla quDits $k$ and $k+1$; and each $N_k$ on data qubit $k$ and ancilla quDit $k$. We will also discuss proposals with slightly different constraints, e.g.~allowing tripartite interaction. The interactions will always be reduced to direct neighbors according to the double-chain topology (dotted ellipses on Fig.~\ref{fig:chains}).

The main text is meant to progressively introduce the main ideas. Details about variations can be found in appendix.


\subsection{Approximate GHZ reservoir through ancilla state conditioning}\label{ssec:statecond}

\subsubsection{Assuming correlated ancilla evolution}\label{sssec:sc:step0}

As a preliminary discussion, let us start with relaxing the locality constraint on the $M_k$ and assume that we have an operator implementing synchronous jumps of all the ancillas. 
 To further simplify, assume that the ancillas are confined to the space  $\text{span}\{\q{gg...g},\q{ee...e}\}$ thanks to some (not further specified) mechanism, and consider the ancilla dissipators:
\begin{equation}\label{eq:step0Mks}
M_{1} = \sqrt{\kappa_u} \; \q{ee...e}\qd{gg...g}  \quad , \quad  M_{2} = \sqrt{\kappa_d} \; \q{gg...g}\qd{ee...e} \; ,\footnote{In fact, thanks to both the departure and arrival states being orthogonal, we may in principle replace \eqref{eq:step0Mks} by a single operator $M = M_1 + M_2$; it is doubtful whether this would be easier to implement.}
\end{equation}
with one ancilla $k$ per data qubit $k$ and jump rates $\kappa_u \ll \kappa_d$. To finalize this preliminary construction, each data qubit keeps applying fast local reset dynamics conditioned on its ancilla being in $\q{e}$: 
\begin{equation}\label{eq:step0Nks}
 N_{k} = \sqrt{\kappa_r} \q{e,+}\qd{e,-}_{k,k} \; \quad \text{ for } k=1,2,...,\, n \; .
\end{equation}
The idea of this scheme is that the data qubits are continuously applying the $L_k$ from \eqref{eq:LTV}; but occasionally the ancillas all jump to $\q{e}$ for a short time, during which this triggers resets of each data qubit to $\q{+}$ as dominating dynamics.

Several relevant observations can already be made with the preliminary system \eqref{eq:LTV},\eqref{eq:step0Mks},\eqref{eq:step0Nks}.
\begin{itemize}
\item  The scheme relies on selecting the time scales as follows. Since reset dynamics \eqref{eq:step0Nks} has to overpower \eqref{eq:LTV} (which is always left on, see third item), we need $\kappa_r \gg \kappa_c$. To favor all-qubit resets before ancillas jump down from $\q{ee...e}$, we also mean to take $\kappa_r \gg \kappa_d$. We nevertheless want to take $\kappa_d$ rather large, to avoid spending unnecessary time doing resets to $\q{+}$ which move the state away from $\q{GHZ_+}$. Finally, we need $\kappa_c \gg \kappa_u$ to leave enough time for re-convergence towards $\q{GHZ_+}$ before the next reset. The rate of protection against general perturbations is then set by the slowest rate i.e.~$\kappa_u$. At first sight there is no clear scaling request between $\kappa_d$ and $\kappa_c$, some optimal tuning should be sought. More analysis is provided in Section \ref{sec:analysis1}.
%
%
%
%
%
%
%
 \item Instead of applying data reset conditioned on an ancilla \emph{being in} state $\q{e}$, one could apply data reset conditioned on the ancilla \emph{jumping to} state $\q{e}$.
 \item The proposed construction switches on and off the single-qubit reset through \eqref{eq:step0Nks}, but it leaves the $L_k$ always on. Instead of dominating the $L_k$ by the $N_k$ when ancillas are in $\q{e}$, one could consider to switch off the $L_k$. Because each $L_k$ already involves two data qubits, unlike the reset operator $\q{+}\qd{-}$, conditional switching of the $L_k$ is somewhat harder and will be considered later in the paper.
 \item There is no need to protect the phase of the ancillas in the canonical basis: the ancillas' only role is to establish a \emph{classical correlation} between all qubits resetting. Importantly, this insensitivity to ancilla phase errors will remain true for the other reservoir constructions below, as proved in Appendix. Thus, we do not really need ancilla quDits: we only need classical Dits, or quDits with heavily biased noise protection i.e.~generalizing the qubits with biased noise where bit-flips are heavily suppressed while phase-flips remain rather common \cite{mirrahimi2014dynamically}. For this reason, we can consider that the ancilla populations remain stable at much longer timescales than the data qubits. (Note that the clock of \cite{Verstraete2009} also has this property, although it is not mentioned.)
\item Reasoning about this system is facilitated by the fact that ancilla dynamics is not influenced by the data states. We will make sure to maintain this property in our other constructions.
\end{itemize}

The remaining, major task is to replace \eqref{eq:step0Mks} by Lindblad dynamics with quasi-local operators, achieving essentially the same effect. In a different context, \cite{Verstraete2009} has used a single ``timer'' ancilla, assuming that it is coupled individually to each data qubit. This would mean, in the setup just described, to use the same unique ancilla $\q{e}$ in each of the $N_k$. Such operators would still be bipartite only; and tripartite only in the context of \cite{Verstraete2009} or if they were used to switch our $L_k$ on and off. However, it would require a single ancilla to be physically connected to all the $n$ data qubits. With the following synchronization mechanism using $n$ ancillas, we cover the more scalable setting where each subsystem is connected only to a few neighbors in a double-chain layout.\footnote{Note that even with a unique timer ancilla connected to each data qubit, \emph{exact} GHZ stabilization for large $n$ remains impossible. Indeed, since each dissipation operator would involve bi- or tripartite interactions only, our slightly generalized version of the \cite{ticozzi2014steady} no-go still applies.}

\begin{figure}
	\begin{center}
\includegraphics[width=150mm, trim=15 160 80 25, clip]{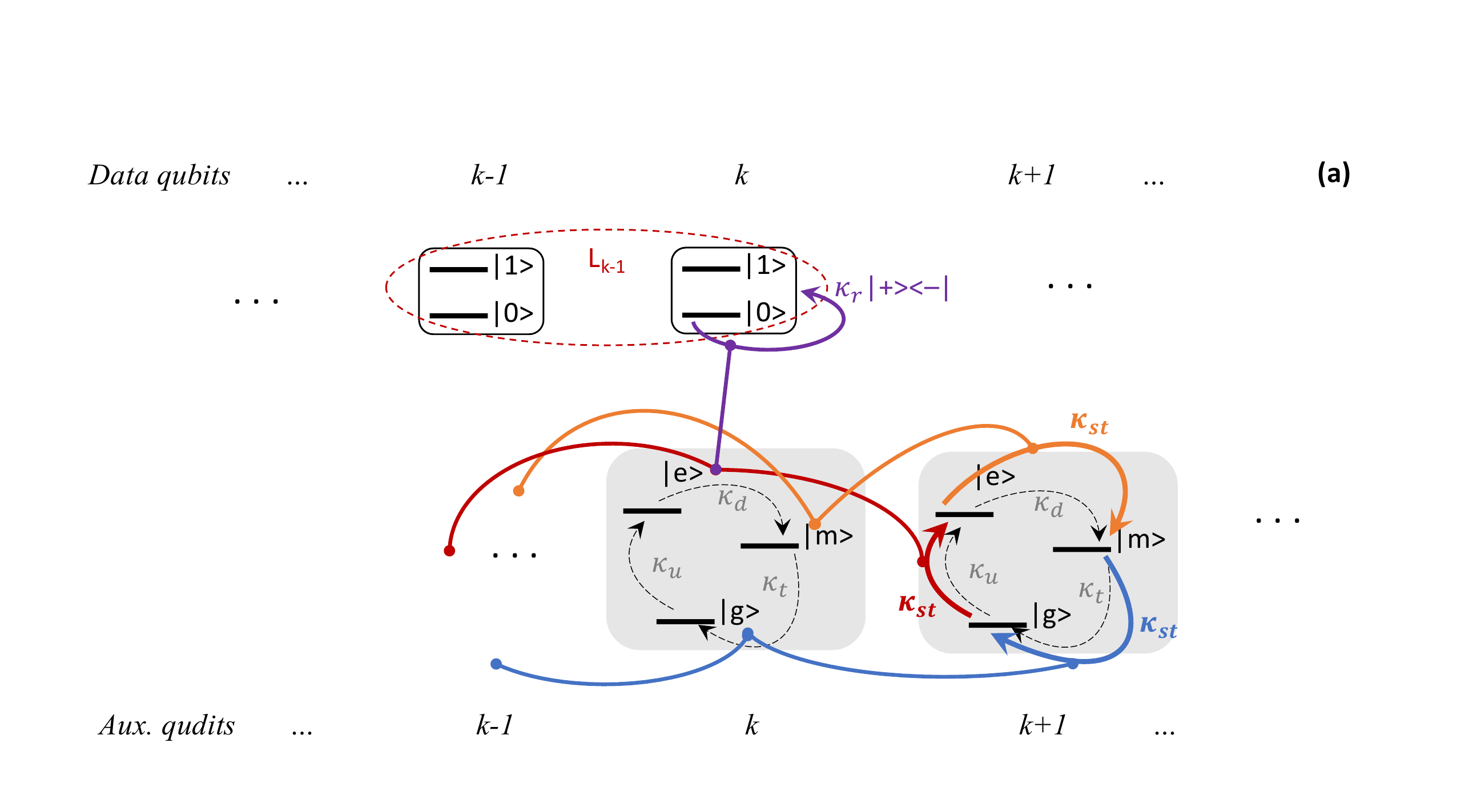}
\includegraphics[width=150mm, trim=10 160 80 20, clip]{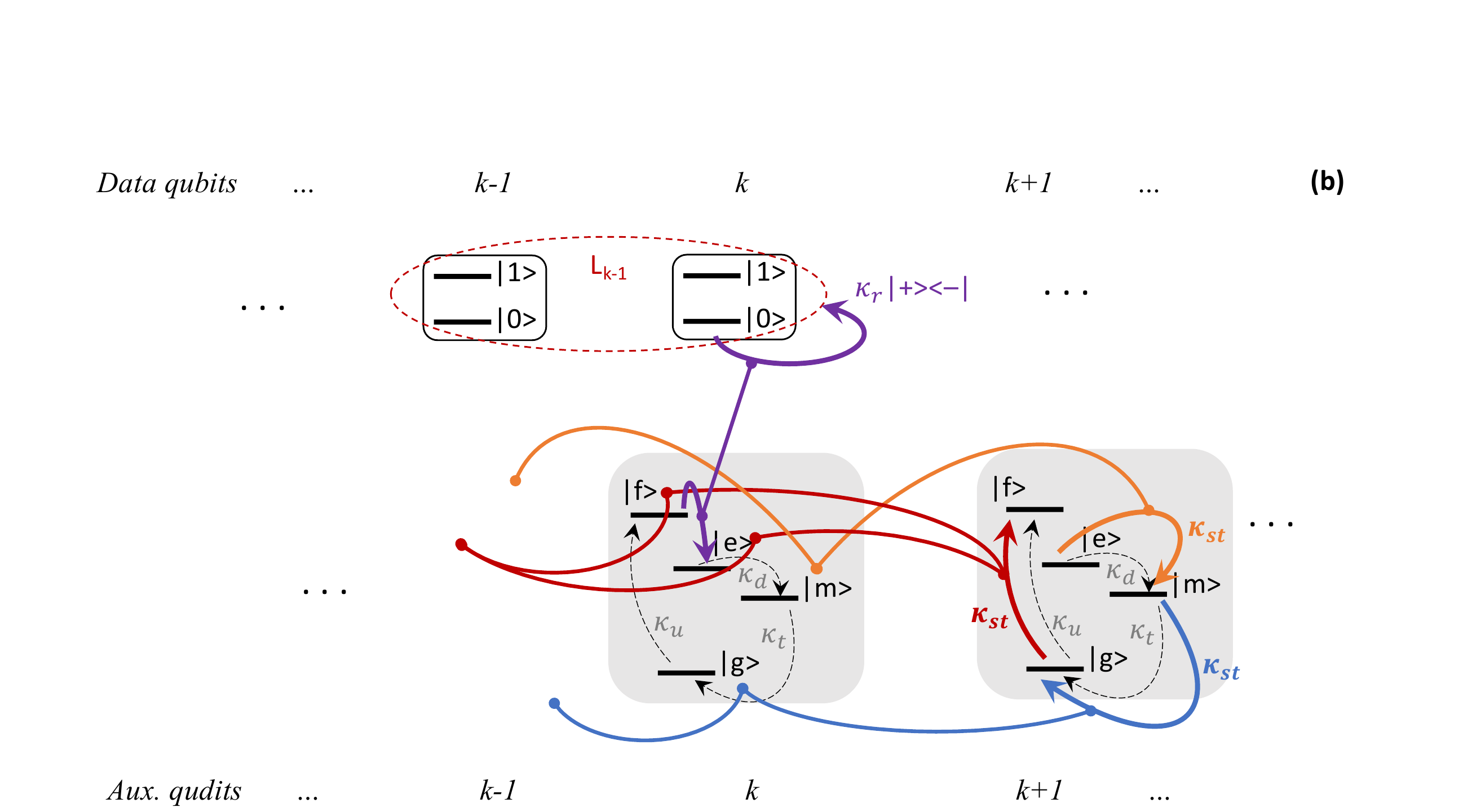}
	\end{center}	
	\caption{Architecture of the correlated ancillas clock inducing synchronized resets of the data qubits to $\q{+}$. 
		\textbf{(a)} Reset with state conditioning (Section \ref{ssec:statecond}):  each data qubit applies a fast reset channel ($\kappa_r$) conditioned on its ancilla being in state $\q{e}$. The spontaneous ($\kappa_u, \kappa_d, \kappa_t$) and neighbor-stimulated ($\kappa_{st}$) ancilla jumps are meant to approximately induce a well-synchronized cycle $\q{gg...g } \rightarrow \q{ee...e} \rightarrow \q{mm...m} \rightarrow \q{gg...g}  \rightarrow ... \; ,$ with random transition times but little time spent on $\q{ee...e}$. Each channel only involves pairwise interactions between neighboring subsystems. To avoid clutter, interactions are shown only for a small part of the system. 
		\textbf{(b)} Reset with jump conditioning (Section \ref{ssec:jumpcond}): each data qubit applies a fast reset when its ancilla jumps to $\q{e}$. In this architecture involving only pairwise interactions, the ancilla jumps are meant to approximately induce a well-synchronized cycle $\q{gg...g } \rightarrow \q{\zeta_1\zeta_2...\zeta_N} \rightarrow \q{mm...m} \rightarrow \q{gg...g}  \rightarrow ... $ where each $\zeta_i \in \{e,f\}$. The additional level $\q{f}$ could be dropped if tripartite interactions among neighbors were allowed.}\label{fig:clockwork}
\end{figure}

\subsubsection{Constructing a correlated ancillas clock}\label{sssec:sc:step1}

Our proposal for replacing \eqref{eq:step0Mks} by dynamics involving local interactions on a chain requires the ancillas to be qutrits, with levels $\q{g},\q{e}$ and $\q{m}$. The idea (see Figure \ref{fig:clockwork}.a) is that each ancilla has a relatively small probability to change its state \emph{spontaneously} towards the ``next'' state in a cyclic way $\q{g} \rightarrow \q{e} \rightarrow \q{m} \rightarrow \q{g}  \rightarrow  ...$, but a relatively large probability to get \emph{stimulated} to the ``next'' state if a neighboring ancilla has this value. The qutrit structure is necessary to introduce directionality in this cycle. Indeed, if the mechanism was implemented with ancilla \emph{qubits}, then an ancilla that has jumped spontaneously say from $\q{g}$ to $\q{e}$ would be attracted back towards $\q{g}$ by its neighbors at the same time as attracting them towards $\q{e}$; on a chain of ancillas, the boundary between ancillas in the $\q{g}$ and $\q{e}$ states would thus move either way at the same rate, suggesting that it would be hard to synchronize the whole chain. In contrast, with the \emph{qutrit} structure, if all ancillas are in $\q{g}$ and one of them jumps to $\q{e}$, then this ancilla has only very low probability to spontaneously jump to the next state $\q{m}$, while the neighbor ancillas have a high probability to join it on level $\q{e}$, attracting in turn their own neighbors, and so on; thus, the ancillas would essentially follow the cycle
\begin{equation}\label{eq:step1cycle}
\q{gg...g } \rightarrow \q{ee...e} \rightarrow \q{mm...m} \rightarrow \q{gg...g}  \rightarrow ... 
\end{equation}
with very little time spent on other states, if the stimulated jump is sufficiently dominating.

With subsystems arranged as a chain, the corresponding jump operators could be:
\begin{eqnarray}\label{eq:step1synch}
M_{k,sp} &=& \sqrt{\kappa_u} \q{e}\qd{g}_k + \sqrt{\kappa_d} \q{m}\qd{e}_k +  \sqrt{\kappa_t} \q{g}\qd{m}_k \\ \nonumber
M_{k,st+} &=& \sqrt{\kappa_{st}}\, ( \q{ee}\qd{ge} +  \q{mm}\qd{em} + \q{gg}\qd{mg}  )_{k,k+1} \\ \nonumber 
M_{k,st-} &=& \sqrt{\kappa_{st}}\, ( \q{ee}\qd{eg} +  \q{mm}\qd{me} + \q{gg}\qd{gm}  )_{k-1,k}
\end{eqnarray}
for each $k=2,3,...,n-1$, and one of the two last channels dropping for ancillas $k=1$ and $k=n$. The indices $_{sp}$ or $_{st}$ distinguish spontaneous or neighbor-stimulated processes, while $_+$ or $_-$ indicate stimulation by the left or right neighbor. The structure \eqref{eq:step1synch} is just one proposal and admits several degrees of freedom which seem general for such clock-systems.
\begin{itemize}
	\item Like for Section \ref{sssec:sc:step0}, we assume that the classical bit value of the ancillas is protected efficiently. This is because quantum phases among canonical states of the ancillas play no role, as ancillas only need to preserve classical correlations; see details in Appendix. This also underlies the following points.
	\item We have written each dissipation operator as a coherent sum of three terms; $M_{k,sp}$ has full rank and thus features no spurious dark states, similarly for the other operators. In principle one could also take e.g. $M_{k,st} = M_{k,st+} + M_{k+1,st-}$, with an additional Hamiltonian to avoid a dark state associated to e.g.~$M_{k,st} (\q{ge}-\q{eg})_{k,k+1} = 0$. This would yield a lower number of dissipation channels, but usually engineering a single coherent dissipator is harder than engineering them separately. Conversely, it is equally valid to split each jump operator e.g.~$M_{k,sp}$ into three separate jump operators. 
	\item It is not essential at all to have the same rates $\kappa_{...}$ for each $k$ for instance. Indeed, synchronized working only requires that the various transitions summarized under $\kappa_{st}$ happen at a very fast rate compared to the others. The average time to perform one cycle  \eqref{eq:step1cycle} is then in first approximation given by 
$(\frac{1}{\sum_k \kappa_{u,k}} + \frac{1}{\sum_k \kappa_{d,k}} + \frac{1}{\sum_k \kappa_{t,k}})$, since each ancilla has a probability to spontaneously launch the transition to the next clock state.
	\item Note that the channels \eqref{eq:step1synch} ensure the target behavior for the whole ancilla Hilbert space, not assuming initialization in a suited subspace as was done in the sentence before \eqref{eq:step0Mks}.
\end{itemize}

The engineered reservoir would thus combine \eqref{eq:step1synch} with the conditional reset \eqref{eq:step0Nks} and the subspace stabilization channels \eqref{eq:LTV}. The selection of the various rates must ensure, roughly:
\begin{enumerate}
	\item ancillas behave as an almost synchronized clock: $\kappa_{st} \gg \kappa_d, \kappa_t, \kappa_u$ 
	\item data qubits do a reset with high probability before ancillas leave $\q{e}$:  $\kappa_r \gg  \kappa_d$	
	\item reset dynamics dominates \eqref{eq:LTV} when both are applied together: $\kappa_r \gg \kappa_c$
	\item resets stop at a well synchronized time with respect to the evolution  \eqref{eq:LTV} : $\kappa_{st} \gg \kappa_c$.	
	\item data qubits have ample time to converge with \eqref{eq:LTV} after each reset round: $\frac{1}{\kappa_c} \ll (\frac{1}{\kappa_t}+\frac{1}{\kappa_u})$
	\item reset periods take up a small fraction of cycle time: $\frac{1}{\kappa_d} \ll (\frac{1}{\kappa_t}+\frac{1}{\kappa_u})$.
\end{enumerate}
Altogether, this suggests the timing guidelines:
\begin{eqnarray}\label{eq:step2atc}
&& \left\{ \frac{1}{\kappa_t} \;,\; \frac{1}{\kappa_u} \right\} \sim T_1  \quad \gg \quad  \left\{ \frac{1}{\kappa_c} \;,\; \frac{1}{\kappa_d} \right\} \sim T_2 \quad \gg  \quad \left\{ \frac{1}{\kappa_r}  \; , \; \frac{1}{\kappa_{st}} \right\} \sim T_3  \; .
\end{eqnarray}
By construction, the scheme gives a limited fidelity even in absence of external perturbations. Fidelity lost due to resets pushing the state away from $\q{GHZ_+}$ can be roughly estimated as the typical ``away from GHZ'' portion of a full cycle and should thus be of order $(\tfrac{1}{\kappa_d}+\tfrac{1}{\kappa_c}) \;/\; (\tfrac{1}{\kappa_d}+\tfrac{1}{\kappa_t}+\tfrac{1}{\kappa_u}) \sim T_2/T_1$. Inaccuracy in resetting to $\q{++...+}$ should add errors of order $(\frac{\kappa_c}{\kappa_r}+\frac{\kappa_c}{\kappa_{st}}+\frac{\kappa_d}{\kappa_r}) \sim T_3/T_2$. In turn, arbitrary external perturbations with characteristic time $T_0$ will, at worst, be rejected according to the slowest reservoir timescale and thus induce errors of order $T_1/T_0$.  At fixed extremal values $T_3$ and $T_0$, the tradeoff between these error contributions will fix the optimal values of $T_1$ and $T_2$. Since an improvement by a factor $C$ on $T_3/T_0$ will have to be factored into three timescale separations, we may expect the error to only improve by $C^{1/3}$. A finer performance analysis, with the dependence on $n$, is provided in Section \ref{sec:analysis1}.\vspace{3mm}

\noindent \emph{Remark 2:} It may be worth noting that the two fastest rates, involved in $T_3$, are of a very different nature: while $\kappa_r$ involves a quantum jump conditioned on an ancilla value, $\kappa_{st}$ only involves classical synchronization. In this sense, the constraint of fast $\kappa_r$ can be considered as harder to achieve.

\subsubsection{Possibility to switch the GHZ stabilizers}\label{sssec:sc:step2}

In the above setup, one might wonder if instead of dominating  \eqref{eq:LTV} with the reset dynamics when the ancilla is in $\q{e}$, one could not switch \eqref{eq:LTV}  off conditioned on ancilla states. This would enable to drop the requirement $\kappa_r \gg \kappa_c$, and by Remark 2, it would possibly enable higher values of $\kappa_c$ to push the state towards $\q{GHZ_+}$. We next mention two such constructions with the ancilla-clock architecture, to highlight the associated issues. More efficient constructions are presented further below, in particular using data qutrits.

\paragraph*{Tripartite:} A first possibility would be to admit tripartite interactions, thus directly conditioning each $L_k$ on the state of an associated ancilla. In this way, each ancilla $k$ would be associated to a \emph{pair of adjacent data qubits} $(k,k+1)$, unlike in the previous scheme. This choice also works for the $N_k$ operators. Indeed, we can strictly exclude the possibility to apply $L_k$ and $N_k$ simultaneously on the same data qubit with the following tripartite interactions:
\begin{eqnarray*}
\tilde{L}_k &=& (\q{g}\qd{g}+\q{m}\qd{m})_k \otimes L_k \quad \text{ for } k=1,2,...,n-1 \; ,\\
\tilde{N}_k &=& \sqrt{\kappa_r} \q{ee}\qd{ee}_{k-1,k} \otimes \q{+}\qd{-}_k  \quad \text{ for } k=2,3,...,n-1 \; ,\\
&& \tilde{N}_1 = \sqrt{\kappa_r} \q{e}\qd{e}_{1} \otimes \q{+}\qd{-}_1 \;\; , \;\; \tilde{N}_n =  \sqrt{\kappa_r} \q{e}\qd{e}_{n-1} \otimes \q{+}\qd{-}_n \; ,
\end{eqnarray*}
with $M_k$ as in \eqref{eq:step1synch}, $L_k$ as in \eqref{eq:LTV}. The fidelity lost due to resets pushing the state away from $\q{GHZ_+}$ is still of order $(\frac{1}{\kappa_c}+\frac{1}{\kappa_d})/T_1$. However, the advantage could be that, as we drop the constraint $\kappa_c \ll \kappa_r$, we can take larger $\kappa_c$ such that the error is dominated by just $(\tfrac{1}{\kappa_d}) \;/\; (\tfrac{1}{\kappa_d}+\tfrac{1}{\kappa_t}+\tfrac{1}{\kappa_u})$. The $\kappa_c$ which is allowed bigger now, as well as the new $\kappa_r$, involve tripartite interactions though, which may be more limiting in practice than the timescale separation requirement.

\paragraph*{Bipartite:}The stricter requirement of bipartite interactions can be met at the cost of additional levels. An efficient solution using \emph{data qutrits} is presented in Section \ref{sec:3level}. Sticking to data qubits and adding levels to the \emph{ancillas}, we could imagine:
\begin{eqnarray*}
N_k\; , M_k && \text{ as in \eqref{eq:step0Nks},\eqref{eq:step1synch}} \\
\tilde{L}_{1,k}/\sqrt{\kappa_{cr}} & = &  \q{m_1}\qd{m}_{k+1}\q{0}\qd{0}_k  + \q{m_2}\qd{m}_{k+1}\q{1}\qd{1}_k +  \q{g_1}\qd{g}_{k+1}\q{0}\qd{0}_k  + \q{g_2}\qd{g}_{k+1}\q{1}\qd{1}_k \\
\tilde{L}_{2,k}/\sqrt{\kappa_{ch}} & = &  \q{m,0}\qd{m_1,1}_{k}  + \q{m,1}\qd{m_2,0}_{k} +  \q{g,0}\qd{g_1,1}_{k}  + \q{g,1}\qd{g_2,0}_{k}\\
\tilde{L}_{3,k}/\sqrt{\kappa_{ch}} & = &  \q{m,0}\qd{m_1,0}_{k}  + \q{m,1}\qd{m_2,1}_{k} +  \q{g,0}\qd{g_1,0}_{k}  + \q{g,1}\qd{g_2,1}_{k}\, .
\end{eqnarray*}
The ieda is that to condition and obtain bipartite operators, we split the action of $L_k$ into two parts. The new levels $m_1,m_2$ (or $g_1,g_2$) of the ancilla serve to transmit the levels $0,1$ of data qubit $k$ to its neighbor $k+1$. This transmission will induce an additional ``downtime'' of order $\tfrac{\kappa_{cr}}{\kappa_{ch}}$ during which the data are entangled with the ancilla. Moreover, the quantum coherence between those ancilla states must now be protected. Both effects are kept in check as the transition through those states is supposed to be fast. Nevertheless, this does not look too practical.

The expected benefit of switching off the $L_k$ was to allow taking larger $\kappa_c$, irrespective of $\kappa_r$. With the above construction, and assuming that $\kappa_{st}$ acting on classical degrees of freedom of the ancillas constitutes no limitation, we can indeed take large $\kappa_{ch}$. However, we must take $\kappa_{cr}$ an order of magnitude lower, and the latter will dominate the convergence rate associated to the effective $L_k$ dissipation channel. 
\vspace{3mm}

In both the tripartite and bipartite schemes, adding more ancilla levels would enable for instance to insert short timeouts between applying \eqref{eq:LTV} and applying resets $\q{+}\qd{-}$.  Such timeouts could mitigate the limited synchronization of the ancillas, trading off increased accuracy for an additional downtime before converging back towards $\q{GHZ_+}$, with again an optimal tradeoff to be sought. However, we stop adding complexity into the reservoir now and consider different architectures.


\subsection{Approximate GHZ reservoir through ancilla jump conditioning}\label{ssec:jumpcond}

One may imagine setups where the data qubits undergo operations conditioned not on the ancilla \emph{being in} a given state, but rather on the ancilla \emph{jumping to} a state. The working of such reservoirs, as represented on Figure \ref{fig:clockwork}.b,  is very similar to the ones with ancilla state conditioning.

The idea motivating such schemes is that resets conditioned on ancilla jumps will be automatically synchronized at the moment of the clock transition. Hence, there is no need to wait for resets to ``very likely have happened'' and then switch them off synchronously. A dual way to look at this operation is that data qubits signal to the ancillas when their resets are done, since these are correlated with an ancilla jump; hence, no need for the ancillas to wait until resets have ``likely'' happened. Concretely, the advantage is that the equivalent of the discussion around \eqref{eq:step2atc} will no longer involve the contribution of $1/\kappa_d$ in $T_2$. However, the reservoir still needs three different timescales.

More details about such schemes can be found in Appendix.\\

The attentive reader may find our analysis until now, based on separately applying either resets or a repetition of $L_k$ operators on all the data qubits, somewhat too pessimistic for large $n$. Indeed, the stabilization of $\q{GHZ_+}$ with resets and $L_k$ operators in fact follows a spatially organized evolution along the qubit chain. This leads to our next type of proposal.


\subsection{Spatio-temporal GHZ wave reservoir}\label{ssec:wave}

Note that $L_k$ assigns to data qubit $k+1$ the logical value of data qubit $k$. Hence, starting from $\q{++...+}$ and applying each jump operator $L_k$ once, in a random order, would be insufficient for stabilizing $\q{GHZ_+}$. However, applying the $L_k$ once in increasing order from $k=1$ to $k=n-1$, would result in $\q{GHZ_+}$. We may thus want to favor such a ``wave'' process, both in the $L_k$ and in the resets.

The corresponding schemes feature two adaptations. First, only the extremal ancilla $1$ undergoes spontaneous jumps to trigger the clock evolution. The other ancillas follow, through stimulated jumps only, in the order of the chain. In Section \ref{ssec:statecond} and Section \ref{ssec:jumpcond}, the clock transition is also propagating along the chain, but there is no particular place for the wave to start, so parts of it could propagate in a direction opposite to the natural direction implied by the $L_k$. Second, the modified reservoir shall contain a mechanism to apply the $L_k$ in increasing order of $k$, instead of in random order as in the previous schemes. The benefits of this spatio-temporal organization are expected to increase with chain length $n$.

Again, many variations are possible. Unfortunately, inducing more order on the $L_k$ leads to the same operational issues as mentioned in Section \ref{sssec:sc:step2}: either we must allow tripartite interactions, or we must split the $L_k$ with coherent ancilla sublevels. By now the reader should be able to devise associated Lindblad operators on their own. A few concrete proposals are described in appendix. Their analysis would be closer to the proposal of Section \ref{ssec:ZeQutrit}.

\section{Architectures with data qutrits}\label{sec:3level}

We now turn towards a completely different way of enlarging the Hilbert space. Indeed, we consider the \emph{data} subsystems to be qutrits ($Q=3$). Their logical space of interest is still the subspace $\text{span}\{\q{0},\q{1}\}$ and we still target the same state $\q{GHZ_+} = (\q{00...0}+\q{11...1}) \; / \; \sqrt{2}$, but the auxiliary levels $\q{2}$ are used for reservoir operation.

It is remarkable that the sole addition of level $\q{2}$ makes the whole ancillas construction unnecessary, while at the same time allowing stronger performance. Although everything acts on data qutrits now, we keep using the decoherence operator letters $L_k,M_k,N_k$ to distinguish similar roles to those with ancillas.

\subsection{A GHZ wave with data qutrits only}\label{ssec:ZeQutrit}

\begin{figure}
\begin{center}
\includegraphics[width=150mm, trim=15 310 80 25, clip]{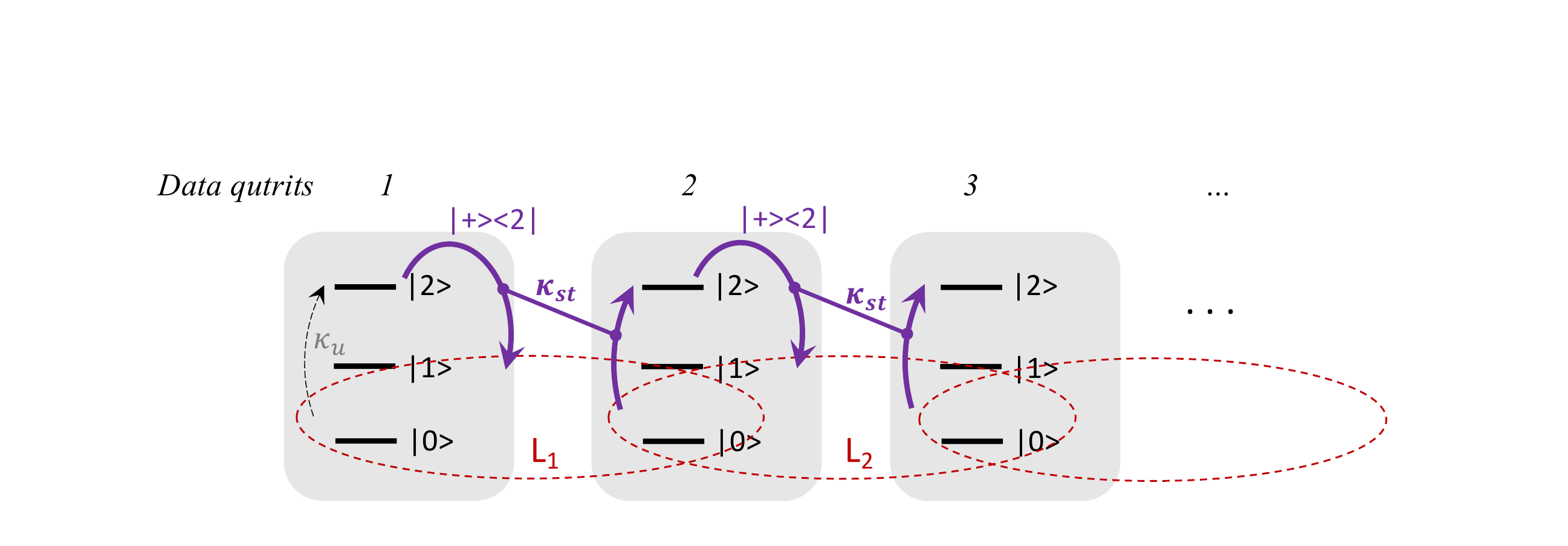}
\end{center}
\caption{Illustration of the dynamics behind the scheme for stabilizing approximately $\q{GHZ_+}$ using a chain of data qutrits, according to the Lindblad operators described by \eqref{eq:waveQ3}. \label{fig:ZeQutrit}}
\end{figure}

The use of qutrits enables a GHZ-stabilizing wave, in the same spirit as Section \ref{ssec:wave}, yet now conditioning the two intended operations $\q{+}\qd{-}_k$ and $L_k$ with just bipartite interactions. We next give the details of this implementation which seems the most natural one. There are two main ideas, illustrated on Figure \ref{fig:ZeQutrit}. First, as a replacement of the ancillas, the reset wave is propagated by the jump down from $\q{2}$ on data qutrit $k$ triggering a jump up to $\q{2}$ on data qutrit $k+1$. Second, as a new feature, the two intended operations are well separated since $L_k$ will have no action as soon as qutrits $k$ and/or $k+1$ are on level $\q{2}$. Explicitly, the following reservoir operators would do the job:
\begin{eqnarray}
\nonumber L_k && \text{ as in \eqref{eq:LTV}} \\
\label{eq:waveQ3}
M_{0,r} &=& \sqrt{\kappa_u} \q{2}\qd{-}_1 \quad , \quad M_{0,i} = \sqrt{\kappa_u} \q{2}\qd{+}_1 \\
\nonumber 
N_{k,r} &=& \sqrt{\kappa_{st}} \q{+,2}\qd{2,-}_{k,k+1}  \quad , \quad  N_{k,i} = \sqrt{\kappa_{st}} \q{+,2}\qd{2,+}_{k,k+1} \; , \\ \nonumber
&& N_{k,v} = \sqrt{\kappa_{st}} \q{+,2}\qd{2,2}_{k,k+1} \quad , \quad \text{ for } k=1,2,...,n-1 \\ \nonumber
N_{n} &=& \sqrt{\kappa_{st}} \q{+}\qd{2}_n \; .
\end{eqnarray}
The intended evolution is to launch a wave by exciting qutrit 1 to $\q{2}$, irrespective of its initial state (hence two operators $M_{0,...}$). This automatically switches off $L_1$. Then qutrit 1 resets from $\q{2}$ to $\q{+}$, while exciting qutrit 2 to $\q{2}$ (operators $N_{1,...}$). This keeps $L_1$ off and switches off $L_2$ as well. Next, qutrit 2 resets to $\q{+}$ while exciting qutrit 3 to $\q{2}$ (operators $N_{2,...}$). This switches back on $L_1$ but not $L_2$; and so on. The last qubit of the chain just resets on its own with $N_n$.

The reader may want to note the following details similar to the ancilla-based reservoirs.
\begin{itemize}
\item The values of the rates may vary as a function of $k$, only their orders of magnitude matter.
\item The $N_{k,...}$ operators are split in three parts in order to ensure possible transition to $\q{+,2}$ for any initial state of the second qubit --- thus anywhere in $\text{span}\{\q{0},\q{1}\}$, but also if it was already excited to $\q{2}$ --- while avoiding any dark states.
\item While there are no classical ancillas anymore, it remains true that the quantum phase associated to the auxiliary \emph{level} $\q{2}$ of any qutrit is unimportant. In other words, a qutrit will never have to be in a coherent superposition of $\q{2}$ with some other levels.
\end{itemize}

This reservoir perfectly synchronizes all stabilizing operations without wasting time. In particular, we can now take  $\kappa_c$ arbitrarily large thanks to the switching on and off of the $L_k$. In fact, regarding timescales, we just have to ensure that reset waves are repeated at a slow rate compared to the data-stabilizing operations, thus:
$$ \frac{1}{\kappa_u} \sim T_1  \quad \gg  \quad \left\{ \frac{1}{\kappa_c}  \; , \; \frac{1}{\kappa_{st}} \right\} \sim T_3  \; , $$
with no intermediate $T_2$ anymore. For an improvement by a factor $C$ on $T_3/T_0$ with $T_0$ the characteristic timescale of perturbations, we may thus now expect the error to improve by $C^{1/2}$. If the operators \eqref{eq:waveQ3} associated to data qutrits are a realistic option, then this looks like our most efficient GHZ reservoir.


\section{Performance analysis: ancilla-clock based schemes}\label{sec:analysis1}

We next analyze more quantitatively and formally the performance of the proposed reservoirs. In particular, this involves determining optimal values for the various rates $\kappa_{...}$. We provide both approximate analytic results and numerical simulations. Full-system numerical results are limited to low values of $n$ due to the exponential growth of Hilbert space dimension with $n$ and $m$, even when taking into account the classical nature of the ancillas. The approximate analytic results thus serve to gain better insight into the scaling with larger $n$.

The present section focuses on the ancillas clock architecture, more precisely the state-conditioning scheme of Section \ref{sssec:sc:step1}. The wave-based scheme with data qutrits is analyzed in Section \ref{sec:analysis2}.

Throughout the analysis sections, we use ``configuration'' to denote a possibility for the system, e.g.~qubits being in $\q{++...+}$ or just having undergone the jump $L_k$. The term ``state'' will rather denote the distribution over configurations (quantum state $\rho$ or probability distribution $p$ associated to a Markov chain).


\subsection{Behavior of the clock ancillas}\label{sec:an:clockancillas}

As already mentioned, and detailed in appendix, we have designed the ancillas' evolution to not be influenced by the data qubits. Moreover, any quantum coherences between ancilla levels are irrelevant for the reservoir working, and when starting with a classical probability distribution over ancilla levels without quantum coherences, the system stays so. We can thus first analyze the subsystem composed of the ancillas only, treating it like a classical Markov chain with probability distribution $p$ over all possible configurations of the ancillas.\\

Transitions in the ancilla clock are governed by the Lindblad equation
\begin{equation}
\tfrac{d}{dt} \rho^a_t =  \sum_{\substack{k \\ j=sp,st+,st-}}  M_{k,j} \rho^a_t {M_{k,j}}^\dagger - \tfrac{1}{2}\, (\; {M_{k,j}}^\dagger M_{k,j} \rho^a_t + \rho^a_t {M_{k,j}}^\dagger {M_{k,j}}  \;)  \
	\end{equation}
on the ancillas' reduced density matrix $\rho^a_t$, with dissipation operators $M_{k,j}$ given by \eqref{eq:step1synch}. Equivalently, assuming $\rho^a_t$ diagonal in the canonical basis at $t=0$, and thus for all $t\geq 0$ (see appendix), this can be represented as a classical Markov chain:
\begin{equation}\label{eq:an:MCX}
\tfrac{d}{dt} p_t = \sum_{\substack{k \\ j=sp,st+,st-}} \; A_{k,j} \, p_t \; =:  A \; p_t \; .
\end{equation}
Consider the state-conditioning scheme of Section \ref{sssec:sc:step1}, the vector $p_t \in \mathbb{R}^{3^n}$ here is thus a distribution over the $3^n$ possible elements constituting the set $\{\q{g},\q{e},\q{m}\}^n$. We will use $p_X$ to denote the population on configuration $X \in \{ \q{g},\q{e},\q{m} \}^n$, for instance $p_{gg..g}$ denotes the population on configuration ``all ancillas in $\q{g}$''. Each transition matrix $A_{k,j}$ in \eqref{eq:an:MCX}, and thus a fortiori the total transition matrix $A$, have non-negative off-diagonal elements, and columns summing to zero. Translating the effect of the operators $M_{k,j}$ given by \eqref{eq:step1synch}, the transition matrices $A_{k,j}$ write as follows:
\begin{eqnarray}\label{eq:an:MCXb}
A_{k,sp} &=& \kappa_u \, (\q{e}\qd{g}-\q{g}\qd{g} )_k + \kappa_d (\q{m}\qd{e}-\q{e}\qd{e})_k + \kappa_t (\q{g}\qd{m}-\q{m}\qd{m})_k \\
\nonumber A_{k,st+} &=& \kappa_{st} \, ( \q{ee}\qd{ge} +  \q{mm}\qd{em} + \q{gg}\qd{mg}  )_{k,k+1} \\
\nonumber && - \kappa_{st} ( \q{ge}\qd{ge} +  \q{em}\qd{em} + \q{mg}\qd{mg}  )_{k,k+1} \\
\nonumber A_{k,st-} &=& \kappa_{st} \, ( \q{ee}\qd{eg} +  \q{mm}\qd{me} + \q{gg}\qd{gm}  )_{k-1,k} \\
\nonumber && - \kappa_{st} ( \q{eg}\qd{eg} +  \q{me}\qd{me} + \q{gm}\qd{gm}  )_{k-1,k} \; .
\end{eqnarray}
Here we kept using the quantum notation for basis vectors with implicit tensor-identity, e.g.~$\q{e}\qd{g}_k - \q{g}\qd{g}_k$ denotes the $3^n \times 3^n$ matrix inducing (at a unit rate) transitions from $\q{e}$ to $\q{g}$ on ancilla $k$, while maintaining the other ancillas' values.\\

The goal now is to compute the steady state of this Markov chain \eqref{eq:an:MCX},\eqref{eq:an:MCXb}, approximately, assuming $\kappa_{st} \gg \kappa_d \gg \kappa_u,\kappa_t$ as we have requested at the design stage\footnote{Strictly speaking, we only need $\kappa_d, \kappa_u,\kappa_t \ll \kappa_{st}$ and $\min(\kappa_u,\kappa_t) \ll \kappa_d$. However, there seems to be no particular incentive for taking $\kappa_u \gg \kappa_t$ or opposite, so for simplicity we here take them of the same order.}. We hence define 
$$\frac{\max(\kappa_{t}, \kappa_{u})}{\kappa_{d}} = \frac{T_2}{T_1} = \epsilon_1  \ll 1 \;\;\; \text{ and } \;\;\; \frac{\kappa_{d}}{\kappa_{st}} = \frac{T_3}{T_2} = \epsilon_2 \ll 1 \; ,$$
and we further assume $n \epsilon_1, \, n \epsilon_2 \ll 1$. The appendix provides a detailed proof to essentially establish the following results.
\begin{Proposition}\label{approx}
Consider the Markov chain \eqref{eq:an:MCX},\eqref{eq:an:MCXb}.\newline 
(a) The steady state population which is not on $\q{gg...g},\q{ee...e},\q{mm...m}$ is of order $O(\epsilon_1 \epsilon_2 \, n^2)$.\newline
(b) In the limit $\epsilon_2 \rightarrow 0$, we have the steady-state populations:
		\begin{align}\label{eq:approxAS}
		p_{gg..g} = \frac{1}{1+\frac{\kappa_{u}}{\kappa_{d}} + \frac{\kappa_{u}}{\kappa_{t}}} \;\; , \quad
		p_{mm..m} = \frac{1}{1+\frac{\kappa_{t}}{\kappa_{u}} + \frac{\kappa_{t}}{\kappa_{d}}}  \;\; , \quad
		p_{ee..e} = \frac{1}{1+\frac{\kappa_{d}}{\kappa_{t}} + \frac{\kappa_{d}}{\kappa_{u}}} \; .
		\end{align}				
\end{Proposition}
In particular thus, $p_{ee..e} \ll 1$.\vspace{4mm}

The jump-conditioning scheme of Section \ref{sssec:jc:step1} admits a similar analysis of the ancilla-clock, now including levels $\q{f}$. Since the intermediate time $T_2 = 1/\kappa_c$ there does not involve the ancillas, we directly define 
$$\epsilon = \frac{\max(\kappa_{u},\kappa_d,\kappa_t)}{\min(\kappa_{st},\kappa_f)} \;\; = \;\; \frac{T_3}{T_1} \;\; \ll 1$$
according to \eqref{eq:step2btc}. Then Proposition \ref{approx}(a) remains valid, with $\epsilon = T_3/T_1$ replacing $\epsilon_1 \epsilon_2 = T_3/T_1$. The steady state will have population of order $O(1)$ on each of $\q{gg...g},\q{ee...e},\q{mm...m}$, and in fact Proposition \ref{approx}(b) remains unchanged, just with $\epsilon_2=\epsilon$, and $\kappa_d$ of the same order as $\kappa_u,\kappa_t$.\\

On the basis of these results, we can approximate the ancillas clock as jumping from $\q{gg...g}$, to $\q{ee...e}$, then $\q{mm...m}$, and so on, with transition rates and a steady state distribution characterized by Proposition \ref{approx}(b). The possibility to achieve very large $\kappa_{st}$ is further encouraged by the fact that the corresponding transitions involve no operations on the truly quantum part of the system, namely the data qubits; having to treat purely classical degrees of freedom may facilitate achieving faster transition rates.

The scaling in $n^2$ in Proposition \ref{approx}(a) can be seen as the consequence of two phenomena. First, the propagation of the synchronization over the ancillas chain takes a typical time $n/\kappa_{st}$. Second, the fact that any of the $n$ ancillas can spontaneously trigger a transition means that the expected time between two such perturbations of the synchronization procedure scales like $1/(n \kappa_d), 1/(n \kappa_u), 1/(n \kappa_t)$. For an optimal working point, it may thus seem wise to decrease the $\kappa_d$, $\kappa_u$, $\kappa_t$ with increasing $n$, in order to moderate this $1/n$ scaling of the expected time between clock transitions. Before deciding on this, we now estimate the corresponding steady state for the data qubits.



\subsection{Data qubits evolution}\label{sec:AncBasedMC}

In order to obtain quantitative results on the data qubits, our analysis involves several model simplifications.
\begin{itemize}
\item First, we  consider that the ancilla clock only goes through $\q{gg...g},\q{ee...e},\q{mm...m}$. The influence of other ancilla configurations on GHZ fidelity estimate is discussed at the very end.
\item Second, we introduce (small) design modifications such that data qubits evolution can be treated like a \emph{classical Markov chain}. More precisely, while the quantum state $\rho(t)$ evolves through non-orthogonal states, the Markov chain will model the (hypothetical) output signal associated to the dissipation operators, which is a classical variable. The design modification ensures that the associated dynamics is Markovian on those output signals, i.e. it indeed evolves autonomously without further depending on $\rho(t)$.
\item Related to the previous point, we measure the fidelity to GHZ as the proportion of state on a particular configuration of the Markov chain. This is a pessimistic bound, since not all other configurations of the Markov chain are orthogonal to $\q{GHZ_+}$. However, since it is trivial to achieve 50\% fidelity with $\q{GHZ_+}$ (e.g.~just take the configuration $\q{00...0}$), it seems legitimate to discard as ``bad'' all the configurations which are not doing significantly better than this.
\item Finally, we will make several approximations in the analysis of the classical Markov chain in order to evaluate its steady state.
\end{itemize}
The second point requires more precise information, which we provide next.

\subsubsection{Markov chain definition}\label{ssec:MC1definition}

The idea is to build a classical Markov chain, over a finite number of configurations, related to the transitions that the dissipation channels would induce in a ``jump''-type unraveling of the Lindbladian dynamics. We will associate a configuration of the Markov chain to a set of (hypothetical) output signal values associated to jump detections. The main issue is to ensure an evolution of this Markov chain which does not further depend on the quantum state.

Concretely, let $Q_k$ denote a generic Lindblad operator like the $L_k, M_k, N_k$ introduced previously. The Lindbladian decoherence associated to $Q_k$ can be viewed as the average over different purity-preserving evolutions, which would be distinguished by a hypothetical output associated to $Q_k$. In particular, in the so-called ``jump stochastic master equation unraveling" of Lindbladian decoherence \cite{gardiner2004quantum}, the channel $Q_k$ is associated to a Poisson process $q_k(t)$  reporting detections of ``quantum jumps''. The Poisson process is determined by expectation
$$
\mathbb{E}(dq_k(t)) = \text{trace}(Q_k \, \rho(t) \, Q_k^\dagger) \; dt  \; ,
$$
with associated state evolutions:
\begin{eqnarray*}
\text{for } q_k(t+dt)-q_k(t) = 1 &:& \rho_{t+dt} \;=\; (Q_k \, \rho(t) \, Q_k^\dagger)  \; / \; \text{trace}(Q_k \, \rho(t) \, Q_k^\dagger) \\
\text{for } q_k(t+dt)-q_k(t) = 0 &:& \rho_{t+dt} \;=\; (V_0 \, \rho(t) \, V_0^\dagger) \; / \; \text{trace}(V_0 \, \rho(t) \, V_0^\dagger)  \\
&& \text{with } V_0 = I + \tfrac{dt}{2} (  \text{trace}(Q_k \, \rho(t) \, Q_k^\dagger)  \, I - Q_k^\dagger Q_k ) \; .
\end{eqnarray*}
Here $I$ denotes the identity operator. In the following, we also consider the case where a single detection signal $q_\mu$ does not distinguish from which operators $Q_{\mu_k}$ a jump is coming. This situation is governed by the following equations:
\begin{equation}\label{eq:stoch:1}
\mathbb{E}(dq_\mu(t)) = {\textstyle \sum_k} \text{trace}(Q_{\mu_k} \, \rho(t) \, Q_{\mu_k}^\dagger) \; dt  \; ,
\end{equation}
\begin{eqnarray*}\label{eq:stoch:2}
\text{for } q_\mu(t+dt)-q_\mu(t) = 1 &:& \rho_{t+dt} \;=\; \frac{{\textstyle \sum_k} Q_{\mu_k} \, \rho(t) \, Q_{\mu_k}^\dagger}{{\textstyle \sum_k} \text{trace}(Q_{\mu_k} \, \rho(t) \, Q_{\mu_k}^\dagger)}\\ \nonumber
\text{for } q_\mu(t+dt)-q_\mu(t) = 0 &:& \rho_{t+dt} \;=\; (V_0 \, \rho(t) \, V_0^\dagger)  \; / \; \text{trace}(V_0 \, \rho(t) \, V_0^\dagger)  \\  \nonumber
&& \text{with } V_0 = I + \tfrac{dt}{2} {\textstyle \sum_k} (  \text{trace}(Q_{\mu_k} \, \rho(t) \, Q_{\mu_k}^\dagger)  \, I - Q_{\mu_k}^\dagger Q_{\mu_k} ) \; .
\end{eqnarray*}
The deterministic evolution described by the Lindblad equation is equivalent to the average evolution, when such detectors are present but their output signal is not recorded. The principle of ``unraveling'' is to view this in converse: while there are no detectors actually present, we reason in terms of hypothetical detection results whose expectation describes the engineered reservoir evolution.
Here, we push this one step further, by designing a specific system architecture which can be studied as a Markov chain on the (hypothetical) signals $q_\mu(t)$ alone. More precisely, \emph{consider a signal $q(t)$ listing the various detections that have happened}, e.g.
$$q(t) =  \mu_1,\, \mu_2,\, \mu_1,\, \mu_3 \; .$$
if up to time $t$ we have seen first a detection on $\mu_1$, then on $\mu_2$, then on $\mu_1$ again, and finally on $\mu_3$ and nothing more. Our aim is to describe the evolution of $q(t)$ like a classical Markov chain.

In order to set up such model, the statistics $\mathbb{E}(dq_\mu(t))$ for the future evolution of the $q_\mu(t)$ should only depend on $q(t)$. Furthermore, to be useful, knowing $q(t)$ should give us clear information about $\rho(t)$. We ensure these by imposing a model with the two following, somewhat stronger properties:
\begin{itemize}
\item[(i)]  The $\mathbb{E}(dq_\mu(t))$ are independent of $\rho(t)$.
\item[(ii)]  $V_0$ is always proportional to identity, such that in absence of any detection the state $\rho(t)$ does not change.
\end{itemize}
These properties are not trivial and we now show how to apply them for the three types of Lindblad operators acting on data qubits:  error channels, resets to $\q{+}$, and two-qubit correlation operators $L_k$.

\paragraph*{Error channels:}  In line with usual quantum computing assumptions, we consider independent bit-flip and phase-flip errors on each qubit, associated respectively to decoherence operators:
\begin{equation}\label{eq:00MC1}
E_{k,1} = \sqrt{\kappa_x}(\q{0}\qd{1}+\q{1}\qd{0})_k \quad , \quad  E_{k,2} = \sqrt{\kappa_z}(\q{0}\qd{0}-\q{1}\qd{1})_k \; , \quad k=1,2,...,n \; .
\end{equation}
These error channels naturally satisfy the properties (i) and (ii) mentioned above, with a detector $q_\mu$ associated to each individual operator. Indeed, since $E_{k,s}^\dagger E_{k,s} = \kappa_s\, I$  proportional to the identity for $s \in \{ x,z \}$, we have 
$$\mathbb{E}(dq_{k,s}(t)) = \kappa_s \text{trace}(\rho(t)) =\kappa_s \quad \text{and} \quad V_0 = I \; .$$
Note that with a loss operator, $E_{k} = \q{0}\qd{1}_k$, this would not be as trivial and some adaptation would be required. This adaptation is in fact strictly analogous to the qubit reset channels discussed next.

\paragraph*{Reset channels:} Consider a reset operator $N_k = \sqrt{\kappa_r} \q{+}\qd{-}_k$. The associated jump detection signal is associated to $\; \mathbb{E}(dq_{N_k}(t)) = \kappa_r \qd{-} \rho(t) \q{-}_k \; $ which does not satisfy condition (i). However, this issue can be solved by adding a no-reset operator. Indeed, consider along the lines of \eqref{eq:stoch:1} that
\begin{equation}\label{eq:00MC2}
q_{k,+} \text{ is associated indistinguishably to both }  N_{k,r} = \sqrt{\kappa_r} \q{+}\qd{-}_k \;\;\; \text{ and }\;\;\;  N_{k,i} = \sqrt{\kappa_r} \q{+}\qd{+}_k \; . 
\end{equation}
Then $N_{k,r}^\dagger N_{k,r} + N_{k,i}^\dagger N_{k,i} = \kappa_r\, I$ is proportional to identity, so properties (i) and (ii) hold.

As a parenthesis, when discarding the other qubits, the Lindblad equation with both $N_{k,r}$ and $N_{k,i}$ writes
$$\tfrac{d}{dt} \rho(t) = \kappa_r (\,\q{+}\qd{+} - \rho(t)\,) \; ,$$
while with $N_{k,r}$ alone the off-diagonal components $\qd{+}\rho\q{-}$ and $\qd{-}\rho\q{+}$ decay twice more slowly. There is thus a true difference between those two models at the Lindbladian level. For our purpose this is not an issue.

\paragraph*{Two-qubit correlation channels:} The operator $L_k$ is conceptually similar to a reset operator, letting qubit $k+1$ jump to the value of qubit $k$. Hence, with no surprise, properties (i) and (ii) can be satisfied only if we modify the setting. Namely, instead of just applying $L_k$, we will assume that, along the lines of \eqref{eq:stoch:1}, we have
\begin{equation}\label{eq:00MC3}
q_{k,L} \text{ associated to }  L_{k} = \sqrt{\kappa_c} (\q{00}\qd{01}+\q{11}\qd{10})_{k,k+1} \;\;\; \text{ and }\;\;\;  \tilde{L}_{k} = \sqrt{\kappa_c} (\q{00}\qd{00}+\q{11}\qd{11})_{k,k+1} \; . 
\end{equation}
Like for the reset channel, this ensures properties (i) and (ii) thanks to $L_k^\dagger L_k + \tilde{L}_k^\dagger \tilde{L}_k = \kappa_c\, I$ proportional to identity, and this \emph{does} (somewhat) modify the model proposed in Section \ref{sec:ancillas}. In terms of system operation, the $\tilde{L}_k$ appears unnecessary. However, it simplifies the analysis by allowing us to treat the whole system as a classical Markov chain on jump detection signals.

\paragraph*{Markov chain:}  Having defined the output signals $q_{k,x},\, q_{k,z},\,q_{k,+},\,q_{k,L}$ with associated Lindblad operators in \eqref{eq:00MC1},\eqref{eq:00MC2},\eqref{eq:00MC3}, we are all set for describing our classical Markov chain. We make this explicit description for the case of state-conditioning; the same approach holds for jump-conditioning.

\begin{figure}
\includegraphics[width=165mm, trim=5 90 50 78, clip]{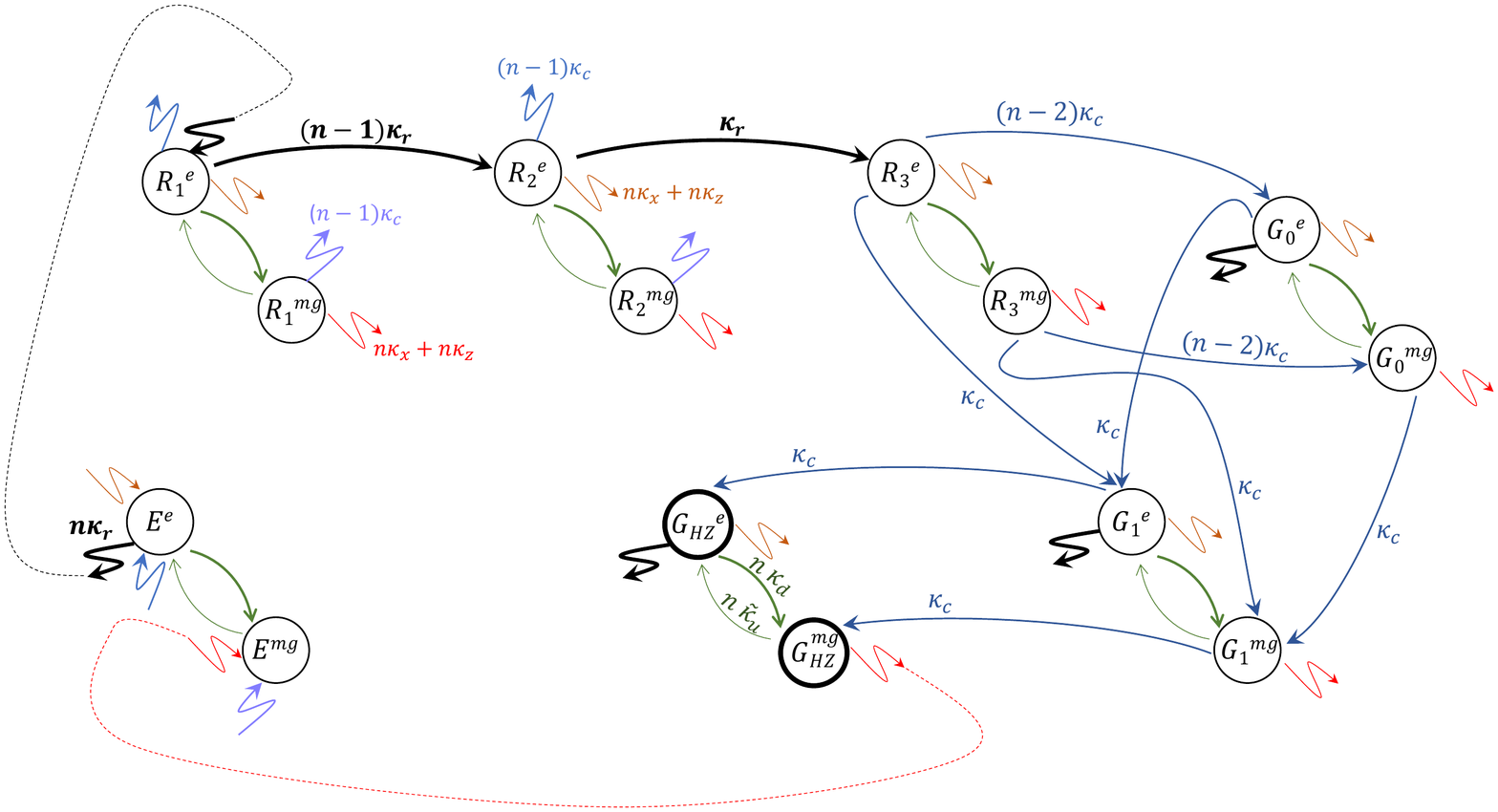}
\caption{Representation of the Markov chain based on hypothetical jump detection outputs for $n=3$ data qubits in the ancilla-state-conditioning architecture. To avoid clutter, wiggly output arrows indicate a connection of all these transitions to the input arrow of the same shape and color; a few dotted lines illustrate these connections. For instance, the six outgoing red arrows all indicate a transition towards configuration $E^{mg}$; its rate is the sum of the bit-flip and phase-flip rates of the $n$ qubits. Green arrows indicate the summarized transitions of the ancillas clock. Thick black arrows indicate reset detections, associated to $\{k+\}$, and are possible only with ancillas in $\q{ee...e}$. All other detections occur irrespectively of the Markov chain configuration, the respective transitions are only a consequence of the grouping of output signals. For instance, the blue arrows are associated to detections $\{k\,L\}$. From configurations $R_1$ and $R_2$, i.e.~when only part of the qubits have reset, this leads to configuration $E$. On configuration $G_{HZ}$, such detection induces no transition. On $R_3$, detecting $\{1L\}$ leads to $G_1$, while detecting $\{k\,L\}$ with $k \neq 1$ (thus here $\{2L\}$) would lead to $G_0$.  For $n>3$, the Markov chain would look similar with just additional states $R_k^e$ and $R_k^{mg}$ before $R_n^e$ and $R_n^{mg}$ on the top row,  and additional states $G_k^e$ and $G_k^{mg}$ before $G_{HZ}^e$ and $G_{HZ}^{mg}$ on the lower row. \label{figa:00MCn}}
\end{figure}

We recall that a configuration of the classical Markov chain would be described by a value of $q(t)$, i.e.~an ordered list of jump detections, like
$\; q(t) = \{1L\}, \{2x\}, \{5+\} \;$
if up to time $t$ we have observed first a jump on $q_{1,L}$ from \eqref{eq:00MC3}, then a jump on $q_{2,x}$ from \eqref{eq:00MC1}, then a jump on $q_{5+}$ from \eqref{eq:00MC2}, and nothing more. Thus, $q(t)$ can take a countable infinity of configurations. We reduce the Markov chain to a finite number of configurations by grouping the values of $q(t)$. This principle can be followed at several degrees of precision, depending on how accurately we want the reduced Markov chain to estimate the associated $\rho(t)$. We propose to use the following reduction:
\begin{itemize}
\item Configuration $R_{\ell}$, for $\ell=1,2,...,n$: groups all $q(t)$ ending with a sequence composed of $\geq 1$ times $\{k_1+\}$, $\{k_2+\}$,..., and $\{k_{\ell}+\}$, for some fixed and differing qubit indices $k_1,...,k_{\ell}$,  preceded by detections different from $\{\cdot +\}$.

In particular, $R_1$: all $q(t)$ ending with $\{k+\}$ for some fixed $k$, possibly repeated several times, and preceded by detections different from $\{\cdot +\}$.

\emph{Examples:} $q(t) = ... , \{1 L\},\, \{3+\}$  belongs to $R_1$ ;   $q(t) = ... , \{2\,x\},\, \{3+\},\, \{1+\}, \{3+\}$  belongs to $R_2$;   $q(t) = ... \{3+\},\, \{1+\}, \{4+\}, \{3+\}, \{2+\} \{1+\}, \{1+\}$  belongs to $R_n$ for $n=4$.

\item Configuration $G_0$:  groups all $q(t)$ ending with a sequence composed of $\geq 1$ times $\{1+\}$, $\{2+\}$,..., and $\{n+\}$, followed by any number of detections $\{k\; L\}$ with $k \neq 1$.

\emph{Examples:}   $q(t) = ... \{3+\},\, \{1+\}, \{4+\}, \{3+\}, \{2+\} \{1+\}, \{1+\}, \{2L\}, \{3L\}, \{3L\}$  belongs to $G_0$ for $n=4$.

\item Configurations $G_{\ell}$, for $\ell=1,2,...,n-2$:   all $q(t)$ ending with a sequence composed of $\geq 1$ times $\{1+\}$, $\{2+\}$,...,  and $\{n+\}$, followed by a sequence of $\{k\; L\}$ containing the ordered subsequence $\{1 L\},\{2 L\},...,\{\ell L\}$ but not the ordered subsequence $\{1 L\},\{2 L\},...,\{(\ell+1) L\}$.

In particular, configuration $G_1$:  all $q(t)$ ending with a sequence composed of $\geq 1$ times $\{1+\}$, $\{2+\}$,...,  $\{n+\}$, followed by a sequence of $\{k\; L\}$ containing $\{1\; L\}$, but no $\{2\;L\}$ after $\{1\; L\}$.

\emph{Examples:}   $q(t) = ... \{3+\},\, \{1+\}, \{4+\}, \{3+\}, \{2+\} \{1+\}, \{1+\}, \{2L\}, \{1L\}, \{3L\}, \{2L\}$  belongs to $G_2$ for $n=4$ ;   $q(t) = ... \{3+\},\, \{1+\}, \{4+\}, \{3+\}, \{2+\} \{1+\}, \{1+\}, \{2L\}, \{1L\}, \{3L\}, \{3L\}$  belongs to $G_1$ for $n=4$.

\item Configuration $G_{HZ}$:  all $q(t)$ ending with a sequence composed of $\geq 1$ times $\{1+\}$, $\{2+\}$,...,  $\{n+\}$, followed by a sequence of $\{k\; L\}$ containing the ordered subsequence $\{1\; L\},\{2\;L\},...,\{(n-1)\; L\}$. 

\emph{Examples:}   $q(t) = ... \{3+\},\, \{1+\}, \{4+\}, \{3+\}, \{2+\} \{1+\}, \{1+\}, \{2L\}, \{1L\}, \{3L\}, \{2L\}, \{3L\}$  belongs to $G_{HZ}$ for $n=4$.

\item Configuration $E$: groups all $q(t)$ of any different form.

\emph{Examples:} $q(t) = ... \{3\,x\}$;   $q(t) = ... \{2\, x\} ,\{2L\}, \{1L\}, \{3L\}$ ;   $q(t) = ...  \{3\, z\}, \{2+\},\{1L\}$.

\item Ancillas clock:  finally, each of the above configurations is split in two, depending if the ancillas clock is on $\q{ee...e}$ or on $\{\q{gg...g},\q{mm...m}\}$.

\emph{Examples:}  $q(t) = ...  \{3\, z\}, \{2+\},\{1L\}$ belongs to $E^e$  or to $E^{mg}$.
\end{itemize}
An example of the resulting Markov Chain for $n=3$ is shown on Figure \ref{figa:00MCn}. Since the transition rates on $q(t)$ are independent of the configuration (up to resets being allowed only when ancillas are in $\q{ee...e}$), the transitions in this Markov Chain come down to a counting argument on ``grouped outputs'' configurations. For the ancillas, since we do not distinguish $\q{mm...m}$ and $\q{gg...g}$ anymore, we define the summarized rate $\tilde{\kappa}_u = (\frac{1}{\kappa_u}+\frac{1}{\kappa_t})^{-1}$ for jumping from $\{\q{gg...g},\q{mm...m}$ to $\q{ee...e}$; this preserves the same population on $p_{ee...e}$ as in Proposition \ref{approx} and the same mean time to go around one cycle $\q{gg...g} \rightarrow \q{ee...e} \rightarrow \q{mm...m} \rightarrow \q{gg...g}$.\\

The Markov chain on $q(t)$ \emph{before grouping}, with a countable infinity of states, \emph{exactly} reflects how $\rho(t)$ evolves with \eqref{eq:00MC1},\eqref{eq:00MC2},\eqref{eq:00MC3}. In contrast, the reduced Markov Chain obtained after grouping, illustrated in Figure \ref{figa:00MCn}, is not as faithful: different representatives (from the countably infinite chain) of a given grouped configuration can correspond to different $\rho(t)$ and even to different fidelities to GHZ. With the reduced Markov chain, we settle for a conservative viewpoint. More precisely: When the reduced Markov chain reaches $G_{HZ}^e$ or $G_{HZ}^{mg}$, it is guaranteed that $\rho(t)$ has reached $\q{GHZ_+}\qd{GHZ_+}$ indeed, and we consider this as the only successful configurations. This involves for instance the following pessimistic approximations:
\begin{itemize}[noitemsep,topsep=0pt]
\item[-] If a bit-flip $\{3\, x\}$ is detected when the Markov chain was in $G_{HZ}^{mg}$ and thus $\rho(t)$ was on $\q{GHZ_+}\qd{GHZ_+}$, then observing next the single jump $\{2L\}$ would in fact correct back the value of qubit 3, bringing back $\rho(t)$ towards $\q{GHZ_+}\qd{GHZ_+}$; instead, in the reduced Markov chain of Figure \ref{figa:00MCn} which does not wish to retain the full history of any signal, the bit flip leads to configuration $E^{mg}$ whatever came before, and from there the whole stabilization sequence has to be traversed again until ending up on $G_{HZ}^e$ or $G_{HZ}^{mg}$. In other words, in the full Markov chain, \emph{some} $q(t)$ ending with $\{3\, x\},\{2L\}$ would correspond to $\q{GHZ_+}$; but in the reduced Markov chain, all the $q(t)$ ending with $\{3\, x\},\{2L\}$ are grouped in $E$, assuming pessimistically that those all correspond to no success.
\item[-] Similarly, the sequence e.g.~$q(t) = ...  \{2+\},\, \{1+\}, \{2+\}, \{1L\}$  would only need to be completed by $\{3+\}, \{2L\}$ in order to obtain $\rho(t) = \q{GHZ_+}\qd{GHZ_+}$; however, the reduced Markov chain of Figure \ref{figa:00MCn} does not distinguish this configuration from e.g.~$q(t) = ...\{1x\}  \{2+\},\, \{3+\}, \{2+\}, \{1L\}$, and accordingly when complementing this by $\{3+\}, \{2L\}$, we must end up in a reduced configuration that is not labeled as successful, placing this $q(t)$ in $E^{mg}$ or $E^e$. This corresponds to neglecting the possibility of evolving reset- and $L_k$-waves concurrently; an analysis improving this point is given in Section \ref{sec:analysis2}.
\end{itemize}
The pessimistic viewpoint on the reduced Markov chain gives, at least, a lower bound on GHZ fidelity, while avoiding an apparently cumbersome $n$-dependent analysis following all the possible ``partial corruptions of the state''. Somewhat more complicated and precise intermediate models are possible, for instance saying that a $\kappa_c$ action on $R_k^e$ brings us to $R_{k-2}^e$ as only two qubits are affected; yet this would not change the main trends. The wave-based analysis in Section \ref{sec:analysis2} aims for more precise numbers. The only non-pessimistic approximations made here are about ancilla evolution, namely (i) assuming that ancillas are always perfectly synchronized and (ii) grouping $\{ \q{mm...m} , \q{gg...g} \}$ as a single configuration, with an effective transition rate back to $\q{ee...e}$. The effect of imperfect ancilla synchronization is analyzed further below.

\subsubsection{Markov chain analysis}\label{ssec:MC1analysis}

The final step to evaluate the performance of the scheme is to compute the steady state of the Markov chain described in Section \ref{ssec:MC1definition}. For simplicity, we assume at this point an equal bit-flip and phase-flip rate $\kappa_x = \kappa_z =: \kappa_p/2$.
In addition to the three timescales $T_1 \gg T_2 \gg T_3$ as described in \eqref{eq:step2atc}, we thus have a fourth one $\tfrac{1}{\kappa_p} = T_0 \gg T_1$, corresponding to perturbations being slower than the reservoir stabilization rate.

Looking at the dominating terms, the argument leading to a steady state close to $\q{GHZ_+}$ goes roughly as follows. The total population $p_{e}$ on $\q{ee...e}$, irrespective of the data qubits situation, is of order $O(T_2/T_1)$. Moreover, any configuration of type $\cdot^{e}$ receives at most $O(n\kappa_u)$  population from configurations of type $\cdot^{mg}$. Therefore, a majority of the population $p_e$ is concentrated on $R_n^{e}$, the only configuration from which one cannot leave at the fastest rate $\kappa_r$ (see Figure \ref{figa:00MCn}). All other configurations have a population at least $O(T_3/T_2)$ smaller, thus of order at most $O(T_3/T_1)$. Next, we observe that each $R_k^{mg}$, with arrivals $n\kappa_d\, p_{R_k^e}$ and departure at a rate $O((n-1)\kappa_c)=O(n\kappa_d)$, must have steady state population of the same order as $R_k^e$. In particular, $p_{R_k^{mg}}=O(T_3/T_1)$ at most for $k<n$ and  $p_{R_n^{mg}}=O(T_2/T_1)$. With a similar argument, and knowing that the $G_k^{e}$ have population at most $O(T_3/T_1)$, we observe that the $G_k^{mg}$ all have population of the same order as $R_n^{mg}$, thus of order $O(T_2/T_1)$. The target configuration $G_{HZ}^{mg}$  then features arrivals $\kappa_c O(T_2/T_1) = O(1/T_1)$ and leaks at a rate $\tilde{\kappa}_u = O(1/T_1)$, meaning that its population is $O(1)$. There remains to observe that $E^{mg}$ cannot have population of order 1 if $\kappa_p \ll \kappa_u$, to conclude that $G_{HZ}^{mg}$ is the only state with population of order 1, and thus necessarily close to 1.

An exact steady state analysis can be carried out with simple algebraic means. The full expressions are provided in the proof in appendix, while the (more readable) result to leading orders in $T_k / T_{k-1}$ is summarized in the following statement.

\begin{Proposition}\label{prop:llp1}
The Markov chain defined in Section \ref{ssec:MC1definition} has all its steady-state population on $G_{HZ}^{mg}$, up to terms of order $T_k/T_{k-1}$. More precisely, assuming $n T_k / T_{k-1}$ to remain small, and up to factors of order $\tfrac{n-1}{n}$, we have:
\begin{equation}\label{eq:LLL}
p_{G_{HZ}^{mg}} + p_{G_{HZ}^{e}} \;\; \approx \;\; 1 - \tfrac{\kappa_p}{\tilde{\kappa}_u} - n(n-1)\tfrac{\tilde{\kappa}_u}{\kappa_c}-\tfrac{\tilde{\kappa}_u}{\kappa_d} - n \ln(n) \tfrac{\kappa_c + \kappa_d}{\kappa_r} \;  + O\left((\tfrac{T_k}{T_{k-1}})^2\right) \; .
\end{equation}
\end{Proposition}
\begin{proof} See appendix \hfill $\square$ \end{proof}

Proposition \ref{prop:llp1} bounds the GHZ-stabilizing performance with terms of order $(T_k/T_{k-1})$, ordered according to $T_1/T_0$, $T_2/T_1$, $T_3/T_2$. All these ratios must be small to ensure a good fidelity to $\q{GHZ_+}$. The reservoir engineering will usually be constrained by the observed error rate $1/T_0$ and the maximally achievable engineered reservoir rate $1/T_3$. The timescales $T_1$ and $T_2$ should be chosen between these two extremes to optimize the performance. The expression confirms that a gain by a factor $C$ on $T_0/T_3$ shall be split up into gains of a factor $C^{1/3}$ on each of the error terms. Concretely, by taking the expression \eqref{eq:LLL} as true, the best setting would be
\begin{eqnarray*}
\tilde{\kappa}_u & \simeq & \kappa_p \; (\kappa_r/\kappa_p)^{1/3} \; 1/(n^{2/3}\ln(n)^{1/3})\\
\kappa_d & \simeq & \kappa_p \; (\kappa_r/\kappa_p)^{2/3} \; 1/(n^{5/6} \ln(n)^{2/3}) \\
\kappa_c & \simeq & \kappa_p \; (\kappa_r/\kappa_p)^{2/3} \;  n^{1/6}/\ln(n)^{2/3} \; ,
\end{eqnarray*}
which yields an error scaling roughly like $\left(n^{7/2} \ln(n)\; \kappa_p / \kappa_r \right)^{1/3}$.  This is of course quite approximate, given all the approximations made on the way. 

\subsubsection{Adding the effect of imperfect ancilla synchronization}\label{sssec:impancmt}

For a full estimate of fidelity to $\q{GHZ_+}$, there remains to consider ancilla population outside the subspace spanned by $\q{gg...g},\q{ee...e},\q{mm...m}$. While this population is of second order, namely at most $T_3/T_1$ on what we have called ``main transition configurations'' (see appendix), \emph{in a general Markov chain} such transition configurations could nevertheless have a dominating impact on the steady state. In appendix, we show how for the present case, the first-order effect of imperfect ancilla synchronization amounts to
$$ \text{replacing } \;\frac{n \ln(n)}{\kappa_r}\; \text{ in \eqref{eq:LLL} by } \;\frac{n \ln(n)}{\kappa_r} + \frac{n-1}{\eta_2 \kappa_{st}}\;,$$
where $1/(\eta_2 \kappa_{st}) = O(1/\kappa_{st})$ is the characteristic time at which the ancillas clock synchronizes.

\subsubsection{Simulation results (ancillas-clock-based architecture)}\label{ssec:simu:ancillas}

Our simulations consider the reservoir model \eqref{eq:step1synch},\eqref{eq:step0Nks},\eqref{eq:LTV}. In absence of concrete constraints on the two rates to be set fastest, we take $\kappa_r = \kappa_{st}$. The error model considers bit-flip and phase-flip errors according to Lindblad operators \eqref{eq:00MC1} on each data qubit individually, with same rate $\kappa_x = \kappa_z = \kappa_p/2$; and no errors on the ancillas. We then set up the full Lindblad equation and compute its steady-state, evaluating its fidelity to $\q{GHZ_+}$ in presence of the perturbation.

\begin{figure}
	\begin{center}
		\includegraphics[width=12cm]{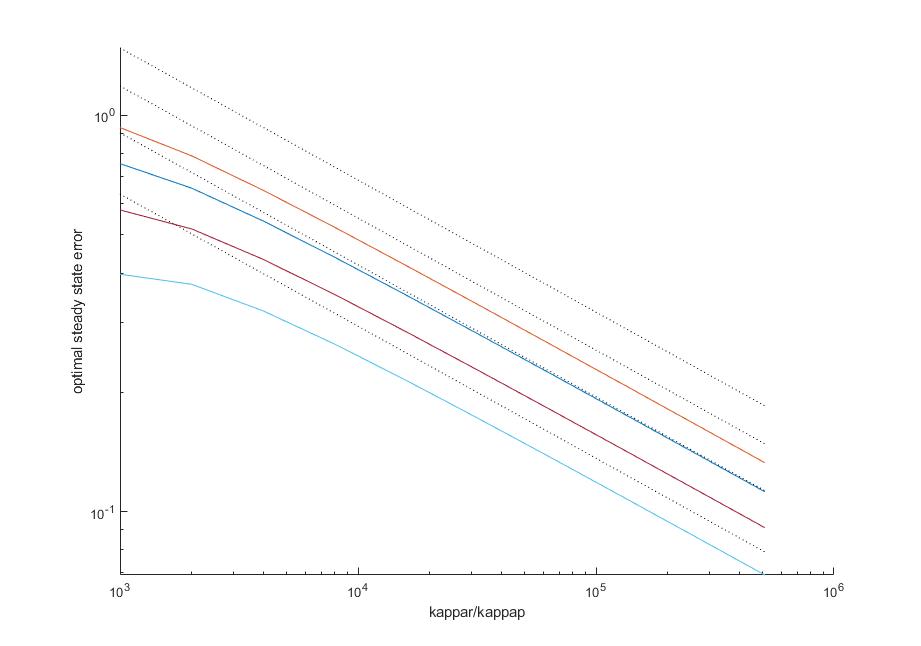}
	\end{center}
	\caption{Steady state error corresponding to the ancilla clock based GHZ reservoir \eqref{eq:step1synch},\eqref{eq:step0Nks},\eqref{eq:LTV} subjected to individual bit-flips and phase-flips at rate $\kappa_p/2$. The number of data qubits is $n=3,4,5,6$ from bottom (lowest error) to top (largest error).
		\label{fig:OptPerfancilla}}
\end{figure}

The ratio $\kappa_{r}/\kappa_p$ between the most extreme rates limits the best achievable fidelity. For several values of this ratio and for $n=3,4,5,6$, we have explored values of the intermediate rates $\tilde{\kappa}_u$, $\kappa_{d}$ and $\kappa_{c}$ around their theoretical best settings as computed in the previous sections. Figure \ref{fig:OptPerfancilla} reports for each $n$ the results corresponding to the local optimum found in this way on all intermediate rates. The simulation values (colored full lines) and the theoretical values of our approximate analysis (dotted black lines) feature the same error scaling in $(\kappa_p / \kappa_r)^{1/3}$ for higher values of $\kappa_{r}/\kappa_p$. Regarding absolute values, the theory appears to be pessimistic by a constant factor at higher values of $\kappa_{r}/\kappa_p$, and further misses out for lower values of $\kappa_{r}/\kappa_p$ and for larger $n$. This is expected from the approximations made in the analysis of Section \ref{sec:AncBasedMC}. Finally, one can note that the $(\kappa_p / \kappa_r)^{1/3}$ scaling makes it difficult to reach high fidelities to GHZ.

\section{Performance analysis: scheme based on data qutrits}\label{sec:analysis2}

This section analyzes the performance of the scheme based on data qutrits, as presented in Section \ref{sec:3level}. This also illustrates the analysis in the framework of ``GHZ-stabilizing waves'' propagating along the qutrit chain.

Like in Section \ref{sec:AncBasedMC}, for easing the theoretical analysis, we slightly adapt the model of Section \ref{ssec:ZeQutrit} in order to obtain classical Markov chains based on hypothetical output detections.  We propose an approximate analysis in two ways. The first one checks how a reset wave (going through level $\q{2}$) propagates along the whole chain unperturbed by $L_k$ events, then an $L_k$ subsequence leads to $\q{GHZ_+}$. This is easier to follow, and pessimistic by a small factor. Indeed, reset wave and GHZ-stabilizing wave can in principle propagate in parallel. A second analysis hence considers the characteristic time for those two waves in parallel; it involves estimating essentially how to cross a 2-dimensional lattice.


\subsection{Qutrit wave description with classical Markov chains}\label{ssec:an2mod}

The procedure is similar to Section  \ref{sec:AncBasedMC}, to which we refer the reader for more details and justifications.

The qutrit level $\q{2}$ plays a role analogous to the classical ancillas of Section \ref{sec:ancillas}. Each qutrit only jumps to and from this level $\q{2}$ by involving states which are orthogonal to it. In particular:
\begin{itemize} 
\item Consider an initial state $\rho(0)$ of the full system where $\qd{1} \rho(0) \q{2}_k = \qd{0} \rho(0) \q{2}_k = 0$ for all qutrits $k$, i.e.~there are no quantum coherences involving level $\q{2}$. Then this property remains true for $\rho(t)$ for all $t\geq 0$. The proof is trivial, just by inspecting the Lindblad equation for those components.
\item Consider a model where for each qubit we only differentiate whether it is in level $\q{2}$ or in the ``logical'' subspace span$\{\q{0},\q{1}\}$, with populations thus denoted $p_{x_1,x_2,...}$ and each $x_i$ taking the value either $2$ or $\ell$ (``logical''); in other words, the levels $\q{0},\q{1}$ are considered as a refined subdivision of the level $\q{\ell}$, and we currently discard this refinement. Then the model of Section \ref{ssec:ZeQutrit}, aggregated in this way on $\{\q{\ell},\q{2}\}^n$, follows an autonomous classical Markov chain $\tfrac{d}{dt} p = A \, p$ with:
\begin{eqnarray}\label{eq:Q3MC1}
\qd{2}_1 A \q{\ell}_1  &=& \kappa_u  \,  I_{2^{n-1}} \; , \\  \nonumber
\qd{\ell}_n A \q{2}_n  &=& \kappa_{st}  \, I_{2^{n-1}} \; , \\  \nonumber
\qd{\ell,2}_{k-1,k} A \q{2,\ell} &=& \kappa_{st}  \, I_{2^{n-2}} \;\;\; \text{ for } k=2,3,...,n \; ;
\end{eqnarray}
here $I_x$ denotes an identity matrix of dimension $x$. The other off-diagonal elements of $A$ are zero, and the diagonal is fixed to ensure zero column sums. This Markov property holds thanks to the internal state $\q{0}$ or $\q{1}$ of $\q{\ell}$ having (by design) no effect on the aggregated jump probabilities involving $M_{...}$ and $N_{...}$ in \eqref{eq:waveQ3}, while the  $L_k$ only change the internal state of $\q{\ell,\ell}_{k,k+1}$, and hence have no effect at this point.
\end{itemize}
These observations imply that, in order to model the full process with a classical Markov chain, we need a particular procedure only when treating the effect of the $L_k$, and possibly of the perturbation. 

Form there, the first point is to associate a (virtual) detector monitoring when jump operators are applied, in a Poisson process unraveling of the Lindblad master equation. We gather in an output signal $q(t)$ the sequence of detection events. In order for this output signal to undergo Markovian dynamics, not further conditioned on e.g.~timing and its back-action on the precise state $\rho(t)$, we treat the error channels and $L_k$ ike in Section \ref{sec:AncBasedMC}. In particular, we modify the setting by assuming that we have two operators $L_k$ and $\tilde{L}_k$ whose jumps we do not distinguish in the outputs, see \eqref{eq:00MC3}. The probability of detecting a jump with $L_k$ or $\tilde{L}_k$ indistinguishably is then the same for any state of type $\q{\ell,\ell}_{k,k+1}$, allowing us to discard the exact evolution of $\rho(t)$. With this adaptation, we can rigorously reduce the Lindblad equation to a classical Markov chain on the signal $q(t)$. The signal $q(t)$ can be any sequence of the following detection events (with any possible repetitions and of arbitrary length):
\begin{itemize}
\item[] $\{ k\, L\}$: jump with $L_k$ or $\tilde{L}_k$, for $k=1,2,...,n-1$  (rate $\kappa_c$)
\item[] $\{k \, +\}$: jump with any of the $N_{k,...}$ indistinguishably for $k=1,2,...,n$  (rate $\kappa_{st}$)
\item[] $\{U\}$:  jump with any of the $M_{0,...}$ indistinguishably  (rate $\kappa_u$)
\item[] $\{ k \, E \}$:  jump with any of the error operators on qutrit $k$ indistinguishably  (rate $\kappa_p$) \; .
\end{itemize}

The second point is to reduce the resulting exact Markov chain, which takes place on a countable infinity of possible $q(t)$, into a Markov chain over a finite set of configurations. For this, we group into one configuration all the signals $q(t)$ ending with a particular property. 
An exact model reduction should be possible, but it would still involve a too large number of possibilities to be practical for analysis. We propose a grouping which involves an approximation on modeling transition rates and on estimating fidelity to $\q{GHZ_+}$, but it appears to capture the dominant effect and compares reasonably well to simulations. The definition of reduced Markov chain configurations is somewhat different in the two analyses and is detailed in Appendix.

\subsection{Results of approximate analysis}\label{ssec:an2modb}

We have carried out two somewhat different analyses, one meant to provide a lower bound on the fidelity and the other meant to estimate it more closely. Computations explained in appendix then yield the following result.

\begin{Proposition}\label{prop:WaveFResult1}
Denote $\kappa_u/\kappa_{st} = \epsilon \ll 1$, $\kappa_c/\kappa_{st} = \gamma = O(1)$ and $\kappa_p/\kappa_u = \epsilon_p \ll 1$.
\begin{itemize}
\item The Markov chain analysis described in Appendix \ref{app:aq3:a} estimates a $\q{GHZ_+}$ state population
\begin{equation} \label{eq:WaveFResult1a}
p_{GHZ+}  \geq  1-n\epsilon_p-n\epsilon-\tfrac{(n-1)\epsilon}{\gamma} + o(\epsilon,\epsilon_p) \; .
\end{equation}
\item The Markov chain analysis described in Appendix \ref{app:aq3:b} estimates a $\q{GHZ_+}$ state population
\begin{equation} \label{eq:WaveFResult1b}
p_{GHZ+}  \simeq 1 - n \epsilon_p - n \epsilon + o(\epsilon,\epsilon_p) \; ,
\end{equation}
where we have assumed $\gamma=1$.
\end{itemize}
\end{Proposition}
The two estimates differ by less than a factor 2 on the error.

A realistic design constraint would be an upper bound on the ratio $\max(\kappa_{st},\kappa_c) / \kappa_p$  between maximally achievable reservoir rates and perturbation rate, which translates into and upper bound on 
\begin{equation}\label{eq:AnW:OptConstraint}
\frac{\max(1,\gamma)}{\epsilon\epsilon_p}.
\end{equation}
Thanks to having less timescales compared to the ancilla-based architectures of Section \ref{sec:analysis1}, an improvement by a factor $C$ on this bound now enables an improvement by $\sqrt{C}$ on both $\epsilon$ and $\epsilon_p$ and thus on the dominant error. In fact, we can compute the optimal tuning according to the estimates of Proposition \ref{prop:WaveFResult1}.
\begin{itemize}
\item In \eqref{eq:WaveFResult1a}, for $\epsilon,\epsilon_p$ fixed it is beneficial to increase $\gamma$, hence $\gamma < 1$ cannot be optimal. Once $\gamma \geq 1$, the constraint on \eqref{eq:AnW:OptConstraint} requires to modify $\epsilon,\epsilon_p$ if we further increase $\gamma$, and it turns out that the overall effect is disadvantageous; thus, we should take $\gamma=1$. This makes sense intuitively, as there seems to be no reason in our wave reservoir to slow down either $\kappa_c$ or $\kappa_{st}$ below the maximally achievable rate.

The optimal value of $\kappa_u$ can then be computed on the basis of variables $\epsilon,\epsilon_p$ with a Lagrangian involving the constraint $\epsilon \epsilon_p - \kappa_p / \kappa_c = 0$. Standard computations lead to the optimal values
$$ \epsilon = \sqrt{\frac{n \kappa_p}{(2n-1) \kappa_c}} \; \quad , \quad \epsilon_p = \sqrt{\frac{(2n-1)\kappa_p}{n\kappa_c}} \; .$$
The corresponding performance is
$$p_{GHZ+} \simeq 1 - 2 \sqrt{n (2n-1) \frac{\kappa_p}{\kappa_c}} \; .$$
As anticipated, the error decreases by $\sqrt{C}$ for an improvement by a factor $C$ on  $\max(\kappa_{st},\kappa_c) / \kappa_p$. It scales essentially linearly in $n$.
\item In \eqref{eq:WaveFResult1b}, we observe a somewhat better scaling with $n$. The optimal setting is (by symmetry) to take 
$$ \epsilon = \epsilon_p = \sqrt{\tfrac{\kappa_p}{\kappa_c}} \; $$
with a corresponding performance
\begin{equation}\label{eq:4conc:perf}
p_{GHZ+} \simeq 1 - 2 n \sqrt{\tfrac{\kappa_p}{\kappa_c}} \; .
\end{equation}
The suggested optimal setting for $\kappa_u$ is thus independent of $n$, the error estimate still linear in $n$ and improved by a factor $\sqrt{2}$ with respec to \eqref{eq:WaveFResult1a}.
\end{itemize}


\subsection{Simulation results (qutrit-based scheme)}\label{ssec:simu:qutritwave}

\begin{figure}
\begin{center}
\includegraphics[width=12cm]{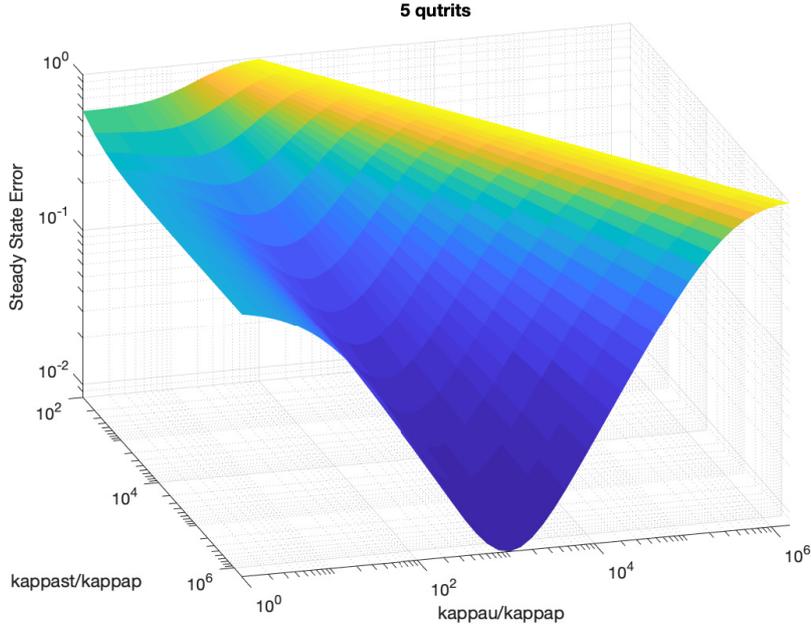}
\end{center}
\caption{Simulation results of the steady state error with respect to the target state $\q{GHZ_+}$, for the qutrit wave reservoir architecture and error channels $P_{k,j}$ with $k=1,...,6$ and $j=1,...,n$ described at the beginning of Section \ref{ssec:simu:qutritwave}. We here vary $\kappa_u$ and $\kappa_{st}=\kappa_c$ (in units of the perturbation rate $\kappa_p$) for an illustratively fixed $n=5$.\label{fig:3D5qt}}
\end{figure}

\begin{figure}
\begin{center}
\includegraphics[width=12cm]{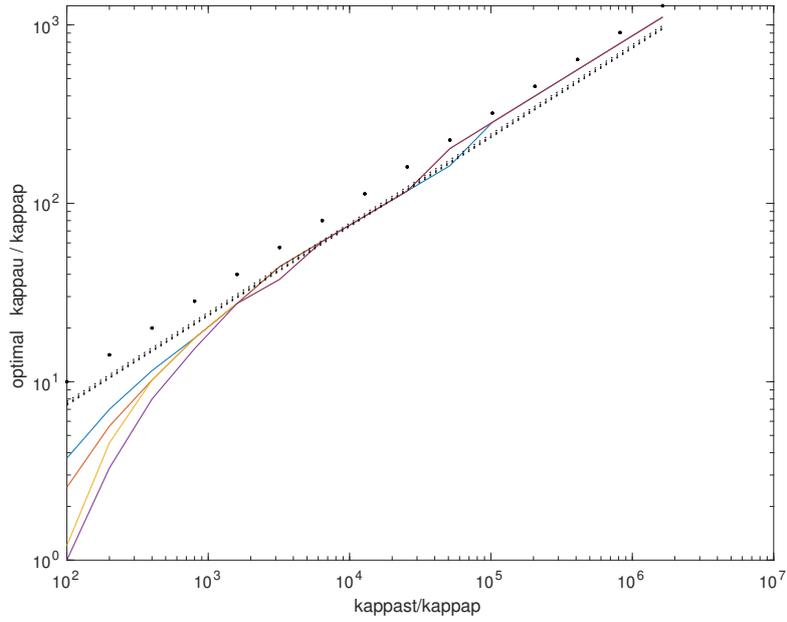}
\end{center}
\caption{Optimal value of $\kappa_u$ (in units of $\kappa_p$), leading to the smallest steady state error with respect to $\q{GHZ_+}$ under the same conditions as Fig.\ref{fig:3D5qt}, here as a function of $n$ (colors) and of maximal reservoir power $\kappa_{st}=\kappa_c$  (in units of $\kappa_p$). Full colored lines are simulation results, dotted black lines represent the analysis according to estimate \eqref{eq:WaveFResult1a}, black dots represent the analysis according to estimate \eqref{eq:WaveFResult1b} and thus independent of $n$. The number of qutrits is $n=3,4,5,6$.
\label{fig:OptKun}}
\end{figure}

\begin{figure}
\begin{center}
\includegraphics[width=12cm]{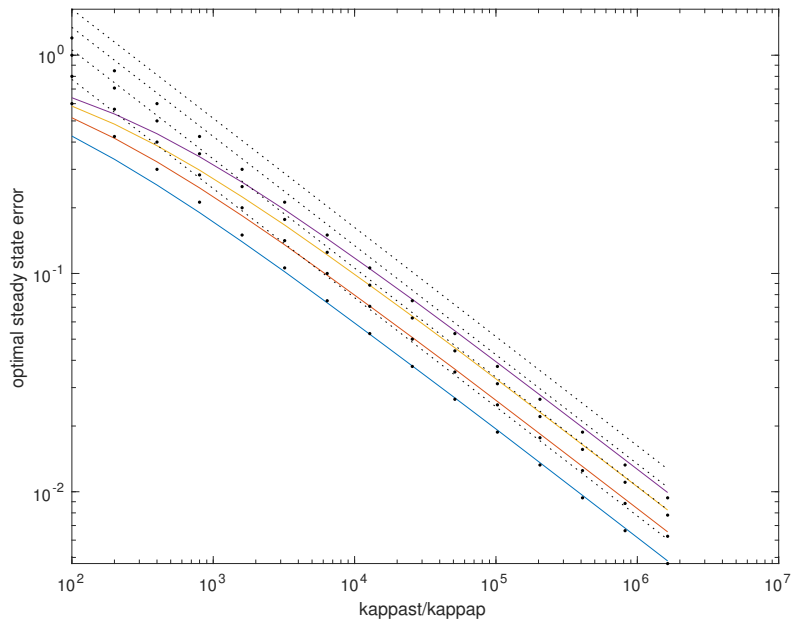}
\end{center}
\caption{Steady state error corresponding to the setting of Figure \ref{fig:OptKun}. The number of qutrits is $n=3,4,5,6$ from bottom (lowest error) to top (largest error).
\label{fig:OptPerf}}
\end{figure}

The simulation results consider the reservoir model \eqref{eq:waveQ3} with $\kappa_c = \kappa_{st}$, since taking both at the maximal achievable jump rate seems best. We do not assume the presence of some $\tilde{L}_k$ as we did for the analysis. As an error model, our target is on each qutrit independently a channel drawing it towards the fully mixed state $\rho = \frac{\q{0}\qd{0}+\q{1}\qd{1}+\q{2}\qd{2}}{3}$. For simplicity and in absence of any concrete physical model in mind, we implement this with the $6n$ jump operators:
\begin{eqnarray*}
&& P_{k,1} = \sqrt{\kappa_p} \q{0}\qd{1} \;\; , \quad   P_{k,2} = \sqrt{\kappa_p} \q{0}\qd{2} \;\; , \quad  P_{k,3} = \sqrt{\kappa_p} \q{1}\qd{2} \;\; ,  \\
&& P_{k,4} = \sqrt{\kappa_p} \q{1}\qd{0} \;\; , \quad   P_{k,5} = \sqrt{\kappa_p} \q{2}\qd{0} \;\; , \quad  P_{k,6} = \sqrt{\kappa_p} \q{2}\qd{1}  \; .
\end{eqnarray*}
We then set up the full Lindblad equation and compute its steady-state, evaluating its fidelity to $\q{GHZ_+}$ in presence of the perturbation.

We have performed simulations varying the parameter $\kappa_{st}/\kappa_p$ and exploring values of the intermediate rate $\kappa_u$, for $n=3,4,5,6$. Beyond this number, the exponential scaling in $n$ leads to too expensive simulations on a laptop. Figure \ref{fig:3D5qt}  illustrates how the steady state error scales with $\kappa_u$ and $\kappa_{st}$ for fixed $\kappa_p$, and here for $n=5$. Other chain lengths show essentially the same behavior. A valley of optimal $\kappa_u$, leading to minimal error, is clearly visible.

Figure \ref{fig:OptKun} shows the optimal value of $\kappa_u$, leading to this minimal error, as a function of $n$ and $\kappa_{st}/\kappa_p$. The simulation values (colored full lines) indicate a below-sampling-step dependence on $n$ over these few values, as predicted by the analysis according to estimate \eqref{eq:WaveFResult1b}, except at low values of $\kappa_{st}/\kappa_p$. Quantitatively, the simulation optima are in good agreement with the theoretically computed optimal setting.  Figure \ref{fig:OptPerf} shows the corresponding steady state error. It confirms that our analytical formulas are close to the true error. More precisely, both our formulas capture the main trend for large values of $\kappa_{st}/\kappa_p$ and overestimate the error at low values, as expected; the result of \eqref{eq:WaveFResult1a} somewhat overestimates the true error at almost all values, while the estimate \eqref{eq:WaveFResult1b} appears to be quite close to the ``truth'' (i.e. full simulation values) once $\kappa_{st}/\kappa_p > 1000$.




\section{Conclusion}

The aim of this paper is to propose a static reservoir engineering design, using quasi-local operators only, which approximately yet \emph{globally} stabilizes the $n$-partite GHZ state $(\q{00...0}+\q{11...1})/\sqrt{2}$,  i.e.~ensuring convergence towards this target from any initial state of the whole system. We thus protect the state against any kind of noise.

Starting from the impossibility result on exact stabilization in \cite{ticozzi2014steady}, we propose architectures to stabilize an approximate GHZ state with a time-independent Lindbladian master equation based on ancillary logic, assuming a chain-like interconnection possibility amongst subsystems involving bi-partite or tri-partite operators. The ancillary logic is based either on auxiliary $d$-level subsystems, or on auxiliary levels in the data subsystems. The ancilla subsystems need never be in quantum superposition and hence we have not discussed the effect of noise on them, but this is no fundamental issue. We propose an approximate theoretical performance analysis based on a Markov chain whose possible configurations are types of \emph{jump detection sequences}. This method, modulo slight adaptations of the reservoir model, circumvents the subtleties of evolving quantum states and jump operators with nonorthogonal components. Simulations for low $n$ confirm the validity of our analysis, and the performance of the scheme, when the relevant timescale separations hold. At best, the error scales linearly with $n$, see e.g.~formula \eqref{eq:4conc:perf}. This appears to be some fundamental limit, since each of the $n$ data subsystems could undergo a phase-flip, transforming $(\q{00...0}+\q{11...1})/\sqrt{2}$ into $(\q{00...0}-\q{11...1})/\sqrt{2}$, while distinguishing the latter two states requires fully nonlocal information, thus propagation of information along the chain of length $n$.
In future work, the $n$-dependence of the error might be improved by modifying the interconnection topology, using e.g.~$d$-dimensional lattices or small-world networks. This may link quantum reservoir engineering to the literature on network-dependence in classical synchronization.

We view these results as a first step towards quantum reservoir engineering designed to replace classical logic / automata in quantum information protocols. Our architectures can also be viewed as evolved variations of the dissipative quantum computing protocols as proposed in \cite{Verstraete2009} for instance, with several new specificies. \newline
 $\bullet$ Our proposals are fully jump-based, introducing timescale separations and directionality with significant performance benefits. \newline
 $\bullet$ While the automaton logic based on ``ancilla subsystems'' is more standard, we also propose a novel way of encoding automaton logic, namely using \emph{auxiliary levels on the data subsystems.} In particular, just having a qutrit at every data subsystem appears to enable more efficient stabilization than with a full logical ancilla at each site. This way of controlling evolution with auxiliary levels thus appears quite powerful and may be worth exploring in more general automaton contexts.\newline
$\bullet$ Finally, it is worth mentioning that our way of conditioning maintains some operations to be applied repetitively, instead of just once. Concretely, instead of a sequence based on CNOT gates which could be implemented in an automaton, and which would generate $\q{GHZ+}$ if applied once but deconstruct it if applied more, we use the jump operators $L_k$ from \cite{ticozzi2014steady} (see formula \eqref{eq:LTV}) which keep correcting bit-flip errors continuously.

This last property means that we keep correcting bit-flip type errors on the data at a rate $\kappa_c$, much faster than the phase flips for which we have constructed the whole scheme with protection rate of order $\kappa_u \ll \kappa_c$. Slow correction of phase-flip with fast correction of bit-flips appears particularly relevant for biased-noise qubits like the cat qubits \cite{mirrahimi2014dynamically}, for which our scheme would thus feature much better performance than with standard depolarizing noise.

Regarding analysis methods, we have introduced a model modification in order to then treat the dynamics rigorously as a classical Markov chain on dissipation channel outputs, instead of investigating the full Lindblad equation. While this does (slightly) modify the actual system, it allows for a simplified treatment which gives significant quantitative results. This method may be worth considering for other sytems.\\

\paragraph*{Acknowledgments} The authors want to thank Lorenza Viola, Francesco Ticozzi, Mazyar Mirrahimi, Pierre Rouchon, Ivan Bardet and Christophe Vuillot for stimulating discussions. This work has been supported by the ANR project HAMROQS and by Plan France 2030 through the project ANR-22-PETQ-0006.

\bibliographystyle{plain} 
\bibliography{refsVM}
\newpage


\section{Appendix A: Generalizing the impossibility result of \cite{ticozzi2014steady}}

In \cite{ticozzi2014steady}, it is proven that a static Lindblad equation cannot stabilize a GHZ state on $n$ qubits if it contains no decoherence operator involving at least $n/2$ qubits. In fact the result also holds for subsystems with $Q>2$ levels, and in presence of ancilla subsystems. We here provide the proofs of these generalizations.

For the case of quQits, i.e.~subsystems of $Q$ levels with $Q>2$, the method is the same as in \cite{ticozzi2014steady}.

\begin{Proposition}
The dynamics \eqref{eq:L0} with only data subsystems of dimension $Q \geq 2$ cannot globally asymptotically stabilize the GHZ state $(\q{00...0}+\q{11...1})/\sqrt{2}$ on $n$ subsystems if each Lindblad operator involves interaction of at most $<n/2$ subsystems.
\end{Proposition}
\begin{proof}
We use two results reported in \cite{ticozzi2014steady}. For a target state $\q{\psi_d}$, denote $\rho_k$ the associated reduced state on the subsystems on which one of the dissipation operators $L_k$ acts nontrivially. Denote $\mathcal{H}_{(k)} = \text{support}(\rho_k \otimes \text{Identity})$ and let $\mathcal{H}_0 = \cap_k \mathcal{H}_{(k)}$. Then:
\begin{itemize}
\item $\q{\psi_d}$ can be globally asymptotically stabilized with the dissipation operators $L_k$ if and only if $\mathcal{H}_0=\text{span}(\q{\psi_d})$  (Theorem 1 in \cite{ticozzi2014steady}).
\item if $\mathcal{H}_0 \supset \text{span}(\q{\psi_d})$, then $\q{\psi_d}$ can be globally asymptotically stabilized by adding a quasi-local Hamiltonian $H$ to the dissipation operators $L_k$,  if this Hamiltonian can satisfy (i) $H \q{\psi_d} = 0$  and (ii) $e^{i H t} \q{\phi}$ leaves $\mathcal{H}_0$ for all other $\q{\phi} \in \mathcal{H}_0$  (Proposition 3 in  \cite{ticozzi2014steady}).
\end{itemize}
Applying this to $\q{\psi_d} = (\q{00...0} + \q{11...1})/\sqrt{2}$ when the state space consists of $\text{span}\{\q{0},\q{1},...,\q{Q-1}\}^{\otimes n}$ involves the exact same reasoning as in \cite{ticozzi2014steady} for $Q=2$, namely: 
\begin{itemize}
\item With bipartite interactions, $\mathcal{H}_{(k)} = \text{span}(\q{00}_{j,\ell},\q{11}_{j,\ell}) \otimes \text{span}\{\q{0},\q{1},...,\q{Q-1}\}^{\otimes (n-2)}$ where $j,\ell$ are the pair of quQits associated to decoherence operator $L_k$. Then $\mathcal{H}_0 = \text{span}\{\q{00...0},\q{11...1}\}$ so there is no way to globally asymptotically stabilize this state with only quasi-local decoherence operators $L_k$.
\item Having $H \q{\psi_d} = 0$ requires 
\begin{equation}\label{eq:opusi}
H \q{00...0} = - H \q{11...1} \; .
\end{equation} 
But if each term in $H$ only acts on $<n/2$ subsystems, then each term on the left of \eqref{eq:opusi} contains $>n/2$ subsystems on $\q{0}$ while each term on the right contains $>n/2$ subsystems on $\q{1}$; i.e.~the left and right hand side of \eqref{eq:opusi} are orthogonal, leaving as only possibility that $H \q{00...0} = H \q{11...1} =0$. But in this case, the state $ (\q{00...0} - \q{11...1})/\sqrt{2} \in \mathcal{H}_0$ also remains invariant. \hfill $\square$
\end{itemize}
\end{proof}

The case with ancillas requires some adaptation, because we do not need to stabilize a pure state over the whole Hilbert space: only the data subsystems must converge towards $\q{\psi_d} = (\q{00...0} + \q{11...1})/\sqrt{2}$, while the ancillas need not converge. We thus here provide a more specific proof for the target state $\q{GHZ_+}$, whose gist one may seek to extract and extend to general settings.

\begin{Proposition}
The dynamics \eqref{eq:L0} with $n$ data qubits $(Q =2)$ and auxiliary subsystems, cannot globally asymptotically stabilize the GHZ state $(\q{00...0}+\q{11...1})/\sqrt{2}$ if each operator involves interaction of at most $<n/2$ data subsystems.
\end{Proposition}
\begin{proof}
Note that if the dynamics \eqref{eq:L0} has a solution with data subsystems remaining on $\q{GHZ_+}$ for all times, then it must have at least one steady state of the form $\bar{\rho} = \q{GHZ_+}\qd{GHZ_+} \otimes \bar{\rho}_{aux}$, even if this steady state is not globally attractive. Indeed, since the dynamics \eqref{eq:L0} is time-independent, a solution $\rho(t)$ always implies a solution $\rho_T(t) =\tfrac{1}{T}\int_0^{T} \rho(t+s)\, ds$ and in the limit $T \rightarrow +\infty$ the latter becomes time-independent.

We use for the data Hilbert space the following specific basis:
\begin{eqnarray} \label{eq:Ebasis}
\q{0+} &:=& (\; \q{000...00} + \q{111...11}\; ) \;/ \sqrt{2} \\ \nonumber
\q{0-}  &:=& (\; \q{000...00} - \q{111...11}\; ) \;/ \sqrt{2} \\ \nonumber
\q{1+} &:=& (\; \q{000...01} + \q{111...10}\; ) \;/ \sqrt{2} \\ \nonumber
\q{1-}  &:=& (\; \q{000...01} - \q{111...10}\; ) \;/ \sqrt{2} \\ \nonumber
... \\ \nonumber
\q{S+} &:=& (\; \q{100...00} + \q{011...11}\; ) \;/ \sqrt{2} \\ \nonumber
\q{S-}  &:=& (\; \q{100...00} -\q{011...11}\; ) \;/ \sqrt{2} \; ,
\end{eqnarray}
where $S=2^{n-1}-1$. Note that we order the two terms on each line such that the second term contains more qubits on $\q{1}$. Thus, $\q{s-} = Z_k \q{s+}$ for all $s \in 0,1,...,S$ and a phase-flip operator $Z_k$ on a \emph{data qubit $k \in 1,2,...,n$ which equals $\q{1}$ in the second term of $\q{s\pm}$}. In particular, $\q{GHZ_+} = \q{0+} = Z_k \q{0-}$ for any $k \in 1,2,...,n$ and for instance, $\q{3-} := (\q{00011} - \q{11100}) / \sqrt{2} = Z_k \q{3+}$ for $k \in 1,2,3$. By construction, for each $s \in 0,1,..,S$, there are at least $n/2$ such qubits on which $Z_k$ can be applied with this property.

Consider any quasi-local decoherence operator or any quasi-local term in the Hamiltonian of \eqref{eq:L0}, and denote it $D_{\ell}$. Each such $D_{\ell}$ acts non-trivially on strictly less than $n/2$ data qubits. Therefore:
\begin{itemize}
\item Take any $s$ $\in 0,1,...,S$ and denote $K_s$ the set of indices of data qubits which equal $\q{1}$ in the second term of the basis as written in \eqref{eq:Ebasis}. For instance, with $n=5$ and if $s=3$ then $K_s = \{1,2,3\}$. The set $K_s$ contains at least $n/2$ elements, hence at least one qubit $\bar{k}$ on which $D_{\ell}$ acts \emph{trivially}, thus for which $Z_{\bar{k}} D_\ell = D_\ell Z_{\bar{k}}$.
\item Taking the phase flip operator on $\bar{k}$, one checks that for each $\ell$ and for any $s$ $\in 0,1,...,S$:
\begin{eqnarray}\label{eq:app:symprop}
\qd{s\pm} D_{\ell} \q{0-} &=& \qd{s\mp} Z_{\bar{k}} D_{\ell} \q{0-} = \qd{s\mp} D_{\ell} Z_{\bar{k}}  \q{0-} =  \qd{s\mp} D_{\ell} \q{0+}   \; , \\ \nonumber
\qd{0-} D_{\ell} \q{s\pm} &=& \qd{0+} Z_{\bar{k}} D_{\ell} \q{s\pm} = \qd{0+} D_{\ell} Z_{\bar{k}}  \q{s\pm} =  \qd{0+} D_{\ell} \q{s\mp}   \;  .
\end{eqnarray}
Note that the result of these brackets are operators on the ancilla Hilbert space.
\item Take $D_{\ell}$ a decoherence operator. Writing down \eqref{eq:L0} and checking the block-diagonal part \emph{outside} $\q{GHZ_+}\qd{GHZ_+}$, a first condition for $\bar{\rho}$ to be invariant is that $\qd{s\pm} D_{\ell} \q{0+} \bar{\rho}_{aux} =0$ for each $\ell$. From \eqref{eq:app:symprop}, this also implies $\qd{s\pm} D_{\ell} \q{0-} \bar{\rho}_{aux} =0$. In other words, this first condition holds towards keeping $\tilde{\rho} := \q{GHZ_-}\qd{GHZ_-} \otimes \bar{\rho}_{aux}$ invariant as well.
\item For each decoherence operator $D_{\ell}$, denote $A_{\ell} = \qd{0+} D_{\ell} \q{0+}$ and $B_{\ell} = \qd{0+} D_{\ell}$. For each Hamiltonian term $D_{j}$, denote $P_{j} = \qd{0+} D_{j} \q{0+}$ and $Q_{j} = \qd{0+} D_{j}$. Writing down \eqref{eq:L0} and checking the remainder of the components, the second condition for $\bar{\rho}$ to be invariant is that
\begin{eqnarray} \label{eq:epepep}
0 &=& \sum_{\ell} A_{\ell} \bar{\rho}_{aux} A_{\ell}^\dagger - \tfrac{1}{2} A_{\ell}^\dagger A_{\ell} \bar{\rho}_{aux} - \tfrac{1}{2} \bar{\rho}_{aux} A_{\ell}^\dagger A_{\ell} - i \sum_{j}[P_j, \bar{\rho}_{aux}]  \; , \\ \nonumber
0 &=& -\tfrac{1}{2} \, \sum_{\ell} \bar{\rho}_{aux} A_{\ell}^\dagger B_{\ell} - i \sum_j \bar{\rho}_{aux} Q_j \; .
\end{eqnarray}
Now by \eqref{eq:app:symprop}, we also have $A_{\ell} = \qd{0-} D_{\ell} \q{0-}$  and  $P_{j} = \qd{0-} D_{j} \q{0-}$, meaning that the first line holds for keeping $\tilde{\rho}$ invariant as well. 

Similarly, annihilating the second line of \eqref{eq:epepep} multiplied by $\q{s\pm}$, for any $s\neq 0$ or for $\q{s\pm}=\q{0-}$, implies by \eqref{eq:app:symprop} that this line multiplied by $\q{s\mp}$ is annihilated when redefining $A,B,P,Q$ with $\q{0-}$ instead of $\q{0+}$. In other words, again the corresponding conditions for keeping $\bar{\rho}$ invariant and for keeping $\tilde{\rho}$ invariant are satisfied together. 

\item The conditions discussed in the last two items are all the ones for keeping a state invariant, as they are just obtained by annihilating each component of \eqref{eq:L0} applied to the steady state. Thus, from those two items, the conditions for keeping $\bar{\rho}$ invariant imply that we would also keep $\tilde{\rho}=\q{GHZ_-}\qd{GHZ_-} \otimes \bar{\rho}_{aux}$ invariant. It is thus impossible to have $\bar{\rho}$ globally asymptotically stable.  \hfill $\square$
\end{itemize} 
\end{proof}

Note that in this result, we assume nothing about the ancillary subsystems: it may even be a single big subsystem connected individually to all the data subsystems. The same impossibility proof keeps holding when allowing ancilla subsystems \emph{and} $Q>2$ in the data subsystems.

\section{Appendix B: Details of alternative GHZ reservoir proposals}

\subsection{Reservoirs based on ancilla jump conditioning}\label{appsec:jcond}

The description is organized like Section \ref{ssec:statecond}, with data and ancillas now moving at the same time. This has some operational consequences. Indeed, to ensure that ancillas keep jumping around the clock cycle, we must add in parallel to the jump operator updating ancilla and data qubit, a jump operator which updates the ancilla similarly even if the data qubit does not (have to) jump. This can probably be revised at places, but we prefer to make this choice systematically such that again, the evolution of the ancillas in our schemes is not influenced by the data state.

\subsubsection{Resets at ancilla jumps}\label{sssec:jc:step0}

We thus start again with a preliminary description considering ancillas with only two levels $\q{g},\q{e}$.
The main idea is to make the data qubit reset instantaneously when the ancilla \emph{jumps to $\q{e}$}, with channels like:
\begin{eqnarray}\label{eq:1stSolution0}
M_k &=& \sqrt{\kappa_d}\q{g}\qd{e}_k \;\; , \quad  N_{k,1} = \sqrt{\kappa_u} \q{e,+}\qd{g,+}_k \; \; , \;\; N_{k,2} = \sqrt{\kappa_u} \q{e,+}\qd{g,-}_k \; , \\ \nonumber 
L_k && \text{ as in \eqref{eq:LTV}}\; .
\end{eqnarray}
The separation of channels $N_{k,1}$ and $N_{k,2}$ is needed to avoid a dark state, or we could take $N_k=N_{k,1}+N_{k,2}$ with a Hamiltonian breaking the dark state. 

The remaining issue, like in Section \ref{ssec:statecond}, is to synchronize the jumps of all the ancillas. In fact this issue becomes even more essential with ``jump-conditioning''. Indeed, even in an approach like \cite{Verstraete2009} where a single ancilla is assumed to be connected to all the data qubits, we cannot apply \eqref{eq:1stSolution0} verbatim: either, assuming $n$ jump operators, the ancilla will reset just a single data qubit at random; or, assuming a single jump operator, it would now involve all the $n+1$ subsystems at once in a single operator. 

Associating an ancilla with more levels to each data qubit, synchronized jumping can be implemented as follows.

\subsubsection{Correlating the ancilla jumps}\label{sssec:jc:step1}

The idea is the same as in Section \ref{sssec:sc:step1}, namely to induce an ancilla clock cycling essentially through
$$\q{gg...g} \rightarrow \q{ee...e} \rightarrow \q{mm...m} \rightarrow \q{gg...g} \rightarrow...$$
thanks to ancilla jumps stimulated by the neighbors. However, as soon as an ancilla jumps to $\q{e}$, the associated data qubit jumps to $\q{+}$. Various constructions are possible and we just list two, illustrated on Figure \ref{fig:clockwork}b, before commenting on their properties.

\paragraph*{Tripartite:}  Correlating a \emph{stimulated} ancilla jump with a data jump requires, a priori, tri-partite interaction. Admitting such operators, with 3-level ancillas, we could propose a reservoir like:
\begin{eqnarray*}
L_k && \text{ as in \eqref{eq:LTV}} \\
N_{k,spr} &=&  \sqrt{\kappa_u} \q{e+}\qd{g-}_k  \quad , \quad N_{k,spi} =  \sqrt{\kappa_u} \q{e+}\qd{g+}_k \\
M_{k,sp}  &=& \sqrt{\kappa_d} \q{m}\qd{e}_k +  \sqrt{\kappa_t} \q{g}\qd{m}_k \\
N_{k,st+r} &=& \sqrt{\kappa_{st}}\, \q{e}\qd{e}_k \; \q{e+}\qd{g-}_{k+1} \quad , \quad N_{k,st+i} = \sqrt{\kappa_{st}}\, \q{e}\qd{e}_k \; \q{e+}\qd{g+}_{k+1} \\
M_{k,st+} &=&   \sqrt{\kappa_{st}}\,(\q{mm}\qd{em}_{k,k+1} + \q{gg}\qd{mg}_{k,k+1})   \\
N_{k,st-r} &=& \sqrt{\kappa_{st}}\, \q{e+}\qd{g-}_{k}\; \q{e}\qd{e}_{k+1}   \quad , \quad N_{k,st-i} = \sqrt{\kappa_{st}}\, \q{e+}\qd{g+}_{k}\;  \q{e}\qd{e}_{k+1} \\
M_{k,st-} &=&  \sqrt{\kappa_{st}}\,( \q{mm}\qd{me}_{k,k+1} + \q{gg}\qd{gm}_{k,k+1}  ) \; .
\end{eqnarray*}
The indices $_r$ and $_i$ stand for reset and idle on data, $_+$ and $_-$ denote left or right neighbor conditioning, while $_{sp}$ and $_{st}$ distinguish spontaneous or neighbor-stimulated processes.

\paragraph*{Bipartite:} The above scheme can be adapted to bi-partite interactions by adding a fourth ancilla level $\q{f}$, such that each ancilla would transition from $\q{g}$ to $\q{f}$ and then only to $\q{e}$. The idea is that the reset of the associated data qubit happens during the very fast \emph{spontaneous} ancilla jump from $\q{f}$ towards $\q{e}$, thus involving only a bipartite interaction. Meanwhile, to keep essentially the same clock behavior, an ancilla in $\q{g}$ gets attracted to $\q{f}$ as soon as one of its neighbors is \emph{in $\q{f}$ or in $\q{e}$.}

Explicitly, the associated channels could be:
\begin{eqnarray}\label{eq:step2bsynch}
L_k && \text{ as in \eqref{eq:LTV}} \\ \nonumber
M_{k,sp} &=& \sqrt{\kappa_u} \q{f}\qd{g}_k + \sqrt{\kappa_d} \q{m}\qd{e}_k +  \sqrt{\kappa_t} \q{g}\qd{m}_k \\ \nonumber
N_{k,r} &=& \sqrt{\kappa_f} \q{e,+}\qd{f,-}_k \quad \; , \quad  N_{k,i} = \sqrt{\kappa_f} \q{e,+}\qd{f,+}_k \\ \nonumber
M_{k,st1+} &=& \sqrt{\kappa_{st}}\, ( \q{ff}\qd{gf} +  \q{mm}\qd{em} + \q{gg}\qd{mg}  )_{k,k+1} \quad , \quad M_{k,st2+} = \sqrt{\kappa_{st}}\, \q{fe}\qd{ge}_{k,k+1}\\ \nonumber 
M_{k,st1-} &=& \sqrt{\kappa_{st}}\, ( \q{ff}\qd{fg} +  \q{mm}\qd{me} + \q{gg}\qd{gm}  )_{k-1,k} \quad , \quad M_{k,st2-} = \sqrt{\kappa_{st}}\, \q{ef}\qd{eg}_{k-1,k} \; ,
\end{eqnarray}
for each $k=2,3,...,n-1$, and one of the last two channels dropping for $k=1$ and $k=n$. Indices $_1$ and $_2$ distinguish stimulated excitation to $\q{f}$ when a neighbor is in $\q{f}$ or in $\q{e}$. \vspace{3mm}

Except for the rates, discussed below, the schemes' properties are similar to those of Section \ref{sssec:sc:step1}.
\begin{itemize}
\item No quantum coherence at all needs to be protected among ancilla levels: they only need to be correlated classical Dits.
\item The dissipation channels are partly split into several channels to avoid dark states, but a linear combination of e.g. $M_{k,sp}, N_{k,r}, N_{k,i}$ associated to a Hamiltonian can have the same effect, if this appears less difficult for implementation.
\item Conversely, coherences in the channel operators are not essential (except of course in $L_k$) and one might as well separate them into more channels. Dissipation rates need not be equal for every $k$, just their order of magnitude matters.
\item The above constructions are meant to facilitate analysis, thanks to designing ancilla dynamics not being influenced by the data state.
\end{itemize}
Regarding the choice of dissipation rates, let us comment on the bi-partite scheme for a fairer comparison with Section \ref{sssec:sc:step1}. We must have essentially:
\begin{enumerate}
\item ancillas behave as an almost synchronized clock: $\kappa_{st} \gg \kappa_d, \kappa_t, \kappa_u$ 
\item transition through $\q{f}$ maintains clock synchronization on $\q{e}$: $(\frac{1}{\kappa_{st}}+\frac{1}{\kappa_f}) \ll \tfrac{1}{\kappa_d}$ 
\item at a reset, \eqref{eq:LTV} has little time to act until all qubits have reset: $(\frac{1}{\kappa_{st}}+\frac{1}{\kappa_f}) \ll  \frac{1}{\kappa_c}$\item data qubits have ample time to converge with \eqref{eq:LTV} after each reset round:  $\frac{1}{\kappa_c} \ll (\frac{1}{\kappa_d}+\frac{1}{\kappa_t}+\frac{1}{\kappa_u})$.
\end{enumerate}
Altogether, this yields the rough timing guidelines:
\begin{eqnarray}\label{eq:step2btc}
&& \left\{ \frac{1}{\kappa_d} \;,\; \frac{1}{\kappa_t} \;,\; \frac{1}{\kappa_u} \right\} \sim T_1 \quad \gg \quad \frac{1}{\kappa_c}  \sim T_2 \quad \gg  \quad \left\{ \frac{1}{\kappa_f}  \; , \; \frac{1}{\kappa_{st}} \right\}\sim T_3 \; . 
\end{eqnarray}
The fidelity lost due to resets pushing the state away from $\q{GHZ_+}$ is now dominated by $\tfrac{1}{\kappa_c} /  (\tfrac{1}{\kappa_d}+\tfrac{1}{\kappa_t}+\tfrac{1}{\kappa_u})  \sim T_2/T_1$, while the inaccuracy in resetting to $\q{++...+}$ adds an error of order $(\frac{\kappa_c}{\kappa_f}+\frac{\kappa_c}{\kappa_{st}}) \sim T_3/T_2$. Although we do not win an order of magnitude, these are still less error terms than in Section \ref{sssec:sc:step1}, at the cost of an additional ancilla level. Also, compared to Section \ref{sssec:sc:step1}, here $\kappa_f$ somewhat replaces $\kappa_r$, but with fewer constraints. Indeed here, no further data resets happen once every ancilla has jumped to $\q{e}$. Therefore, tightly synchronizing this jump, automatically implies a short reset period and a good reset effect despite the presence of $\kappa_c$.

\subsubsection{Jump operators for the GHZ stabilizers}\label{sssec:jc:step2}

Like in Section \ref{ssec:statecond}, we cannot take $\kappa_c$ too large in the above scheme, because else the $L_k$ would have significant (and deteriorating) effect before all qubits have been reset synchronously. Conditioning the $L_k$ on ancilla states, or in line with the present context on ancilla \emph{jumps}, looks particularly tempting here, as it would seem to remove the intermediate timescale $T_2$. 

A first issue with this is that, like in Section \ref{ssec:statecond}, such conditioning appears to involve either tri-partite interactions or significantly more complicated ancillas. The jump-conditioning context warrants two more points of attention.
\begin{itemize}
\item[1.] Since the $L_k$ preserve the GHZ state, it is a priori beneficial to apply them as often as possible. Indeed, while the lowest rate of protection will be dominated by the characteristic time $T_1$ for phase-flip corrections, the $L_k$ alone are sufficient for correcting bit-flip errors on the data qubits. Having a faster bit-flip correction could be beneficial, in particular considering the existence of physical systems implementing biased noise qubits \cite{mirrahimi2014dynamically} where phase flips are much less likely than bit-flips. Since conditioning the application of $L_k$ on e.g.~an ancilla jump from $\q{e}$ to $\q{m}$ makes the bit-flip correction as slow as $T_1$, one should in turn identify a clear benefit before considering such operation.
\item[2.] As analyzed in the main text, the $L_k$ jumps must be applied \emph{in some order} after each reset to ensure reaching $\q{GHZ_+}$. It is thus not enough to apply them once in random order. Furthermore, before applying $L_k$ we must ensure that \emph{both} associated data qubits have reset. All this organization rather points towards wave-propagation proposals, as we describe next.
\end{itemize}

\subsection{Reservoirs based on ancillas and a propagating wave}

We here present some concrete schemes along the principles of Section \ref{ssec:wave}.

\subsubsection{Tripartite interaction with two timescales}\label{sssec:wave3}

When allowing tripartite interaction, it is possible to design relatively powerful wave-inspired reservoirs based on just two different timescales. This expressly hinges on the observation that both $\q{+}\qd{-}$ and $L_k$ just have to be applied in order from lowest to highest data index $k$. We start with a scheme whose logic is very simple to follow. We then propose a second scheme which appears both more powerful and simpler in terms of resources. In both constructions, we assign one ancilla $k$ to each \emph{pair} of consecutive data qubits $(k,k+1)$.

\paragraph*{Tripartite, jump-conditioning:} Each operation on data is triggered by an ancilla jump. We use ancillas with 4 levels, like in the jump-conditioning process of Section \ref{appsec:jcond}. The idea is entirely sequential: wait for a long time on $\q{GHZ_+}$ (in absence of perturbations), before launching the following jump sequence which should end up in $\q{GHZ_+}$ as fast as possible:
\begin{eqnarray}
\nonumber \text{reset qubit $1$ with $\q{+}\qd{-}$ ;}  & \text{reset qubit $2$ with $\q{+}\qd{-}$} ; & \text{apply $L_1$ ;} \\
\label{eq:seqnaive} & \text{reset qubit $3$ with $\q{+}\qd{-}$ ;} & \text{apply $L_2$ ;} \\
\nonumber ... ; &  \text{reset qubit $k+1$ with $\q{+}\qd{-}$ ;} & \text{apply $L_k$ ;} ...
\end{eqnarray}
Thanks to commutation of operators on distinct subsystems, this sequence is indeed strictly equivalent to applying first $\q{+}\qd{-}$ on each data qubit, then the $L_k$ in the favorable order from $k=1$ to $k=n-1$. Yet it avoids to wait until all resets have been done, before launching the sequence of $L_k$; this is both more efficient and easier to implement locally. This sequence of events could be implemented with an ``ancilla automaton'' using the following operators:
\begin{eqnarray}\label{eq:wave3pj}
N_{1,r12} &=& \sqrt{\kappa_u} \q{e,+,+}\qd{g,-,-}_{1,1,2} \quad \; , \quad  N_{1,r1} = \sqrt{\kappa_u} \q{e,+,+}\qd{g,-,+}_{1,1,2} \\ \nonumber
N_{1,r2} &=& \sqrt{\kappa_u} \q{e,+,+}\qd{g,+,-}_{1,1,2} \quad \; , \quad  N_{1,i} = \sqrt{\kappa_u} \q{e,+,+}\qd{g,+,+}_{1,1,2} \\ \nonumber
N_{k,r} &=& \sqrt{\kappa_{st}} \q{e,+}\qd{f,-}_{k,k+1} \quad \; , \quad N_{k,i} = \sqrt{\kappa_{st}} \q{e,+}\qd{f,+}_{k,k+1} \quad \text{for } k=2,3,...,n-1 \\ \nonumber
M_{k} &=&  \sqrt{\kappa_{st}} \q{g,f}\qd{m,g}_{k,k+1} \quad \text{for } k=1,2,...,n-2 \\ \nonumber
\tilde{L}_{k,r} &=&  \sqrt{\kappa_c} \q{m}\qd{e}_k \otimes (\q{11}\qd{10} + \q{00}\qd{01})_{k,k+1} \quad \; , \;\\ \nonumber
&& \tilde{L}_{k,i} \;=\; \sqrt{\kappa_c} \q{m}\qd{e}_k \otimes (\q{11}\qd{11} + \q{00}\qd{00})_{k,k+1} \; \quad \text{for } k=1,2,...,n-2 \\ \nonumber
\tilde{L}_{n-1,r} &=&  \sqrt{\kappa_c} \q{g}\qd{e}_{n-1} \otimes (\q{11}\qd{10} + \q{00}\qd{01})_{n-1,n} \quad \; , \; \\ \nonumber
&& \tilde{L}_{n-1,i} \;=\; \sqrt{\kappa_c} \q{g}\qd{e}_{n-1} \otimes (\q{11}\qd{11} + \q{00}\qd{00})_{n-1,n} \; .
\end{eqnarray}
Starting with ancillas in $\q{gg..g}$, the sequence is launched by one of the $N_{1,...}$: as ancilla 1 jumps to $\q{e}$ it resets data qubits 1 and 2 towards $\q{++}$. Then the $\tilde{L}_{1,...}$ can act, so ancilla 1 jumps to $\q{m}$ while applying $L_1$ on the data, or projecting onto the subspace on which $L_1$ had to act idle. Ancilla 1 finally jumps to $\q{g}$ under the action of $M_1$, while kicking ancilla 2 towards level $\q{f}$. This triggers via $N_{2,...}$ the reset of data qubit 3 towards $\q{+}$ while ancilla 2 jumps to $\q{e}$, and so on. 

Remarks on some details:
\begin{itemize}
\item For each intended ancilla transition, several operators are needed in order to avoid dark states from the associated data evolution. 
This concerns in particular the presence of both $\tilde{L}_{k,r}$ and $\tilde{L}_{k,i}$. One easily checks that $\tilde{L}_{k,i}$ as well preserves the eigenstates of $\sigma_x^{\otimes n}$, as is required to stabilize a well-defined superposition of $\q{00...0}$ and $\q{11...1}$ with this scheme.
\item The ancilla level $\q{f}$ is introduced just to avoid having ancilla $k-1$, ancilla $k$ and data qubit $k+1$ in a single operator. Indeed, although each of the operators would remain just tripartite in absence of $\q{f}$, together they would require ancilla $k-1$ to have connections to data qubits $k-1$, $k$, and $k+1$. This would possibly imply a significantly harder layout. If such a connection is available, then the level $\q{f}$ could be skipped; in this case, the $M_k$ and $N_{k+1,...}$ for $k\geq 1$ could be merged, so the $N_{k,...}$ would remain only for $k=1$ to launch a clock cycle.
\item The last ancilla $n-1$ needs to trigger no neighbor and thus skips the state $\q{m}$ entirely.
\item Like for the other constructions, the ancillas only encode classical information on their levels.
\end{itemize}

With this strategy, the next operation at site $k$ is triggered as soon as site $k-1$ has finished, circumventing the inefficient waiting times. Hence, the reservoir just relies on taking
\begin{equation}\label{eq:wave3pjp}
\kappa_c, \kappa_{st} \gg \kappa_u \; ,
\end{equation}
while $\kappa_u$ should dominate the typical perturbation characteristic rate $1/T_0$. Thanks to the sequential construction, we need one less timescale separation compared to the ``random order'' solutions of Sections \ref{ssec:statecond} and \ref{appsec:jcond}.

\paragraph*{Tripartite, ancilla qubits:} While the ``automaton'' reservoir implementing \eqref{eq:seqnaive} is easy to understand, it is not the most efficient one. In particular, the data qubits could keep applying the $L_k$ as described in \eqref{eq:LTV} more often, namely as soon as they are not resetting. Besides the potential benefits for bit-flip corrections, this allows significant simplification of the conditioning, reducing the ancillas to qubits ($D=2$). The following reservoir, with one ancilla per data, works in this sense:
\begin{eqnarray}\label{eq:wave3pm}
N_{1,r} &=& \sqrt{\kappa_u} \q{e,+,}\qd{g,-}_{1,1} \quad \; , \quad  N_{1,i} = \sqrt{\kappa_u} \q{e,+}\qd{g,+}_{1,1} \\ \nonumber
N_{k,r} &=& \sqrt{\kappa_{st}} \q{g,e,+}\qd{e,g,-}_{k-1,k,k} \quad \; , \quad N_{k,i} = \sqrt{\kappa_{st}} \q{g,e,+}\qd{e,g,+}_{k-1,k,k} \\ \nonumber
N_{k,v} &=& \sqrt{\kappa_{st}} \q{g,e}\qd{e,e}_{k-1,k} \quad \; \quad \text{for } k=2,3,...,n-1 \\ \nonumber
\tilde{L}_k &=&  \sqrt{\kappa_c} \q{g}\qd{g}_k \otimes (\q{11}\qd{10} + \q{00}\qd{01})_{k,k+1} \; .
\end{eqnarray}
The key idea is to switch off $L_k$ as soon as a reset is performed on data qubit $k$, and to switch it back on once the reset has been done on qubit $k+1$ too.  Thus, when the $N_{1,...}$ launch a reset cycle, they reset qubit 1 and switch off $\tilde{L}_1$ by putting the first ancilla into $\q{e}$. Next, $N_{2,r}$ performs a reset on qubit 2, while at the same time the exchange $\q{g,e}\qd{e,g}_{1,2}$ switches back on $\tilde{L}_1$ and switches off $\tilde{L}_2$; and so on. 

Compared to the previous scheme, \eqref{eq:wave3pm} is thus (possibly) applying the $L_k$ more frequently and repeatedly: (i) while the preceding data qubits are resetting, (ii) while the following data qubits are resetting and (iii) while other qubits are applying their $L_j$. Point (i) has no impact since at this time we are away from GHZ anyways, and the soon-to-happen reset on data qubits $\geq k$ makes their current state (with or without $L_k$ applied) irrelevant. Point (ii) is not detrimental since $L_k$ commutes with all the remaining resets, so all its actions can equivalently be seen as happening after all resets have been completed; with the preceding scheme we used the same argument, but mentioning a single jump with each $L_k$. Finally, point (iii) is fine because starting from $\q{++...+}$ (by the argument for point (ii)), convergence towards $\q{GHZ_+}$ is ensured by applying the \emph{subsequence} $L_1,L_2,...,L_{n-1}$, irrespective of which other $L_j$ the full sequence may contain. Hence, these more frequent $L_k$ can only do better.

We have the same remarks as for the other constructions:
\begin{itemize}
\item Splitting into several operators avoids dark states. In particular, the role of the $N_{k,v}$ is just to ensure global convergence of the ancilla reservoir, i.e.~avoiding to get stuck if by chance several ancillas were in $\q{e}$; we do not care about the associated data action since this situation should nominally never happen.
\item The ancillas only encode classical information on their levels.
\end{itemize}
The rates still just have to satisfy
$$\kappa_c, \kappa_{st} \gg \kappa_u $$
for good performance. The construction \eqref{eq:wave3pm} thus remains essentially as fast as \eqref{eq:wave3pj}, while using only two-level ancillas and applying the $L_k$ more continuously.

\subsubsection{Bipartite interaction}\label{sssec:wave2}

We now address the construction of a scheme with bipartite interactions only. Since conditioning $L_k$ on an ancilla would necessarily imply tri-partite interaction, we face the same options as in the non-wave constructions:
\begin{itemize}
\item[(i)] Either leave the $L_k$ on all the time, like in e.g.~Section \ref{sssec:sc:step1}, while applying the resets in a wave. This implies small fidelity losses associated to applying $L_k$ while data qubit $k$ has reset to $\q{+}$ and data qubit $k+1$ still has to.
\item[(ii)] Or, separate the $L_k$ operator in two steps, like in e.g.~Section \ref{sssec:sc:step2}. This allows to switch off $L_k$, but between these two steps it adds a downtime, during which moreover the ancillas must maintain quantum coherences.
\end{itemize}
Since a scheme of type (ii) appears not too practical, we briefly describe a scheme of type (i). A more efficient scheme with $L_k$ switch-off and bipartite interactions is proposed in Section \ref{sec:3level} when working with data qutrits ($Q=3$).\\

Associating one ancilla qutrit to each data qubit ($Q=2$, $D=3$, $m=n$), we can propose the following reservoir where the resets to $\q{+}$ follow a wave:
\begin{eqnarray}
\nonumber L_k && \text{ as in \eqref{eq:LTV}} \\
\label{eq:wave2allTV}
M_{1} &=& \sqrt{\kappa_u} \q{e}\qd{g} \\ \nonumber
M_{k,r} &=& \sqrt{\kappa_{st}} \q{g,e}\qd{m,g}_{k-1,k}  \quad , \quad  M_{k,i} = \sqrt{\kappa_{st}} \q{g,e}\qd{m,m}_{k-1,k} \; , \\ \nonumber
&& M_{k,v} = \sqrt{\kappa_{st}} \q{g,e}\qd{m,e}_{k-1,k} \quad , \quad \text{ for } k=2,3,...,n \\ \nonumber
N_{k,r} &=& \sqrt{\kappa_{st}} \q{m,+}\qd{e,-}_{k,k}  \quad , \quad  N_{k,i} = \sqrt{\kappa_{st}} \q{m,+}\qd{e,+}_{k,k}  \text{ for } k=1,2,...,n \; .
\end{eqnarray}
The nominal operator sequence for the resets would be $M_1$; $N_{1,r}$ or $N_{1,i}$; $M_{2,r}$; $N_{2,r}$ or $N_{2,i}$; ... ,  while the $L_k$ have a (smaller) probability to act at any time. Details are similar to the other schemes, among others:
\begin{itemize}
\item The two operators $N_{k,...}$ ensure a reset to $\q{+}_k$ whatever the associated data qubit state. The $M_{k,\{i,v\}}$, not mentioned in the nominal sequence, are added to avoid getting stuck at non-nominal ancilla states. An exception to this is ancilla $k=n$, which does not jump down to $\q{g}$ as it excites no next neighbor, and thus nominally applies $M_{k,i}$; an alternative would be to reduce this last ancilla to a qubit.
\item Ancillas only encode classical information.
\end{itemize}
Compared to the tripartite coupling schemes, fidelity is lost when an $L_k$ jump occurs between applying $N_{k,...}$ and $N_{k+1,...}$ in this sequence. Making this event unlikely requires to re-instate a separation of timescales for good reservoir operation:
$$ \kappa_{st} \gg \kappa_{c} \gg \kappa_u \; .$$
The benefit of observing the wave property is that this intermediate timescale will involve no $n$-dependence: after $N_{k,...}$, we just have to wait for $N_{k+1,...}$ to happen, not for all resets to complete.

\section{Appendix C: Details of performance analysis}

\subsection{Exact simplifications of the ancilla-based architecture dynamics}

We here give more details about the way in which the ancilla-based clock dynamics can be reduced, without involving any approximations.

\paragraph*{1. Ancilla evolution is not influenced by data state:} The description of the system architecture quite speaks for itself. The formal property goes as follows.
\begin{itemize}
\item Take the Lindblad equation \eqref{eq:L0} with the corresponding operators;
\item Plug in any state $\rho_t$ of the joint system;
\item Compute the time derivative of the ancillas states, e.g.~trace$\left(\q{ee}\qd{ge}_{k,k+1} \, \tfrac{d}{dt}\rho_t\right)$
\item Observe that those rates only depend on $\rho^{(A)}_t$, the partial trace of $\rho_t$ over the data subsystems.
\end{itemize}
The computations involve no originality and are left for the interested reader. 
We next explain how, furthermore, the ancillas can rigorously be reduced to a purely classical system.

\paragraph*{2. On the phase/coherence of ancillas in the canonical basis:}  We consider the system  \eqref{eq:L0} and the partial trace over data qubits, obtaining a Lindblad equation on ancillas only as explained in the previous paragraph.

Assume that the density matrix $\rho_t$ only has population on the diagonal in the canonical ancillas basis, i.e.:
$$\qd{r} \rho_t \q{s} = 0 \quad \text{for all } r \neq s \in \{g,e,m\} \; .$$
Our claim is that the Lindblad equation preserves this property. This is in fact easy to check: any particular jump operator $M_k$ or $N_k$  maps an ancillas canonical state to an orthogonal ancillas canonical state, so no coherences can ever appear. 

A stronger claim would be that if any coherences are present initially, then they exponentially vanish over time. Although such property can be put in place, it is not at all essential for our analysis, so we leave this proof for the interested reader.

\subsection{Analysis of the ancillas clock Markov chain}

We here provide the detailed analysis towards the results summarized in Section \ref{sec:an:clockancillas}. We thus start by considering the Markov chain defined by \eqref{eq:an:MCX}.

	
\subsubsection{With stimulated jumps only, the string of ancilla qubits exponentially converges to a distribution over $\q{gg...g}$, $\q{mm...m}$ and $\q{ee...e}$.}\label{section2}

In this first item, we thus prove that with only the interaction in $\kappa_{st}$, the distribution $p$ over ancilla states converges, from any initial state, to span$\{\q{gg..g}, \, \q{mm..m}, \, \q{ee..e} \}$.
	
With $\kappa_u=\kappa_d=\kappa_t=0$, we are thus in the situation where the only interactions possible are an ancilla on $g$ attracting a neighbour ancilla from level $m$ to level $g$,  an ancilla on $e$ attracting a neighbour ancilla from level $g$ to level $e$, and  an ancilla on $m$ attracting a neighbour ancilla from level $e$ to level $m$, following the cycle $g \rightarrow e \rightarrow m \rightarrow g...$.
	
The evolution of the ancillas can be seen as a sequence of steps where one ancilla changes at each step. We want to show, with only those interactions, that starting from any configuration, we converge in a finite number of steps towards a state where all the ancillas are either on $\q{gg...g}$, or $\q{mm...m}$, or $\q{ee...e}$. For this, we define a frontier $F_{A,B} $ as a separation between two consecutive ancillas $A$ and $B$ in the chain, with $A,B \in$ $ \{g,e,m\}$ and $A \neq B $. We now demonstrate that the number of frontiers falls to zero in a finite number of steps.

\begin{figure}
\includegraphics[width=130mm, trim=15 310 80 85, clip]{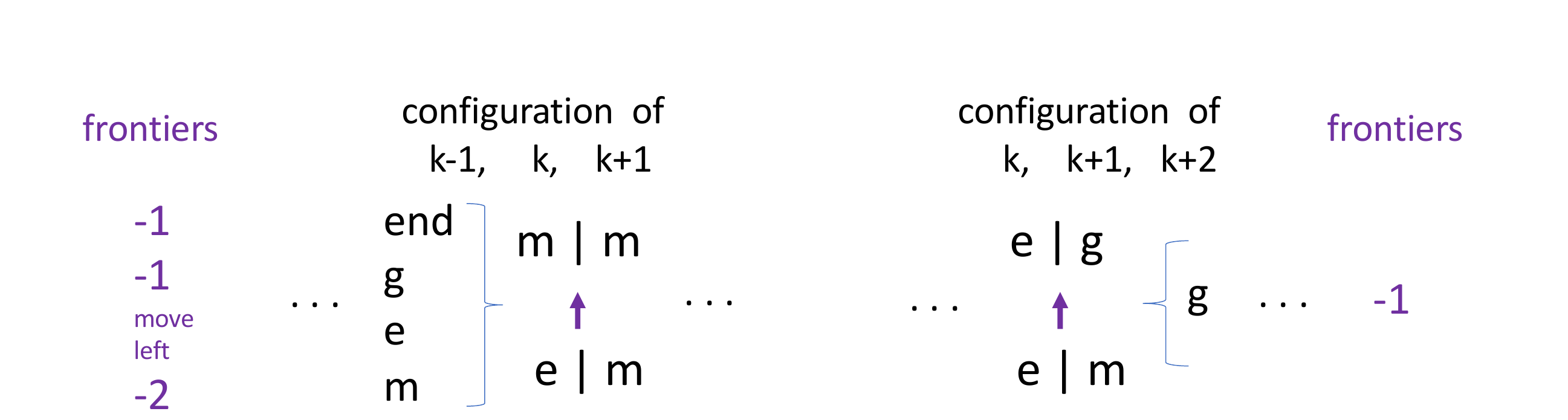}
\caption{Two possible evolutions (left and right) of ``frontier'' $F_{e,m}$ on ancilla qutrits $k$ and $k+1$ (bottom) under only stimulated jump operators $\kappa_{st}\neq 0$. The outer columns (purple) list the associated evolution of the number of frontiers depending on the rest of the chain. The situation for other frontiers is obtained by symmetry. \label{fig:frontiersjump}}
\end{figure}	
	
\begin{itemize}
\item Consider two adjacent ancillas $k,k+1$ forming a frontier $F_{e,m}$. Figure \ref{fig:frontiersjump} lists the implications of all the possible jumps involving those ancillas. 

The jump operator involving those ancillas can only make them jump to $\q{m,m}_{k,k+1}$. If ancilla $k=1$ or if ancilla $k-1$ was in $\q{g}$, then this decreases the number of frontiers by 1. If ancilla $k-1$ was in $\q{m}$, then this decreases the number of frontiers by 2. If ancilla $k-1$ was in $\q{e}$, then the frontier $F_{e,m}$ moves down one index, to ancillas $k-1,k$.

The jump operator involving ancillas $k-1,k$ can switch ancilla $k$ towards $\q{m}$, iff ancilla $k-1$ was on $m$; this decreases the number of frontiers by 2.

Finally, the jump operator involving ancillas $k+1,k+2$ can switch ancilla $k+1$ towards $\q{g}$, iff ancilla $k+1$ was on $\q{g}$; this decreases the number of frontiers by 1.

Altogether: either the number of frontiers strictly decreases, or the frontier $F_{e,m}$ moves towards lower indices.

\item By circular symmetry on the clock levels, the same is true for frontiers $F_{m,g}$ and $F_{g,e}$. By reversal of the index order, jumps involving the ancillas of a frontier $F_{m,e}$, $F_{g,m}$ or $F_{e,g}$ either strictly decrease the number of frontiers, or move this frontier towards \emph{higher} indices.

\item Consider two adjacent ancillas $k,k+1$ on the same level e.g.~$\q{ee}$. Ancilla $k+1$ can only jump towards $m$, iff ancilla $k+2$ was on $m$, thus moving towards lower indices the frontier $F_{e,m}$ which was present at ancillas $k+1,k+2$. Simlarly, ancilla $k$ can only jump towards $m$ by moving the frontier $F_{m,e}$ from indices $k-1,k$ towards $k,k+1$. 

Altogether, and by symmetry for $\q{gg}$ and $\q{mm}$: the number of frontiers does not change, and the frontiers move in the same way as for the previous items. 
\end{itemize}
To summarize, at any jump, either the number of frontiers strictly decreases, or a frontier moves in a unique direction. This process can only go on for a finite number of steps, since frontiers can only move up to the end of the chain before at least one frontier has to disappear. We have thus demonstrated that, in a stochastic jump viewpoint, the number of frontiers must go to zero after a finite number of jumps. Since every jump is a Poisson point process of parameter $\kappa_{st}$, the distribution $p$ over ancilla states converges to span$\{\q{gg...g}, \, \q{mm...m}, \, \q{ee...e} \}$ exponentially.\\

It may be worth nothing that this property would not be as obvious for every interconnection graph among ancillas. Indeed, imagine for instance ancillas interconnected in a cycle, instead of a chain. Then, it is perfectly possible that several frontiers keep moving around the cycle, with thus blocks of qutrits synchronized in $\q{gg...g}$, others in $\q{mm...m}$, others in  $\q{ee...e}$, propagating around the graph. A convergence analysis in this case would require to delve into the respective probabilities of the trajectories, including those with an infinite number of jumps. The ends of the chain thus play a determining role in our simple analysis of exponential convergence.


\subsubsection{With stimulated jumps dominating spontaneous jumps, the string of ancilla qubits exponentially converges towards a unique steady state with little population outside $\q{gg...g}$, $\q{mm...m}$ and $\q{ee...e}$.}\label{section3}

Now consider the full ancillas clock, with $\kappa_{st} \gg \kappa_u,\kappa_d,\kappa_t > 0$. This situation can be seen as a perturbation of the previous case, with the perturbed transition matrix $A(\epsilon) = \kappa_{st} (A_0 + \epsilon A_1)$ where $A_0$ represents the stimulated jumps (Section \ref{section2}) and  $A_1$  represents the spontaneous jumps. Anticipating that we will take $\kappa_d > \kappa_u, \kappa_t$, we thus denote $\kappa_d / \kappa_{st} =: \epsilon$. 

To quickly characterize the steady state of this perturbed transition matrix, we can use one of the results of Theorem S6.1, chapter S6, of \cite{gohberg2005matrix}:

\begin{Proposition}\label{PropAeps}
Take $A(\epsilon)$ a complex matrix-valued function analytic in a domain $\Omega \in \mathbb{C}$ with $r = max_{\epsilon \in \Omega} (rank(A(\epsilon)))$. There exist  $y_{r+1}(\epsilon)$, ...$y_N(\epsilon)$,  some analytic vector-valued functions which constitute a basis for the null space of $A(\epsilon)$,  for all $\epsilon \ge 0$ except for a set of isolated points which consists exactly of those $\epsilon_0$ for wich $rank(A(\epsilon_0)) < r$. For such exceptional $\epsilon_0$, we still have the inclusion  $span\{y_{r+1}(\epsilon_0), ... y_N(\epsilon_0)\} \subset  Ker(A(\epsilon_0))$.
	\end{Proposition}

In our case, we have a rank $r=N-1$ for the perturbed matrix $A(\epsilon)$  for $\epsilon \neq 0$, since our Markov chain is irreducible: indeed, by using the spontaneous jumps of individual ancillas, we can go very simply from any combination of ancilla levels to any other combination of ancilla levels. This matrix thus has a unique steady state. The rank degenerates to $N-3$ for $\epsilon = 0$, as we have shown in Section \ref{section2} that $\text{Ker}(A(0)) = \text{span}\{\q{gg...g}, \, \q{mm...m}, \, \q{ee...e} \}$. Proposition \ref{PropAeps} allows us to say that there exists $y_N(\epsilon)$, an analytic vector-valued function which constitutes a basis for the null space of $A(\epsilon)$,  for all $\epsilon > 0$, and that $span\{ y_N(0)\} \subset  Ker(A(0))$.
	
When $\epsilon$ goes to zero, $y_N(\epsilon)$ thus analytically tends towards $y_N(0) \in \text{span}\{\q{gg..g}, \, \q{mm..m}, \, \q{ee..e} \}$. Concretely: only the configurations $\q{gg...g}, \q{mm...m}, \q{ee...e}$ can have population of order 1 in steady state. We next use this insight to approximately compute the steady state.


\subsubsection{The steady state populations on $\q{gg...g}$, $\q{mm...m}$ and $\q{ee...e}$ are \emph{each} at least an order of magnitude larger than the ones of all the other configurations.}\label{sectionbis}

We next argue that the ancillas subsystem can really be viewed as a synchronized clock on the three levels $\q{gg...g}$, $\q{mm...m}$ and $\q{ee...e}$, by showing that steady state population on \emph{each} of these three configurations is more significant than on the unsynchronized ones. We denote  $\frac{\max(\kappa_{t}, \kappa_{u})}{\kappa_{d}} = \epsilon_1 \ll 1$ and $\frac{\kappa_{d}}{\kappa_{st}}= \epsilon_2 \ll 1$. 

\begin{Definition}
We call \emph{principal configurations} the three configurations $\q{gg...g}$, $\q{mm...m}$ and $\q{ee...e}$. We call \emph{main transition configurations} the configurations, with $1$ or $2$ frontiers, resulting from a single spontaneous jump of any ancilla out of a principal configuration, followed by an arbitrary number of stimulated jumps.
\end{Definition}

 For instance, $\q{emee...e}$ is a main transition configuration resulting from a random jump on ancilla 2 out of $\q{eee...e}$; and $\q{mmmee...e}$ is also a main transition configuration, reached after this ancilla has stimulated jumps of ancillas 1 and 3. In contrast, $\q{egee..e}$ for instance is not a main transition configuration, since a single spontaneous clock jump ouf of $\q{ee...e}$ or $\q{mm...m}$ would involve some symbols $\q{m}$, while a single jump out of $\q{gg..g}$ should involve a single connected string of ancillas which have jumped to $\q{e}$. As we will see, the main transition configurations are the main states enabling a flow from a principal configuration to another. We start with a preliminary result.

\begin{Proposition}\label{prop:bis}
The distribution in steady state satisfies $p_{ee...e} \le O(\epsilon_1)$.
\end{Proposition}
\begin{proof}
The steady state condition on level $\q{me...e}$ writes	
	\begin{equation}\label{maitresse4.521}
	(\kappa_{st}  + (n-1) \kappa_{d} + \kappa_{t}) p_{me...e} =  \kappa_{d} p_{ee...e} + P_1
	\end{equation}	
	where $P_1$ represents all the other populations arriving in $\q{me...e}$, so we have $P_1>0$. Dividing by $(\kappa_{st}  + (n-1) \kappa_{d} + \kappa_{t})$, we obtain that $p_{me...e}$ must be at least of order $\epsilon_2  p_{ee...e}$.
	We can repeat this reasoning on level $\q{mme...e}$ of the steady state condition, yielding:
	\begin{equation}\label{maitresse4.522}
	(\kappa_{st}  + (n-2) \kappa_{d} + 2\kappa_{t}) p_{mme...e} =  \kappa_{st} p_{me...e} + P_2
	\end{equation}	
	where $P_2$ represents all the other populations arriving in $\q{mme...e}$, so we have $P_2>0$.
	Dividing by $ (\kappa_{st}  + (n-2) \kappa_{d} + 2\kappa_{t})$, we see that $p_{mme...e}$ must be at least of the same order as $p_{me...e}$. We can iterate this process until proving that $p_{m...me}$ must be at least of order $\epsilon_2 p_{ee...e}$. From there, the steady state equation on level $\q{mm...m}$ gives	
	\begin{equation}\label{maitresse4.523}
	N \kappa_{t} p_{m...m} =  \kappa_{st} p_{m...me} + P_3
	\end{equation}	
	where $P_3 > 0$ represents all the other populations arriving in $\q{m...m}$.
	Dividing by $N \kappa_{t}$ gives $p_{m...m}$ at least of order $\frac{1}{\epsilon_1 \epsilon_2} \cdot \epsilon_2 p_{ee...e} = \frac{ p_{ee...e}}{\epsilon_1}$. Since $p_{m...m}$ can be at most of order $1$, we must indeed have $p_{ee...e}$ at most of order $\epsilon_1$.
 \hfill $\square$
\end{proof}	

Using this insight, we can compute the order of magnitude of steady state population on the principal configurations.

	\begin{Proposition}\label{prop:7}
		We have $p_{gg...g}$ and $p_{mm...m}$ of order one, and $p_{ee...e}$ is of order $\epsilon_1$.						
	\end{Proposition}
	\begin{proof}
	Using exactly the same method as in Proposition \ref{prop:bis}, we can prove that:
	\begin{itemize}
	\item $p_{mm...m}$ of order 1 $\implies$ $p_{gg...g} \ge O(1)$
	\item $p_{gg...g}$ of order 1 $\implies$ $p_{ee...e} \ge O(\epsilon_1)$
	\item $p_{ee...e} \ge O(\epsilon_1)$  $\implies$ $p_{mm...m} \ge O(1)$
	\end{itemize}
	Furthermore, since the populations must sum to 1, either $ p_{mm...m}$ or $p_{gg...g}$ must be of order one at least. Combining these facts necessarily implies the conclusion.	\hfill $\square$
	\end{proof}
	
Next, we can prove that the main transition configurations have a population an order of magnitude lower than $\q{ee...e}$ in steady state.

	\begin{Proposition}\label{prop:maintransitionconfigs}
		The main transition configurations have a population of order $\epsilon_1 \epsilon_2$.
	\end{Proposition}	
	\begin{proof}
We already know the steady-state populations of the configurations $\q{gg...g}$, $\q{mm...m}$ and $\q{ee...e}$. Using the same reasoning as in Proposition \ref{prop:bis}, starting from the main transition configurations, we can prove that all main transition configurations are at most of order $\epsilon_1 \epsilon_2$: if bigger this would lead to the principal configurations being of an order bigger than what we already proved in Proposition \ref{prop:7}. In the same way, starting from the principal configurations whose populations we know, we can prove that all transition configurations are at least of order $\epsilon_1 \epsilon_2$.
	\hfill $\square$
	\end{proof}

Finally, we prove with the two following propositions that any other configurations have a population of order $o(\epsilon_1 \epsilon_2)$ in steady state.
		
	\begin{Proposition}\label{prop:toto}
		Take $X_k^0$ a configuration with $k$ frontiers, $k \ge 3$. Then  $p_{X_k^0} = o(\epsilon_1 \epsilon_2)$.
	\end{Proposition}
	\begin{proof}
	First, note that the populations of those configurations must be at most of order $(\epsilon_1 \epsilon_2)$: if bigger, then with the same reasoning as above this would lead to the principal configurations being of an order bigger than what we have already proved in Proposition \ref{prop:7}.
	
	We conduct a proof by induction for $k$ going down from $n-1$ to $3$, $n\geq4$.
	
	$\bullet$ Initialization: We look at a configuration $X_{n-1}^0$ with $n-1$ frontiers. No stimulated jump can ever lead to this configuration, so the only way to arrive there is with a spontaneous jump from configurations with $n-1$, $n-2$ or $n-3$ frontiers, denoted $X_{n-1}^{j}$, $X_{n-2}^{j}$ and $X_{n-3}^{j}$ with $j$ spanning the different configurations. Moreover, the stimulated jumps draw $X_{n-1}^0$ onto other configurations, at a rate $\kappa_{st}$ multiplied by the number $s_n \in [1, n-1]$ of ancillas which could undergo a synchronization jump, plus a small probability $s'_n \, \kappa_d$ to leave with spontaneous jumps. This gives the steady state equation for configuration $X_{n-1}^0$:
\begin{multline}\label{maitresse1.1}
	(s_n \kappa_{st} + s'_n \kappa_d) p_{X_{n-1}^0} =  \kappa_{u} [ \sum_{i} p_{X_{n-1}^{i}} + \sum_{j} p_{X_{n-2}^{j}} + \sum_{k} p_{X_{n-3}^{k}}] + \kappa_{d} [ \sum_{i'} p_{X_{n-1}^{i'}} + \sum_{j'} p_{X_{n-2}^{j'}} + \sum_{k'} p_{X_{n-3}^{k'}}] \\
	+ \kappa_{t} [ \sum_{i''} p_{X_{n-1}^{i''}} + \sum_{j''} p_{X_{n-2}^{j''}} + \sum_{k''} p_{X_{n-3}^{k''}}]
	\end{multline}
All the populations on the right side of this equation are of order at most $\epsilon_1 \epsilon_2$, as they are populations of states with at least 1 frontier. We can rewrite		
	\begin{equation}\label{maitresse1.2}
	(s_n \kappa_{st} + s'_n \kappa_d) p_{X_{n-1}^0} =  \kappa_{u} O(\epsilon_1 \epsilon_2) + \kappa_{d} O(\epsilon_1 \epsilon_2) + \kappa_{t} O(\epsilon_1 \epsilon_2)
	\end{equation}
and dividing by $\kappa_{st}$, we get
\begin{equation}\label{maitresse1.3}
p_{X_{n-1}^0} =   o(\epsilon_1 \epsilon_2)
\end{equation}
	
	$\bullet$ Induction: We assume the property true for $k+1,k+2,...n-1$ and show that it is true for $k$, provided $k \ge 3$. 

Once again, the stimulated jumps draw the configuration $X_k^0$ onto other ones, at a rate $\kappa_{st} s_n < n$, as do some spontaneous jumps at a much smaller rate $s'_n \kappa_{d}$. Configurations that can directly jump to $X_k^0$ are either with $k$, $k+1$ or $k+2$ frontiers at a rate $\kappa_{st}$(as synchronization jumps can only lower the number of frontiers, and can lower it by at most 2), or configurations with at least $k-2$ frontiers at a smaller rate $\kappa_{u}$, $\kappa_{d}$ or $\kappa_{t}$; since $k \ge 3$, those configurations with at least $k-2 \ge 1$ frontiers have a population of order $O(\epsilon_1 \epsilon_2)$ or smaller. Thus the steady state equation for configuration $X_{k}^0$ would look like:		
	\begin{equation}\label{maitresse2.1}
	(s_n \kappa_{st} + s'_n \kappa_d)  p_{X_k^0} = \kappa_{st} \sum_{i \neq 0}  p_{X_k^{i}} + \kappa_{st} [\sum_{j} p_{X_{k+1}^{j}} + \sum_{l} p_{X_{k+2}^{l}}] + \kappa_{u} O(\epsilon_1 \epsilon_2) + \kappa_{d} O(\epsilon_1 \epsilon_2) + \kappa_{t} O(\epsilon_1 \epsilon_2)
	\end{equation}
The $k+1$ or $k+2$ frontier states are of order $o(\epsilon_1 \epsilon_2)$ and thus division by $\kappa_{st}$ gives:	
\begin{equation}\label{maitresse2.2}
(s_n + s'_n \epsilon_1)\,  p_{X_k^0} =\sum_{i \neq 0}  p_{X_k^{i}}  + o(\epsilon_1 \epsilon_2) \; .
\end{equation}
There remains to efficiently characterize the connections between various configurations $X_k^0$ and $X_k^i$ in the above equation. 

 \begin{figure}	
\includegraphics[width=160mm, trim=100 290 50 128, clip]{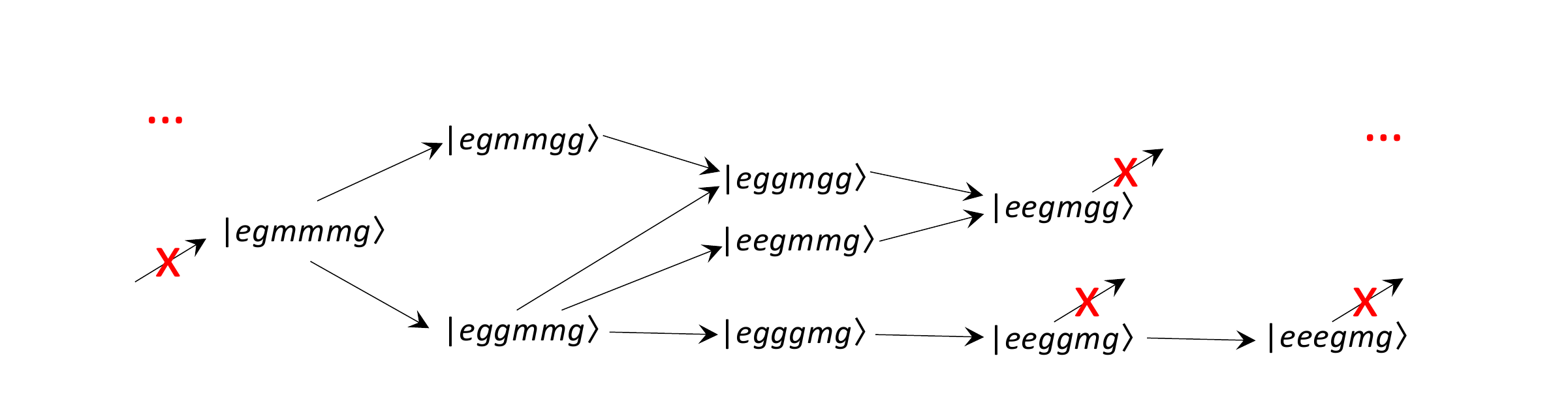}
\caption{Part of the directed graph of possible frontier evolutions maintaining the number of frontiers $k$, illustrated for $k=3$ and $n=6$. Crossed arrows indicate that there is no further incoming or outgoing edge.  \label{fig:frontiersgraph2}}
\end{figure}

\begin{itemize}
\item[-] Since the incoming $p_{X_k^{i}}$ in \eqref{maitresse2.2} result from evolutions with $\kappa_{st}$ only, we resort again to the analysis of frontiers evolution as described in Section \ref{section2}. Thus, for jumps with $\kappa_{st}$ and maintaining the number of frontiers, a frontier $F_{e,m}$ for instance can only move towards lower indices, with unique directions of motions for other frontiers determined by symmetry. Conversely, the incoming states to a configuration $X_k^0$ are thus obtained by ``moving back'' one frontier, e.g.~$F_{e,m}$ towards one higher index.
\item[-] We can build a graph $\mathcal{G}$ whose nodes are the configurations with $k$ frontiers and whose directed edges represent valid frontier motions with $\kappa_{st}$, see Figure \ref{fig:frontiersgraph2}. Thanks to the unique direction of frontier motion, this graph contains no directed cycles. In particular, it features states with no incoming edges (respectively, no outgoing edges), i.e.~where none of the $k$ frontiers can be moved back (respectively, further) anymore.
\item[-] Consider $X_k^0$ any node of $\mathcal{G}$ with no incoming edges. Thus \eqref{maitresse2.2} reduces to $(s_n + s'_n \epsilon_1)\,  p_{X_k^0}= o(\epsilon_1 \epsilon_2)$ or equivalently $p_{X_k^0}= o(\epsilon_1 \epsilon_2)$.

Now when we move to $X_k^0$ a different node of $\mathcal{G}$, we can remove the nodes just treated from the sum over $X_k^i$ in the right-hand side, since these nodes are thus captured by the term $o(\epsilon_1 \epsilon_2)$. Therefore, we remove those nodes (with no incoming edges) from $\mathcal{G}$.
\item[-] The modified graph $\mathcal{G}$ now features new nodes with no incoming edges, for which \eqref{maitresse2.2} reduces to $(s_n + s'_n \epsilon_1)\,  p_{X_k^0}= o(\epsilon_1 \epsilon_2)$. We can thus repeat the above reasoning, until all nodes have been treated, proving that $p_{X_k^0}= o(\epsilon_1 \epsilon_2)$ for any configuration $X_k^0$ with $k$ frontiers.
\end{itemize}
The induction concludes the proof down to $k=3$. \hfill $\square$
\end{proof} 

There remains to treat the configurations with less than 3 frontiers yet which are not main transition configurations. The main transition configurations include \emph{all} the configurations with $k=1$ frontier. For instance, $\q{ee...emm...m}$ is obtained from $\q{ee.ee}$ by spontaneous jump $\q{e}\rightarrow \q{m}$ at the frontier, then stimulated jumps to $\q{m}$ propagating along higher and higher indices only; any other $k=1$ configuration is obtained similarly. The main transition configurations also include some configurations with two frontiers and involving two level types, namely those where the ``inner'' level attract the outer one. For instance, $\q{eeemmmee}$  can be obtained from $\q{ee...e}$ with spontaneous jump $\q{e}\rightarrow \q{m}$ somewhere in the middle, then stimulated jumps to $\q{m}$ of the neightbors. However, the converse configuration $\q{mmmeeemm}$ is not a main transition configuration, since a single spontaneous jump followed by stimulated jumps cannot lead to this situation when starting from a synchronized configuration $\q{gg...g},\q{mm...m}$, or $\q{ee...e}$. Likewise, configurations with 2 frontiers but involving the 3 different levels $\q{g},\q{e}$ and $\q{m}$ are not main transition configurations. We thus conclude our claims by treating these configurations with 2 frontiers.
	
\begin{Proposition}\label{epscarre}
Take $X_2^1$ a state with 2 frontiers and all three levels $\q{g},\q{e}$ and $\q{m}$ present in the configuration (for example $\q{egmm...}$), and $X_2^2$ a state with 2 frontiers, involving only two levels and with the outer ancilla levels attracting the inner ones (for example $\q{...mme...emm...}$). Then $p_{X_2^1}= o(\epsilon_1 \epsilon_2)$ and $p_{X_2^2}= o(\epsilon_1 \epsilon_2)$.
\end{Proposition}
\begin {proof}
We can apply the same reasoning as in the proof of Proposition \ref{prop:toto} for both cases $X_k^0 = X_2^1$ or $X_k^0 = X_2^2$. Configurations jumping towards $X_k^0$ at a rate  $\kappa_{u}$, $\kappa_{d}$ or $\kappa_{t}$ cannot come from a configuration with 0 frontier, so their contribution in the steady state equation is at most of order $\epsilon_1 O(\epsilon_1 \epsilon_2) = o(\epsilon_1 \epsilon_2)$. We then obtain the same steady state equation \eqref{maitresse2.2}, and we can repeat the proof with the graph $\mathcal{G}$ involving moving frontiers. The nodes now are the configurations of the same type as $X_2^1$ or as $X_2^2$ respectively, as one easily checks that stimulated jumps preserving the number of frontiers must also preserve this type of configuration.
\hfill $\square$
\end{proof}


\subsubsection{Final approximate steady state computation}\label{section5}

The preceding propositions lead us to the following approximate computation of the steady state for the ancillas clock in the state-conditioning architecture. We first characterize the population on each of the three main configurations at first order. We then establish the $n$-dependence of the dominant population on other configurations. These are the results reported in the main text as Proposition \ref{approx}.

\begin{Proposition}\label{approx2app}
		In the limit $\epsilon_2 \rightarrow 0$, we have the steady-state populations:
		\begin{align}\label{eq:approxAS}
		p_{gg..g} = \frac{1}{1+\frac{\kappa_{u}}{\kappa_{d}} + \frac{\kappa_{u}}{\kappa_{t}}} \;\; , \quad
		p_{mm..m} = \frac{1}{1+\frac{\kappa_{t}}{\kappa_{u}} + \frac{\kappa_{t}}{\kappa_{d}}}  \;\; , \quad
		p_{ee..e} = \frac{1}{1+\frac{\kappa_{d}}{\kappa_{t}} + \frac{\kappa_{d}}{\kappa_{u}}} \; .
		\end{align}				
\end{Proposition}
\begin{proof}
The limit $\epsilon_2 \rightarrow 0$ is the one where all the population is on the principal configurations $\q{gg...g}$, $\q{mm...m}$ and $\q{ee...e}$. In other words, when being on the configuration $\q{gg...g}$, a spontaneous jump of any of the ancillas to $\q{e}$ will almost immediately lead to the configuration $\q{ee...e}$, and similarly on the other principal configurations.
We thus have the following steady state equations:	
	\begin{align}\label{eq:jumpcondmodelerror3.1.1}
	N\kappa_{d}  p_{ee..e} &= N\kappa_{u}  p_{gg...g} \quad &,& \quad  p_{ee..e} +  p_{mm...m}  +  p_{gg...g}  = 1 \\ \nonumber
	N\kappa_{t}  p_{mm...m} &= N\kappa_{d}  p_{ee..e} \quad &,& \quad  N\kappa_{u}  p_{gg...g} &= N\kappa_{t}  p_{mm...m}
	\end{align}
	which immediately leads to the result. \hfill $\square$
\end{proof}

\begin{Proposition}\label{prop:Ndep1}
Consider a fixed number of ancillas $n$ and small $\epsilon_1, \epsilon_2$ such that $n\epsilon_1 \ll 1$, $n\epsilon_2 \ll 1$. Then
\begin{equation}\label{eq:odregrandeur}
p_{gg..g} + p_{mm..m} + p_{ee..e}  >  1 - \frac{3}{4} (n-1)(\tfrac{3n}{2}+1) \,  \epsilon_1 \epsilon_2 + o(\epsilon_1 \epsilon_2) \; .
\end{equation}
Using the time scales of \eqref{eq:step2atc}, this correponds to a scaling in $\frac{T_3}{T_1} n^2$.		
\end{Proposition}
\begin{proof}
We have shown that $p_{gg..g}$, $p_{mm..m}$ and $p_{ee..e}$ are the only populations of order bigger than $O(\epsilon_1 \epsilon_2)$, and main transition configurations are the only ones featuring populations of order $O(\epsilon_1 \epsilon_2)$. We have to evaluate the latter, for example the main transition configurations between $\q{g...g}$ and $\q{e...e}$. The corresponding part of the ancillas clock Markov chain is represented on Figure \ref{fig:NdepMC}.

\begin{figure}
\includegraphics[width=150mm, trim=10 110 150 50, clip]{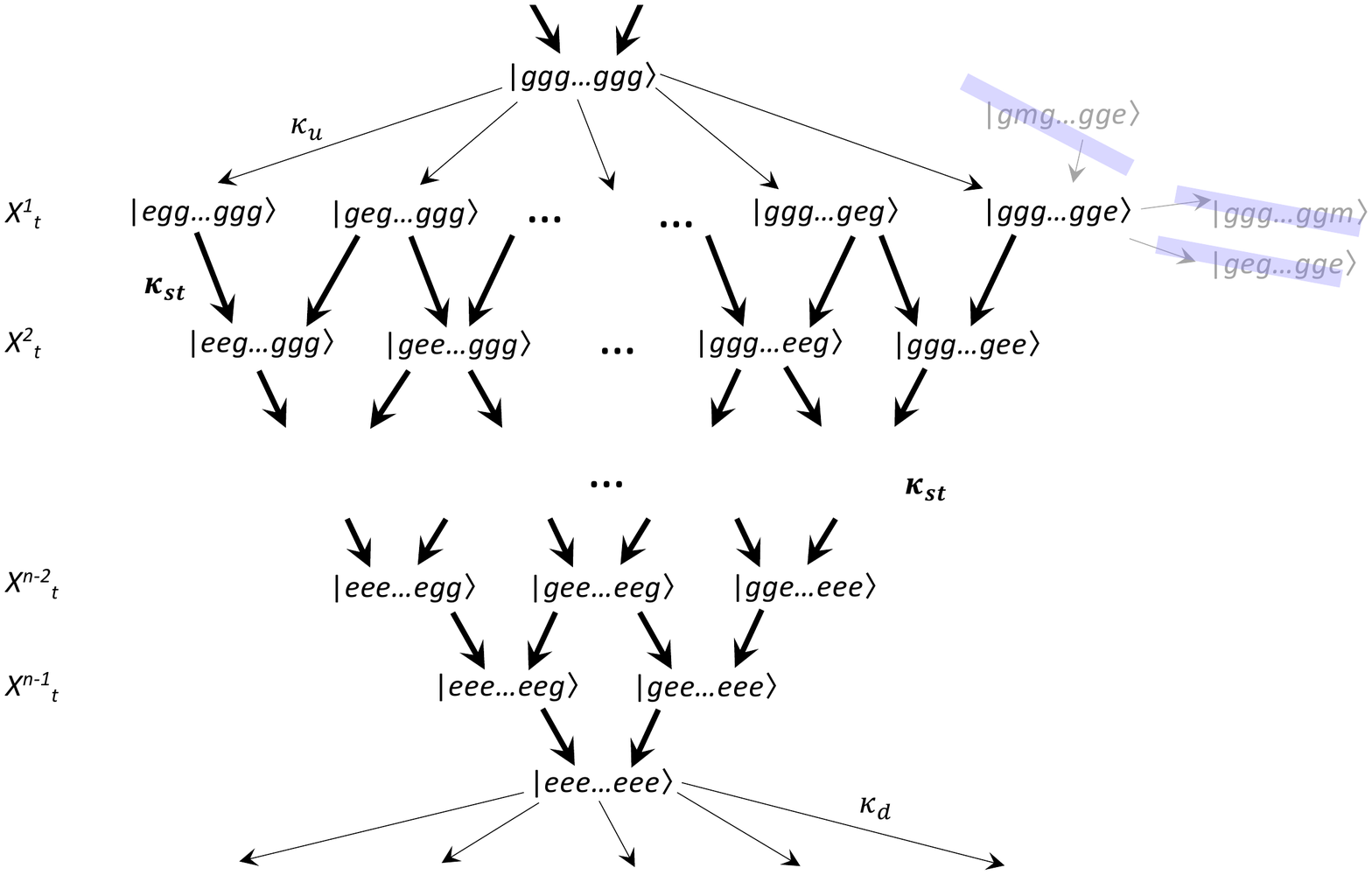}
\caption{Representation of part of the ancillas clock Markov chain, used in the proof of Proposition~\ref{prop:Ndep1}. The main transition configurations are represented in full black, along with the relevant transitions towards the proof's result. A few other configurations and transitions are represented  for illustration, shaded and crossed; indeed, those configurations play a negligible role in the computation and are thus discarded.\label{fig:NdepMC}}
\end{figure}

\begin{itemize}	
\item For the first line, the main transition configurations $X_{t}^1$ with a single ancilla on $\q{e}$, we have the following steady state equation, with $d$ being 1 or 2 depending on if we have a configuration with 1 or 2 frontiers:
\begin{equation}\label{ordereps}
(D \kappa_{st}  + (n-1) \kappa_{u} + \kappa_{t}) p_{X_{t}^1} = \kappa_{u} p_{gg..g} + \kappa_{st} o(\epsilon_1 \epsilon_2) \; .
\end{equation}
Indeed, configurations that can jump onto $X_{t}^1$ are either $\q{g...g}$ under application of the corresponding spontaneous jump to $\q{e}$; or configurations involving $n-2$ times $\q{g}$, one $\q{e}$, and one ancilla on $\q{m}$, whose population is thus $o(\epsilon_1 \epsilon_2)$ by our preceding results. On the other side, $X_{t}^1$ can be left through any corresponding spontaneous jump, or by stimulated attraction of a neighboring $\q{g}$ towards $\q{e}$ by the single $\q{e}$ ancilla. Dividing by $(D \kappa_{st}  + (n-1) \kappa_{u} + \kappa_{t})$, we get
\begin{equation}\label{ordereps2}
p_{{X_{t}}^1} = \frac{\kappa_{u}}{D \kappa_{st}}  p_{gg..g} +  o(\epsilon_1 \epsilon_2) \; .
\end{equation}
\item Now consider the second line, the main transition configurations $X_{t}^2$ with two ancillas on $\q{e}$. We can arrive on $X_{t}^2$ either with one of the $n-2$ ancillas jumping from $\q{m}$ to $\q{g}$, thus coming from a state with population $o(\epsilon_1 \epsilon_2)$; or with (predominantly stimulated) jump from $\q{g}$ onto $\q{e}$ of one of the other 2 ancillas. Leaving $X_t^2$ follows the same scheme as for $X_t^1$. Thus, if $X_{t}^2$  has two frontiers, then its steady state equation writes:
\begin{equation}\label{ordereps3}
	(2 \kappa_{st}  + (n-2) \kappa_{u} + 2 \kappa_{t}) p_{X_{t}^2} = \kappa_{st}  p_{X_{i}^1} + \kappa_{st}  p_{X_{j}^1} + \kappa_{st} o(\epsilon_1 \epsilon_2) \; ,
\end{equation}
where both $p_{X_{i}^1}$ and $p_{X_{j}^1}$ satisfy \eqref{ordereps2} with $D=2$. This leads to
\begin{equation}\label{ordereps4}
p_{X_{t}^2} =  \frac{\kappa_{u}}{2 \kappa_{st}}  p_{gg..g} +  o(\epsilon_1 \epsilon_2)
\end{equation}
for configurations $X_{t}^2$ with 2 frontiers.

If $X_t^2$ has a single frontier, then its steady state equation writes:
	\begin{equation}\label{ordereps5}
	(\kappa_{st}  + (n-2) \kappa_{u} + 2 \kappa_{t}) p_{X_{t}^2} = \kappa_{st}  p_{X_1^1} + \kappa_{st}  p_{X_2^1} + \kappa_{st} o(\epsilon_1 \epsilon_2) \; ,
	\end{equation}
where $p_{X_1^1}$ and $p_{X_2^1}$ satisfy \eqref{ordereps2} with $D=1$ and $D=2$ respectively. This leads to
\begin{equation}\label{ordereps6}
p_{X_{t}^2} =  \frac{3\kappa_{u}}{2 \kappa_{st}}  p_{gg..g} +  o(\epsilon_1 \epsilon_2)
\end{equation}
for configurations $X_{t}^2$ with 1 frontier.
\item We can pursue a similar reasoning to show that the main transition configurations with 2 frontiers all have a population $\frac{\kappa_{u}}{2 \kappa_{st}}  p_{gg..g} +  o(\epsilon_1 \epsilon_2)$ in steady state, while the main transition configurations with 1 frontier have a population  $\frac{(k+1)\kappa_{u}}{2 \kappa_{st}}  p_{gg..g} +  o(\epsilon_1 \epsilon_2)$ when $k$ ancillas are on $\q{e}$.
\item Summing up the populations of the main transition configurations from $\q{g...g}$ to $\q{e...e}$ on all those lines then yields:
\begin{equation}\label{ordereps7}
	p_{t_{ge}} =  \frac{ \kappa_{u}}{2 \kappa_{st}} (n-1)(\tfrac{3n}{2}+1)  p_{gg..g} +  o(\epsilon_1 \epsilon_2)
	\end{equation}
\item Similar properties hold, by circular symmetry, for the other pairs of levels, yielding:
\begin{eqnarray}
\label{eq:ordereps9}
	p_{t_{em}} &=&  \frac{ \kappa_{d}}{2 \kappa_{st}} (n-1)(\tfrac{3n}{2}+1)  p_{ee..e} +  o(\epsilon_1 \epsilon_2) \\
\nonumber
	p_{t_{mg}} &=&  \frac{ \kappa_{t}}{2 \kappa_{st}} (n-1)(\tfrac{3n}{2}+1)  p_{mm..m} +  o(\epsilon_1 \epsilon_2) \; .
\end{eqnarray}
By summing up all these contributions and using Proposition \ref{approx2app}, we obtain
$$p_{t_{ge}} + p_{t_{em}} + p_{t_{mg}} = (n-1)(\tfrac{3n}{2}+1) \, \frac{3 \kappa_u \kappa_t}{2 \kappa_{st}(\kappa_t+\kappa_u)} \; .$$
\end{itemize}
The result follows by definition of $\epsilon_1$ and $\epsilon_2$. \hfill $\square$
\end{proof}


\subsubsection{Similar results hold for the jump-conditioning ancillas clock}\label{sssec:jcac}

Until now our analysis has been restricted to the ancillas clock associated to the state-conditioning scheme of Section \ref{sssec:sc:step1}. We now consider the jump-conditioning scheme of Section \ref{sssec:jc:step1}. The associated ancillas clock features an additional level $\q{f}$ on each ancilla. This level spontaneoulsy jumps down to $\q{e}$ at a rate $\kappa_f$, \emph{irrespective of the associated data state}. More explanation on the latter property is given in Section \ref{sec:AncBasedMC}. Stimulated jumps from $\q{g}$ onto $\q{e}$ occur if a neighboring ancilla is on $\q{e}$ or $\q{f}$. Furthermore, the rates now satisfy the timescale separation
$\; \kappa_f,\; \kappa_{st} \gg \kappa_{d},\kappa_{t},\kappa_{u} \; $.
	
The idea is that this ancillas clock would behave very similarly to the one with state-conditioning, with each ancilla just quickly transitioning through $\q{f}$ on its way from $\q{g}$ to $\q{e}$. The analysis confirming this can be carried out as follows.\\

In section \ref{section2}, we were keeping only the transitions associated to fastest timescales, which here would be $\kappa_f$ and $\kappa_{st}$. Exponential convergence towards span$\{\q{gg...g},\q{ee...e},\q{mm...m}\}$ was proved using an argument on deleting all ``frontiers'' in a finite number of steps. A similar reasoning can be applied with the addition of level $\q{f}$, and by considering that $\q{fe}$ or $\q{ef}$ is \emph{not} a frontier. The essential idea is that any $\q{f}$ jumps to $\q{e}$ irrespectively of the other ancillas, in this way the $\q{f}$ level does not really play a role in the synchronization mechanism.\\

In section 	\ref{section3}, we were analyzing the perturbation of the fast dynamics by the slower one. We have established that only the configurations $\q{gg...g}, \q{mm...m}, \q{ee...e}$ can have population of order 1 in steady state. This result remains true as well. The parameter $\epsilon \ll 1$ now corresponds to $T_3/T_1$ according to \eqref{eq:step2btc}.\\

In section \ref{sectionbis}, we were distinguishing the ``main transition configurations'' with population of order $\epsilon_1 \epsilon_2$, the three principal configurations with higher population (either $O(1)$ or $O(\epsilon_1)$), and all other configurations with population an order lower. A similar analysis can be carried out in presence of level $\q{f}$, with the following modifications.
\begin{itemize}
\item There are only two timescales: a rapid one $\kappa_{st},\kappa_f$, while all the slower rates $\kappa_{u},\kappa_d,\kappa_t$ are of the same order. We thus denote $\epsilon = \max(\kappa_{u},\kappa_d,\kappa_t) \; / \; \min(\kappa_{st},\kappa_f)$.  
\item Consequently, the populations $p_{gg...g}, \; p_{mm...m}, \; p_{ee...e}$ will each be of order 1. In fact, this is already valid in section \ref{sectionbis} if one assumes  $\kappa_{u},\kappa_d,\kappa_t$ of the same order, thus taking $\epsilon_1 = O(1)$. 
\item To prove this result in presence of level $\q{f}$, we can follow a similar reasoning with ``main transition configurations'', where we enlarge the set of such configurations for the transition from $\q{gg...g}$ to $\q{ee...e}$:  they now include all the ``main transition configurations'' identified in section \ref{sectionbis}, plus all configurations obtained from those where an arbitrary number of $\q{e}$ levels is replaced by $\q{f}$, plus all configurations consisting entirely of $k\geq 1$ levels $\q{f}$ and $n-k$ levels $\q{e}$.

This just mirrors the quick transition $\q{g} \rightarrow \q{f} \rightarrow \q{e}$ for each ancilla. The rest of the proof then follows similar lines.
\end{itemize}
As a result, we then obtain that $p_{gg...g}, \; p_{mm...m}, \; p_{ee...e}$ are all of order 1, the ``main transition configurations'' have population of order $\epsilon$, and all other configurations have population $o(\epsilon)$.\\
	
In section \ref{section5}, we were providing the approximate distribution over $\q{gg...g},\q{ee...e},\q{mm...m}$, and the $n$-dependence of the population outside those three configurations.
\begin{itemize}
\item In the present case, the distribution result of Proposition \ref{approx} remains unchanged, with thus $\epsilon_2=\epsilon$ and just $\kappa_d$ of the same order as $\kappa_u,\kappa_t$. 
\item Regarding the $n$-dependence of the population on ``main transition configurations'', we can take the following viewpoint to treat the now different transition from $\q{gg...g}$ to $\q{ee...e}$.

First, consider as irrelevant whether an ancilla is on $\q{e}$ or $\q{f}$, grouping those two levels as some super-level $\q{\zeta} =$ ``$\q{e}$ or $\q{f}$''. With this, we can repeat verbatim the proof of section \ref{section5} and we obtain the same evaluations for the total population on each super-level configuration, e.g.~for $p_{g\zeta gg...g}$. In particular, $p_{\zeta\zeta...\zeta}$ gets the population of order 1 which \eqref{eq:approxAS} attributes to $\q{ee...e}$. More precisely, according to Proposition \ref{prop:Ndep1}, the population not on $\q{gg...g},\q{\zeta\zeta...\zeta},\q{mm...m}$ is of order $O(\epsilon \, n^2)$.

Next, there remains to single out the configuration $\q{ee...e}$ out of the super-level configuration $\q{\zeta\zeta...\zeta}$. For this, let us denote $X_k^{j}$ the configurations with $k$ ancilla on $\q{f}$ and the other ancillas on $\q{e}$, and denote by $p_k$ the sum of all the populations on configurations of type $X_k^{j}$.
\begin{itemize}
\item The steady-state equation for $p_{ee...e}$ writes: 
$$n \kappa_d p_{ee...e} = \kappa_r \sum_j  p_{X_1^j} = \kappa_r p_1 \; ,$$
or equivalently $p_1 = O(n \epsilon)$.
\item Consider the steady-state equation for $p_{X_k^j}$, with $k=1,2,...,n-1$. The outgoing rate is $(k \kappa_r +(n-k) \kappa_d) \; p_{X_k^j}$, namely any of the $k$ ancillas on $\q{f}$ spontaneously jumping towards $\q{e}$ or any of the $n-k$ ancillas on $\q{e}$ spontaneously jumping towards $\q{m}$. The incoming rate is the sum of possible transitions from configurations involving one ancilla on  $\q{g}$, and transitions from states of type $X_{k+1}^{\ell}$. Dropping the former from the equation, we get:
$$(k \kappa_r +(n-k) \kappa_d) \; p_{X_k^j} \; > \; \kappa_r \sum_{\ell \in \mathcal{N}_k^j} \; p_{X_{k+1}^{\ell}} \; ,$$
with the set $\mathcal{N}_k^j$ containing $(n-k)$ elements. Since $p_{X_k^j}$ is already of order $\epsilon$, we will neglect the term in $\kappa_d p_{X_k^j} $ here. Then summing the equation over $j$, we obtain
$$ k\, p_k > (k+1)\, p_{k+1}\; . $$
The factor $(k+1)$ on the right comes from the fact that each of the $(k+1)$ ancillas on $\q{f}$ in a given configuration $X_{k+1}^{\ell}$ can once play the role of the incoming path to some $X_k^j$.
\item Thus iteratively, we have that $p_k = \tfrac{1}{k} p_1$ for $k=1,2,...,n$. Thus 
$$\sum_{k=1}^n p_k = O(\log(n)) p_1 = O(n \log(n) \epsilon) \; .$$
\end{itemize}
This proves that the population not on $\q{gg...g},\q{ee...e},\q{mm...m}$ is still of order $O(\epsilon \, n^2)$.
\end{itemize}

\subsection{Analysis of the data qubits in clock ancilla-based scheme}\label{ssec:app:proof1}

We first provide the proof of Proposition \ref{prop:llp1} about the steady state of the reduced Markov chain as defined in Section \ref{ssec:MC1definition}, around Figure \ref{figa:00MCn}. The reader is referred to this place for recalling notation and description.

\begin{proof}
The steady state equation for $R_k^{mg}$ readily yields
\begin{equation}\label{eq:llp1}
p_{R_k^{mg}} = \frac{\kappa_d}{\kappa_p + \tilde{\kappa}_u + \tfrac{n-1}{n} \kappa_c} \, p_{R_k^{e}} \;\;, \;\;\; \text{ for } k=1,2,...,n \, .
\end{equation}
Next, the steady state equations for $R_k^e$ write
\begin{eqnarray}\label{eq:llp2}
(n \kappa_p + n \kappa_d + (n-1) \kappa_c + (n-1) \kappa_r) p_{R_1^e} &=& n \tilde{\kappa}_u p_{R_1^{mg}} + n \kappa_r \left( p_{e} - {\textstyle \sum_{k=1}^n} p_{R_k^e} \right) \;\;, \\ \nonumber
(n \kappa_p + n \kappa_d + (n-1) \kappa_c + (n-k) \kappa_r) p_{R_k^e} &=& n \tilde{\kappa}_u p_{R_k^{mg}} + (n-k+1) \kappa_r p_{R_{k-1}^e}  \quad \text{ for } k=2,3,...,n\; ,
\end{eqnarray}
where $p_{e}$ is the total population on $\cdot ^{e}$, irrespective of the data qubits situation. Using \eqref{eq:llp1} and recalling that $\; p_{e} = \frac{\tilde{\kappa}_u}{\tilde{\kappa}_u+\kappa_d} \;$, this solves to:
\begin{eqnarray}\label{eq:llp3}
p_{R_n^e} & = & \frac{p_e}{1+\tfrac{a_0}{\kappa_r} + \tfrac{a_0(a_0+\kappa_r)}{2! \, \kappa_r^2} + ... + \tfrac{a_0(a_0+\kappa_r)...(a_0+(n-1)\kappa_r)}{n! \, \kappa_r^n}} \; , \\ \nonumber
p_{R_k^e} & = & \frac{a_0(a_0+\kappa_r)...(a_0+(n-1-k)\kappa_r)}{(n-k)!\, \kappa_r^{n-k}} \; p_{R_n^e} \;\; \text{ for } k=1,2,...,n-1 \; , \\ \nonumber
 \text{with} && a_0 \;\;=\;\; n \kappa_p + (n-1) \kappa_c + n \kappa_d \frac{\kappa_p + \tfrac{n-1}{n} \kappa_c}{\kappa_p + \tfrac{n-1}{n} \kappa_c + \tilde{\kappa}_u} \; .
\end{eqnarray}

Next, writing the steady state conditions for the pair of configurations $G_k^{mg}$, $G_k^e$ leads to the explicit recursion:
\begin{eqnarray}\label{eq:llp4}
b_1 \, p_{G_k^{mg}} & = & \kappa_c\, p_{G_{k-1}^{mg}} + \tfrac{n \kappa_d}{b_0+n\kappa_d} \;  \kappa_c\, p_{G_{k-1}^{e}} \; , \\ \nonumber
(b_0 + n \kappa_d ) p_{G_k^{e}} & = & \tfrac{n\tilde{\kappa}_u}{b_1} \; \kappa_c \, p_{G_{k-1}^{mg}} + \left(1 + \tfrac{n \kappa_d}{b_0+n\kappa_d} \tfrac{n\tilde{\kappa}_u}{b_1} \right) \; \kappa_c \, p_{G_{k-1}^e} \\ \nonumber
 \text{with} && b_0 \;\; = \;\; n \kappa_p + \kappa_c + n \kappa_r \\ \nonumber
&& b_1 \;\; = \;\; \left( n\kappa_p + n \tilde{\kappa}_u \tfrac{b_0}{b_0+n \kappa_d}  + \kappa_c \right) \; ,
\end{eqnarray}
for $k=2,3,...,n-2$. For $k=1$, the expressions \eqref{eq:llp4} hold but replacing $p_{G_{k-1}^{mg}}$ and  $p_{G_{k-1}^e}$ on the right hand side respectively by $(p_{G_0^{mg}}+p_{R_n^{mg}})$ and $(p_{G_0^{e}}+p_{R_n^{e}})$. For $k=0$, the expressions \eqref{eq:llp4} hold but replacing  $p_{G_{k-1}^{mg}}$ and  $p_{G_{k-1}^e}$ respectively by $(n-2) p_{R_n^{mg}}$ and $(n-2) p_{R_n^e}$. This allows, at least in principle, to explicitly compute through to $G_{n-2}^{mg}$ and  $G_{n-2}^{e}$.  Finally, the steady-state equations for $G_{HZ}^{mg}$ and $G_{HZ}^e$ lead to
\begin{equation}\label{eq:llp5}
\left( n \kappa_p + n \tilde{\kappa}_u \, \tfrac{\kappa_r + \kappa_p}{\kappa_r+\kappa_p+\kappa_d} \right) \;\, p_{G_{HZ}^{mg}} \; = \; 
\kappa_c \, p_{G^{mg}_{n-2}} + \tfrac{\kappa_d}{\kappa_d+\kappa_r+\kappa_p} \, \kappa_c \, p_{G^e_{n-2}} \; .
\end{equation}
Note that, while we provide these exact expressions here for completeness, all the terms in $G_k^e$ in fact have no impact on the leading-order computation.

The statement of the Proposition is obtained by concatenating these explicit expressions, keeping only the leading order terms in $T_k / T_{k-1}$ to obtain a more readable result. For low values of $n$ the sums in \eqref{eq:llp3} can be computed explicitly by hand. For large values of $n$, we keep in this sum the first-order term in $a_0/\kappa_r$ and then approximate like $\sum_{k=1}^n 1/k \approx \int_0^n 1/x \; dx = O(\ln(n))$. \hfill $\square$\\
\end{proof}

\begin{figure}
\includegraphics[width=150mm, trim=35 140 70 100, clip]{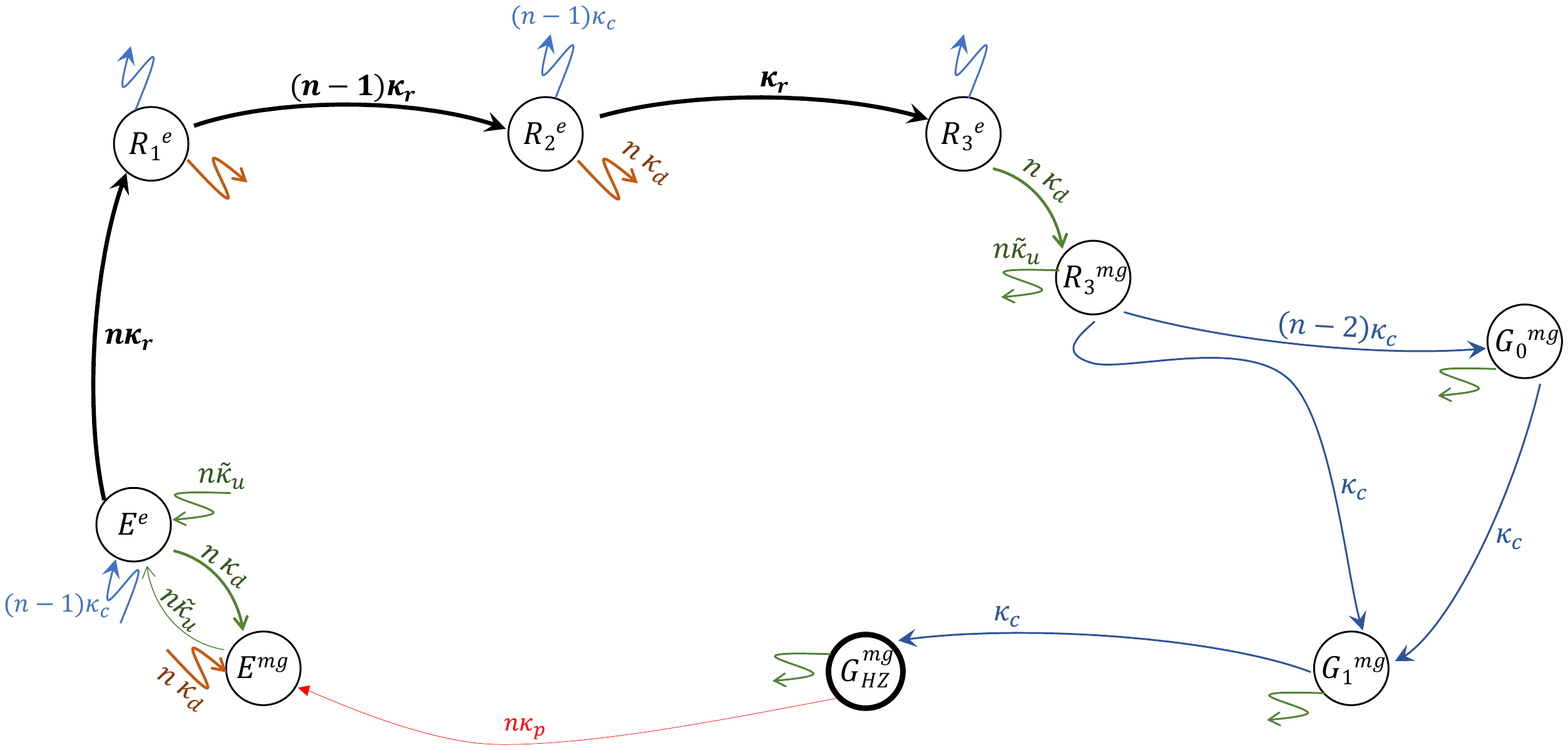}
\caption{Further reduced Markov chain, with effective transition rates covering most relevant transitions, and yielding the same steady-state population on $G_{HZ}$ as reported in Proposition \ref{prop:llp1}.\label{fig:reredMC}}
\end{figure}

The leading-order result reported in Proposition \ref{prop:llp1} is in fact the same as with the further reduced Markov chain represented on Figure \ref{fig:reredMC}. There we have dropped several links, which leads to simplified computation of the steady sate:
\begin{itemize}[noitemsep,topsep=0pt]
\item We only keep the error in $\kappa_p$ leaking from $G_{HZ}^{mg}$, the rest appears negligible.
\item We only keep progress towards $\q{GHZ_+}$ with $\kappa_c$ for the $G_k^{mg}$, dropping those from $G_k^e$.
\item  We reroute the $\kappa_d$ flow from any $G_k^e$ to $E^{mg}$, instead of back to $G_k^{mg}$
\item Similarly, we reroute the $\kappa_d$ flow from $R_k^e$ to $E^{mg}$ for all $k<n$. The corresponding $R_k^{mg}$ then have no input anymore and can be dropped. In other words, this amounts to grouping the configurations $R_k^{mg}$ into $E^{mg}$.
\item We also drop the $G_k^{e}$ and group them into $E^e$. Indeed, with the previous points, these $G_k^e$ make no progress towards $\q{GHZ_+}$ unless transiting through $R_1^e$ again. We thus reroute the corresponding flows towards $E^e$. 
\end{itemize}
Except the first one which involves very low-order terms, all simplifications are clearly pessimistic; you can note for instance that $R_n^e$ is quite pessimistically re-routed with $(n-1)\kappa_c$ towards $E_e$ instead of $G_0^e$ and $G_1^e$. Yet, one can quickly check that all these have no first-order effect on the steady-state fidelity, by re-computing it with this simpler Markov chain. We will use the Markov chain of Figure \ref{fig:reredMC} as a simpler starting point to study the effect of imperfect ancilla synchronization.

\subsubsection{Effect of imperfect ancilla synchronization}\label{sec:imperfsynch}

We next quantify how the transitions through ancilla configurations outside $\q{gg...g},\q{ee...e},\q{mm...m}$ modify the result to first order. We therefore re-introduce all the possible ancilla configurations, and group them as a function of approximately equal effect on the data. Explicitly, our goal is to group in one set all configurations with some ancillas on $g$ and some on $m$; as another one all configurations with some ancillas on $e$ and some on $m$; another one with some ancillas on $e$ and some on $g$; and finally a configuration with ancillas taking 3 different values. We then set up a Markov chain whose configurations are those sets. In order to bound our fidelity estimate from below, we perform this grouping with pessimistic approximations.

\begin{figure}
\includegraphics[width=120mm, trim=20 330 310 98, clip]{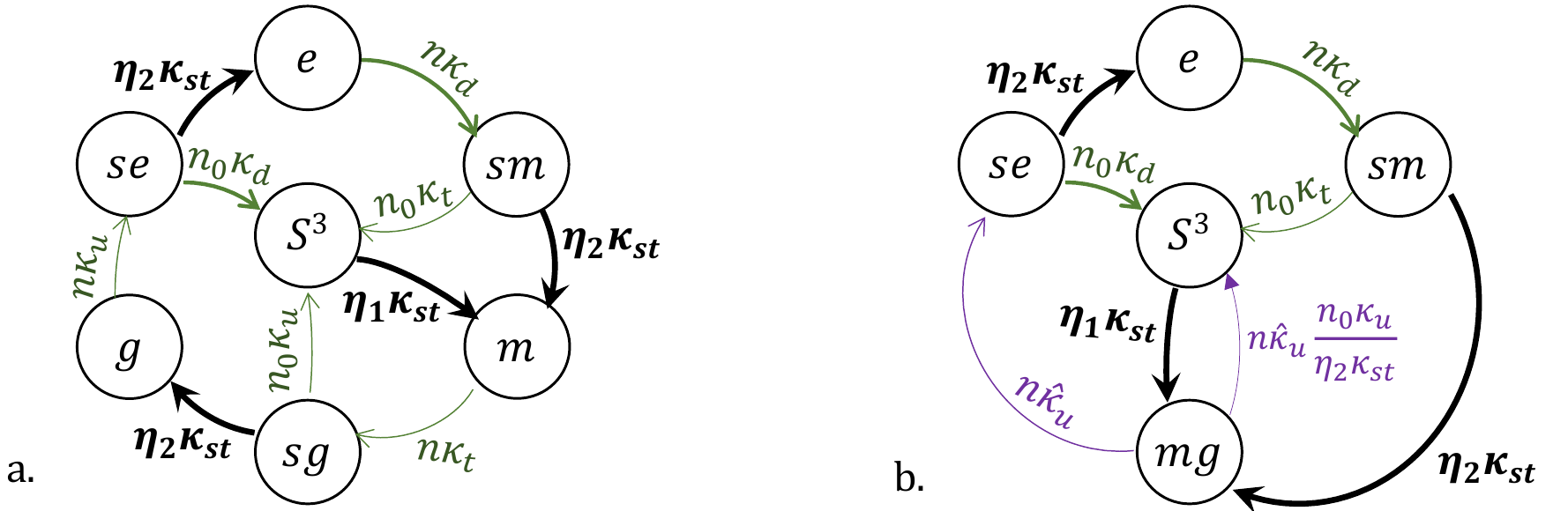}
\caption{\textbf{a.} Reduced ancilla Markov chain, grouping several states in an approximation to the effect of unsynchronized ancillas on data (pessimistic, with $n_0=n-1$). The values of $\eta_1$ and $\eta_2$ are of order 1, probably $n$-dependent, and characterize how quickly the ancillas re-synchronize. \text{b.} Exact reduction of the Markov chain represented on panel \textbf{a.} concerning steady-state computation, by grouping $p_g+ p_{sg}+ p_m =: p_{mg}$.
\label{fig:anca}}
\end{figure}

Regarding the data, the following applies.
\begin{itemize}
\item Whenever ancillas take two different values, there is the risk that a spontaneous jump leads to the situation where they take three different values (thus some on $\q{e}$, some on $\q{m}$, some on $\q{g}$). This is totally outside the intended regime. We have established in Section \ref{section3} that from there the ancillas exponentially converge back to $\q{gg...g},\q{ee...e},\q{mm...m}$ eventually, at a rate proportional to $\kappa_{st}$.
\newline A pessimistic regime would consider that for these ancillas configuration, the data goes erroneous whatever happens, and remains so until ancillas converge back to a single value. Furthermore, since a full reset may then be needed, the worst ancilla value to reach is $\q{mm...m}$, furthest from reset. 
\newline We thus model as $S^3$ all the ancilla configurations taking 3 different values. From there, ancillas evolve towards $E^{m}$ at a rate $\eta_1 \kappa_{st}$ with $\eta_1$ of order 1. For e.g.~ancillas distributed over $\q{e}$ and $\q{g}$ in any configuration, the rate to reach $S^3$ is $(n-1) \kappa_d$, assuming pessimistically that up to $(n-1)$ ancillas may be on $\q{e}$ with the danger to relax to $\q{m}$. This being settled, we now consider other transitions, involving only two different ancilla values.
\item Some ancillas on $\q{m}$ and some on $\q{g}$: The data undergo the same dynamics for all these configurations. So, we must just keep track of the possibility to jump to $S^3$ in this case. With a pessimistic bound $(n-1) \kappa_u$ on the rate towards $S_3$, we can group all these ancilla configurations into one, called  ``$sg$'' (for ``some already $g$'').
\newline The transition from $sg$ to $g$ (thus ``all on $g$'') happens at a rate $\eta_2 \kappa_{st}$ with $\eta_2$ of order 1, while the jump from $\q{mm...m}$ towards $sg$ happens at rate $n \kappa_t$.
\item Some ancillas on $\q{g}$ and some on $\q{e}$: The idea is that --- see Fig.\ref{fig:reredMC} --- these ancilla configurations should happen only during the transition from $\q{gg...g}$ towards $\q{ee...e}$, starting a full reset to $\q{++...+}$. A pessimistic approximation assumes that a full reset is still needed once the ancillas reach $\q{ee...e}$. We can then group all these ancilla configurations into one, called ``$se$'' (for ``some already on $e$''), towards which the ancillas converge at a rate $n \kappa_u$ and from which they leave towards $\q{ee...e}$ at a rate $\eta_2 \kappa_{st}$. The data goes towards $E^{se}$ then $E^e$.
\item Some ancillas on $\q{e}$ and some on $\q{m}$: The nominal chain transitions there between $\q{ee...e}$ hence resets, and $\q{mm...m}$ hence letting $\kappa_c$ terms consecutively take over. To group those ancilla configurations into one, we must discard the place at which data resets still happen along the chain. In this sense, a pessimistic bound assumes that (i) no more useful resets can be guaranteed and (ii) since resets are likely being done still repetitively, any $\kappa_c$ jumps occurring there can lead to an error. 
\newline With this common data behavior, we group the ancilla configurations into ``$sm$'' (for ``some already $m$''). The transition from $sm$ to $m$ (thus ``all on $m$'') happens at a rate $\eta_2 \kappa_{st}$, while the rate of jumps from $e$ towards $sm$ is $n \kappa_d$.
\end{itemize}
This grouping allows us to model a simplified ancilla Markov chain as represented on Figure \ref{fig:anca}.a, with $n_0=n-1$ in this pessimistic approximation. The ancilla configurations $m$, $sg$ and $g$ can be further grouped. Indeed, identifying their steady-state fraction on $sg$ is sufficient for knowing the jump rate to $S^3$, irrespective of data configuration; while for the rest, the data undergo the same dynamics for all those ancilla configurations. We can easily express $p_g, p_{sg}, p_m$ at steady state on Figure \ref{fig:anca}.a as a function of their total population $p_g+ p_{sg}+ p_m =: p_{mg}$. From this we can compute adjusted outgoing rates and further reduce the model to the Markov chain on Figure \ref{fig:anca}.b, where 
$$\hat{\kappa}_u =  \frac{n \kappa_u \kappa_t \eta_2 \kappa_{st}}{n \kappa_u (\eta_2 \kappa_{st}+n_0 \kappa_u + n \kappa_t) + \eta_2 \kappa_{st} n \kappa_t} \; = \tilde{\kappa}_u \, (1-O(\epsilon_1\epsilon_2)) \; .$$\\

\begin{figure}
\includegraphics[width=150mm, trim=20 110 55 118, clip]{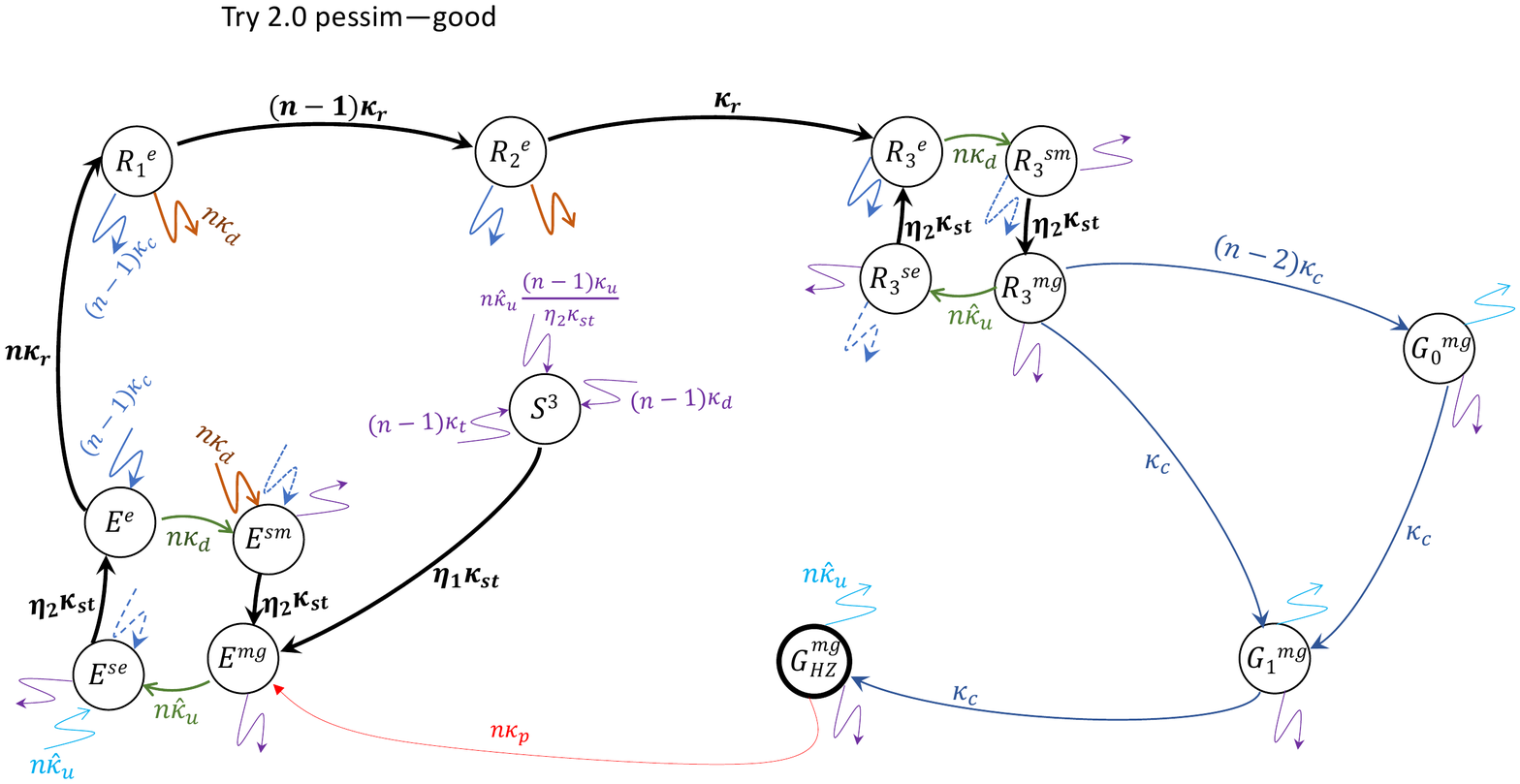}
\caption{Reduced Markov chain on sets of (hypothetical) output signals, taking imperfect ancilla synchronization into account. Pessimistic approximations have been done on the associated data evolution, along the lines of Fig.\ref{fig:reredMC} and as explained in the text.
\label{fig:newdataMC}}
\end{figure}

Thus in principle, to obtain the full Markov chain involving imperfect ancilla synchronization, each pair of configurations $Q^e, Q^{mg}$ on Figure \ref{figa:00MCn} must be replaced by a foursome as depicted on Fig.\ref{fig:anca}.b, with corresponding data evolutions; and adding unsynchronized state $S^3$. The reduction of Figure \ref{figa:00MCn} to Figure \ref{fig:reredMC}, in particular merging $R_k^{mg}$ into $E^{mg}$ for $k<n$ and merging $G_k^{e}$ into $E_e$, allows for important (pessimistic) simplifications of the resulting Markov chain. Note that the merging of various configurations is valid in terms of \emph{steady state populations}, which is what we want to compute. The Markov chain, represented on Figure \ref{fig:newdataMC}, is set up as follows.
\begin{itemize}[noitemsep,topsep=0pt]
\item Configuration $S^3$ flows to $E^m$, merged into $E^{mg}$, at rate $\eta_1 \kappa_{st}$.
\item Data configurations $E^{...}$ need all resets, so a pessimistic bound is obtained by assuming that no useful data evolution is done unless all ancillas are on $e$. Data evolves either towards $S^3$, or from $E^e$ towards $R_1^e$.
\item From data configuration $R_1^e$, we have assumed on Fig.\ref{fig:reredMC} that an ancilla jump directly leads to $E^{mg}$. Accordingly, we here assume that $R_1^e$ leads to $E^{sm}$, unless it undergoes one of the data evolutions like on Fig.\ref{fig:reredMC}. The configurations $R_1^{sm}, R_1^{mg}, R_1^{sg}$ are not needed as the corresponding data is all assumed in $E$. 
\item The same property holds for all $R_k$ with $k<n$.
\item For $G_k^{...}$, we want to avoid resets. We have already assumed in Figure \ref{fig:reredMC} that the data goes into error when ancillas jump to $e$. This is motivated by the fact that subsequently applying a reset to $\q{+}$ on one data qubit anyways leads to $R_1^e$ from both situations, while other jumps are much less likely and only help for $G_k^{...}$. A similar pessimistic bound with imperfect ancilla synchronization is to jump with $n \hat{\kappa}_u$ from $G_k^{mg}$ directly to $E^{se}$, instead of to $G_k^{se}$. The configurations $G_k ^{se}$, $G_k^e$  and $G_k^{sm}$ are thereby merged into the corresponding $E^{...}$ configurations. Further, like in Figure \ref{fig:reredMC}, $G_k^{mg}$ evolves with $\kappa_c$ towards $G_{k+1}^{mg}$; and we finally keep the small external perturbation $\kappa_p$ on $G_{HZ}^{mg}$ only.
\item On $R_n^{...}$, imperfect ancillas synchronization plays no direct role, since data resets have no effect. In principle the data would go to $G_0$ and $G_1$ while keeping ancilla state. Now following the approximations done so far, the flow towards $G_k^{se}$, $G_k^e$  and $G_k^{sm}$ is instead routed to $E^{se}$, $E^{e}$ and $E^{sm}$.\\
\end{itemize}

The steady state of the Markov chain depicted on Fig.\ref{fig:newdataMC} can be computed along similar lines as for the Markov chains involving perfect ancilla synchronization, as represented on Fig.\ref{figa:00MCn} or \ref{fig:reredMC}. Recall that $n_0=n-1$, and that $\hat{\kappa}_u \simeq \tilde{\kappa}_u$ up to second order terms.
\begin{itemize}
\item The ancilla steady state distribution, not influenced by data configuration, can be computed rather directly from Fig.\ref{fig:anca}.b. The dominant orders are:
\begin{eqnarray*}
p_{se} = p_{sm} \simeq \frac{n \hat{\kappa}_u}{\eta_2 \kappa_{st}} \; \left( 1 - \tfrac{\hat{\kappa}_u}{\kappa_d} - \tfrac{n_0 \kappa_d}{\eta_2 \kappa_{st}}  \right)  &\quad ; \quad &
p_{S^3} =  \left( \tfrac{n_0 \kappa_d}{\eta_1 \kappa_{st}} + \tfrac{n_0 \kappa_u}{\eta_2 \kappa_{st}}\,\tfrac{n_0 \kappa_d + \eta_2 \kappa_{st}}{\eta_1 \kappa_{st}}  \right) \; p_{se}
\\
p_e \simeq \frac{\hat{\kappa}_u}{\kappa_d} \, \left( 1 - \tfrac{\hat{\kappa}_u}{\kappa_d} - \tfrac{n_0 \kappa_d}{\eta_2 \kappa_{st}} \right)   &\quad ; \quad &  
p_{mg} \simeq  1-\frac{\hat{\kappa}_u}{\kappa_d} \; .
\end{eqnarray*}
\item We next consider the square of Markov chain configurations $R_n^{mg},\, R_n^{sm},\, R_n^{se}, R_n^e$ on Fig.\ref{fig:newdataMC}. By first expressing all the corresponding steady state populations as a function of $p_{R_n^e}$, we get at dominant orders:
\begin{eqnarray} 
\label{eq:pRnepRnmg} p_{R_n^{mg}} & \; = \; & \frac{n \kappa_d}{n_0 \kappa_c + n\hat{\kappa}_u} \, \tfrac{\eta_2 \kappa_{st}}{\eta_2 \kappa_{st} + n_0 \kappa_c} \\
\nonumber p_{R_n^e} & \; = \; & \frac{\kappa_r}{n \kappa_d + n_0 \kappa_c} \; \left( 1 + \tfrac{n \hat{\kappa}_u}{n_0 \kappa_c} \, \tfrac{n \kappa_d}{n \kappa_d+n_0 \kappa_c} \right) \; p_{R_{n-1}^e}\; .
\end{eqnarray}
The second expression is exactly like in the case with perfect ancilla synchronization.
\item From there, like in the proof of Proposition \ref{prop:llp1} at the beginning of Section \ref{ssec:app:proof1}, using $p_e$ and a set of equations insensitive to imperfect synchronization, we get that
\begin{equation}\label{eq:Rnefull}
p_{R_n^e}\; = \; \frac{\hat{\kappa}_u}{\kappa_d} \, \left( 1 - \tfrac{\hat{\kappa}_u}{\kappa_d} - \tfrac{n_0 \kappa_d}{\eta_2 \kappa_{st}} - \tfrac{\kappa_c + \kappa_d}{\kappa_r} n \ln(n) \, \right) \; .
\end{equation}
Here the imperfect ancilla synchronization thus replaces $\tilde{\kappa}_u$ by $\hat{\kappa}_u$, which is just a second-order effect, and contributes a term in $\tfrac{n_0 \kappa_d}{\eta_2 \kappa_{st}}$, which arises through the modification of $p_e$.
\item Computing the steady state relations for the chain from $p_{R_n^{mg}}$ progressively through the $p_{G_k^{mg}}$, we get exactly like for the perfect-ancilla Markov chain of Figure \ref{fig:reredMC}:
$$p_{G_{HZ}^{mg}} \simeq \frac{n_0 \kappa_c}{n \hat{\kappa}_u} \,  \left( 1 - \tfrac{\kappa_p}{\hat{\kappa}_u} - n(n-1) \tfrac{\hat{\kappa}_u}{\kappa_c} \right) \, p_{R_n^{mg}} \; .$$
\item Using \eqref{eq:pRnepRnmg} and \eqref{eq:Rnefull} in this last result finally yields the conclusion stated in Section \ref{sssec:impancmt}.
\end{itemize}

\subsection{Details of the qutrit-wave analysis, first method}\label{app:aq3:a}

This method considers that, in order to reach $\q{GHZ_+}$, the system must undergo first a sequence of resets through level $\q{2}$ \emph{irrespectively of the occurence of any $L_k$ jumps}, then a sequence of jumps with $L_k$ exclusively and containing the ordered subsequence $L_1, L_2, ... L_{n-1}$. Note that now the error model involves the ``ancilla'' level $\q{2}$ as well, such that a flurry of non-nominal jump sequences can in principle appear.

Like in the previous sections, we simplify the analysis by studying a classical Markov chain whose configurations are sets of (hypothetical) jump detection sequences $q(t)$. We refer the reader to Section \ref{ssec:an2mod} for the definition of the associated signal.

The goal of this first analysis method is to provide a pessimistic bound on fidelity to $\q{GHZ_+}$. We hence define the following Markov chain configurations:
\begin{itemize}
\item $U$: any $q(t)$ ending with $\{ U \}$, followed by an arbitrary sequence of $\{k \, + \}$ or $\{ k\, L\}$ all with $k>1$;
\item $R_k$ for $k=1,2,...,n-1$: any $q(t)$ ending with $\{ U \}$ , followed by an arbitrary sequence of $\{k \, + \}$ or $\{ k\, L\}$ containing the subsequence 
$\{1 \, +\}, \{2 \, +\},... ,\{k \, +\}$ but not the subsequence $\{1 \, +\}, \{2 \, +\},... ,\{(k+1) \, +\}$;
\item $R_n$: any $q(t)$ ending with $\{ U \}$ , followed by an arbitrary sequence of $\{k \, + \}$ or $\{ k\, L\}$ containing the subsequence 
$\{1 \, +\}, \{2 \, +\},... ,\{n \, +\}$ but not the subsequence $\{1 \, +\}, \{2 \, +\},... ,\{n \, +\}, \{1 \, L\}$;
\item $G_k$ for $k=1,2,...,n-2$: any $q(t)$ ending with a sequence like $R_n$, followed by an arbitrary sequence of $\{ k\, L\}$ containing the subsequence $\{1 \, L\}, \{2 \, L\},... ,\{k \, L\}$ but not the subsequence $\{1 \, L\}, \{2 \, L\},... ,\{(k+1) \, L\}$;
\item $G_{n-1}=: G_{HZ}$: same as $G_k$, except the last condition is dropped since $\{ n \, L \}$ does not exist;
\item $E$: any other $q(t)$, i.e. not containing $\{ U \}$ or ending with $\{ k \, E \}$ for some $k$ not followed by $\{ U \}$.
\newline (The reader is invited to check that these are indeed the only remaining possibilities; otherwise see explanations in Section \ref{app:aq3:b})
\end{itemize}
This grouping involves some obviously pessimistic approximations --- see below --- but also a few subtleties. For instance, note that the occurrence of  $\{j \, L\}$ between $\{j \, + \}$ and $\{ (j+1) \, + \}$ would be detrimental to GHZ stabilization, so it seems that we must exclude it from $R_k$. However, this is not necessary, since after $\{j \, + \}$ we cannot have $\{j \, L\}$ before having seen either $\{ (j+1) \, + \}$ or $\{ (j+1) \, + \}$;  the former maintains the ``good'' sequence, the latter leads to $E$ anyways.

\begin{figure}
\begin{center}
\includegraphics[width=110mm, trim=150 90 150 140, clip]{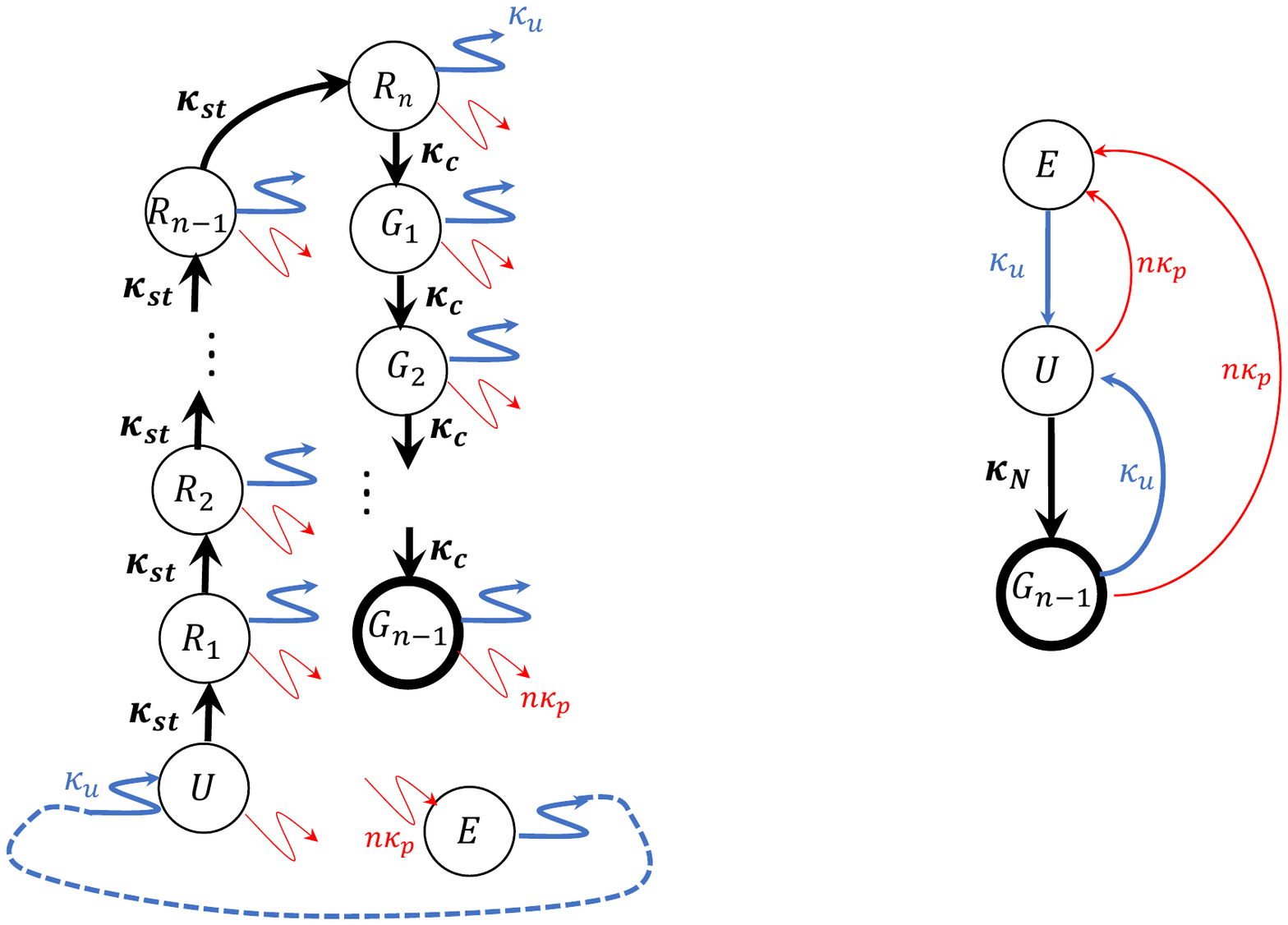}
\end{center}
\caption{Left: The Markov chain used for the analysis of Section \ref{app:aq3:a}. Wiggly arrows indicate that all outflows connect to the inflow of the same color and shape. Right: reduced Markov chain where $1/\kappa_N$ is the expected time to cross the chain on the left from $U$ to $G_{n-1}$, and yielding the same steady state $p_{G_HZ} =p_{G_{n-1}}$ at first order in $\epsilon,\; \epsilon_2$.
\label{fig:WaveAnPart2}}
\end{figure}

We have been pessimistic on several points in this aggregation of output signals. 
\begin{itemize}
\item[] First, we have required for reaching $G_{HZ}$ that we follow first a full reset wave, then a full $L_k$ wave. In principle, the two could propagate together, and an analysis in this way is carried out in the next section. 
\item[] Second, starting from $\q{++...+}$, it is not strictly necessary to have $\{1 \, L\}, \{2 \, L\},... ,\{n-1 \, L\}$ in order to end up on $\q{GHZ_+}$. However, the possible alternatives seem to be few, at the cost of a more complicated analysis, which we will not carry out. \item[] Third, we have considered that, from any configuration, any error brings us into a ``completely useless'' configuration $E$ from which the whole reset-then-$L_k$ wave must be reapplied. This is of course pessimistic, since e.g.~in the output sequence ending with $\{ U \},\{1 \, +\}, \{2 \, +\}, \{4 \, E \}$, the error detection has no lasting effect and will be just erased by pursuing the reset wave. The grouping made above allows us to significantly simplify the analysis, by having a uniform rate $n \kappa_p$ for flowing towards $E$. Moreover, a more precise modeling in fact makes no difference on the leading order. Indeed, we are targeting $p_{GHZ} = 1 - O(\epsilon,\epsilon_2)$; with this, just taking into account a flow at rate $n \kappa_p$ from $G_{HZ}$ towards $E$, together with a rate $\kappa_u$ for leaving $E$ by launching a reset wave, we would already obtain $p_E > n \kappa_p (1-O(\epsilon)) \; / \; \kappa_u  = n \epsilon_2 + o(\epsilon, \epsilon_2)$. The result reported in Proposition \ref{prop:WaveFResult1}, thus with the Markov chain of Fig.~\ref{fig:WaveAnPart2}left, says no worse in terms of $\epsilon_2$ error.
\end{itemize}

The result reported in Proposition \ref{prop:WaveFResult1}, first part, is obtained rather straightforwardly by writing the steady state conditions of the Markov chain depicted on Fig.~\ref{fig:WaveAnPart2}left, in a sequential way starting at configuration $U$:
\begin{eqnarray*}
(\kappa_{st}+ n \kappa_p)\, p_U = \kappa_u (1-p_U) & \Rightarrow & p_U = \tfrac{\kappa_u}{\kappa_p+\kappa_u+\kappa_{st}} \\
(\kappa_{st}+ n \kappa_p+\kappa_u)\, p_{R_1} = \kappa_{st} p_U \\
(\kappa_{st}+ n \kappa_p+\kappa_u)\, p_{R_{k}} = \kappa_{st} p_{R_k} & \Rightarrow & p_{R_k} = \left(\tfrac{\kappa_st}{\kappa_p+\kappa_u+\kappa_{st}}\right)^k \, p_U \\
(\kappa_c + n \kappa_p+\kappa_u)\, p_{R_{n}} = \kappa_{st} p_{R_{n-1}} & \Rightarrow & p_{R_n} = \tfrac{\kappa_st}{\kappa_p+\kappa_u+\kappa_{c}} \, p_{R_{n-1}} \\
\end{eqnarray*}
and similarly
\begin{eqnarray*}
 p_{G_k} =\left( \tfrac{\kappa_c}{\kappa_p+\kappa_u+\kappa_{c}}\right)^k \, p_{R_{n}} & \; , \; &  p_{G_{HZ}} = \tfrac{\kappa_c}{\kappa_p+\kappa_u} \, p_{G_{n-2}} \; .
\end{eqnarray*}
Multiplying this out, we get
$$p_{G_{HZ}} = \left( \tfrac{\kappa_st}{\kappa_p+\kappa_u+\kappa_{st}}\right)^n \left(\tfrac{\kappa_c}{\kappa_p+\kappa_u+\kappa_{c}} \right)^{n-1} \tfrac{\kappa_u}{\kappa_p+\kappa_u} \; .$$
Introducing the notation $\epsilon, \epsilon_2, \gamma$ and keeping only the first order terms in $\epsilon,\epsilon_2$ yields the reported result.\\

Note that the same result would be obtained, at least at first order, as a steady state of the Markov chain represented on Fig.~\ref{fig:WaveAnPart2}right, where we have summarized the whole chain of dominating events, i.e.~with $\kappa_{st}$ and $\kappa_c$, by a single transition at an effective rate $\kappa_N$. This effective rate is computed such that $1/\kappa_N$ corresponds to the expected time for crossing the chain of Fig.~\ref{fig:WaveAnPart2}left from $U$ to $G_{n-1}$, thus
$$\frac{1}{\kappa_N} = \frac{n}{\kappa_{st}} + \frac{n-1}{\kappa_c} \; .$$  
This reduction is possible because jumps out of this chain happen at the same rates $\kappa_p$ and $\kappa_u$ irrespectively of the precise configuration. Our second analysis will hence directly aim at such characteristic times.

\subsection{Details of the qutrit-wave analysis, second method}\label{app:aq3:b}

\begin{figure}
\begin{center}
\includegraphics[width=130mm, trim=40 165 100 120, clip]{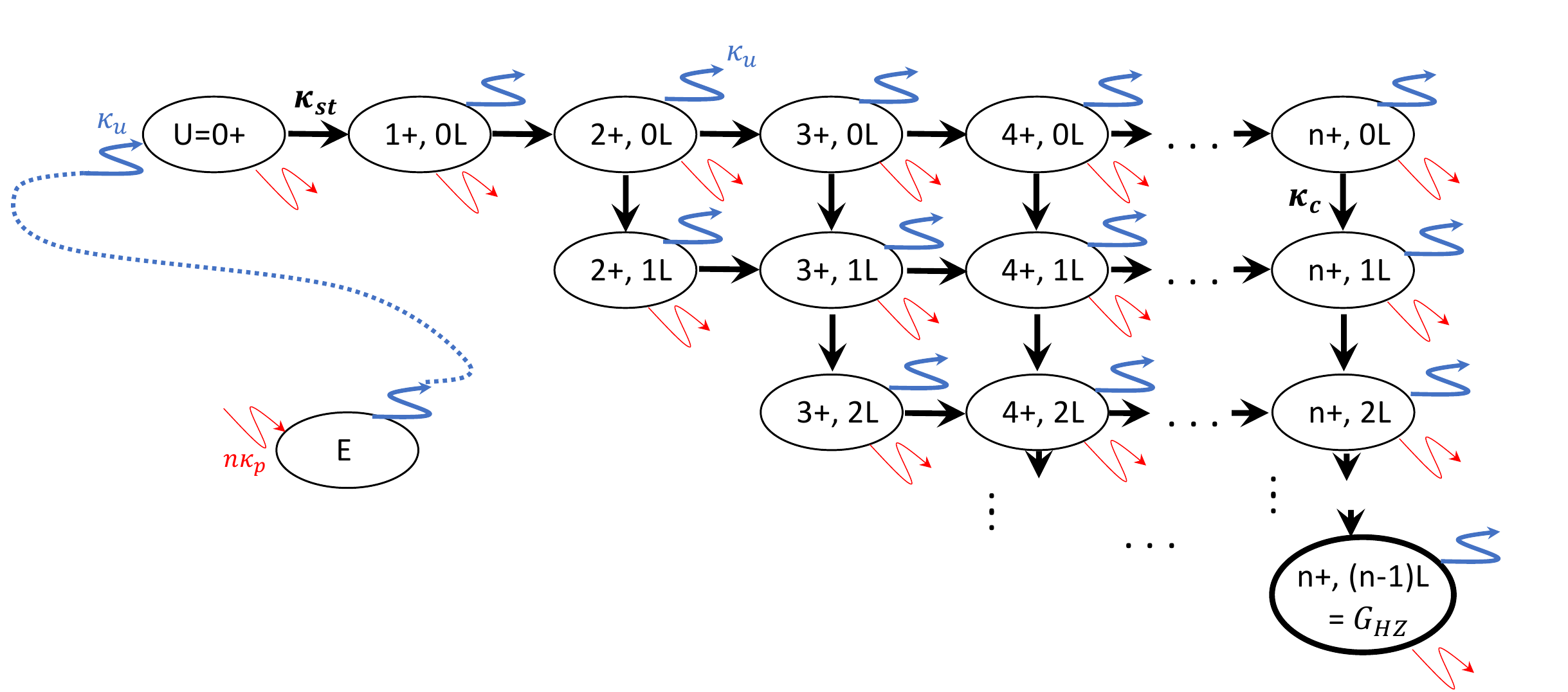}
\end{center}
\caption{Markov chain defined for the ``wave-propagation'' analysis of Section \ref{app:aq3:b}. Like on other figures, outgoing wiggly arrows represent a flow towards the wiggly arrow of the same color and shape. \label{fig:wave3MC}}
\end{figure}

The previous section considers that the wave of resets, propagating through level $\q{2}$, has to finish first before a wave of $L_k$ (or $\tilde{L}_k$) would be launched and reach the GHZ state. In fact there is no reason for the reset wave to finish before the $L_k$ wave starts. We can thus try another analysis to take into account this concomitant propagation, instead of capturing only the success rate of two consecutive waves.

We again simplify the analysis by studying a classical Markov chain whose configurations are sets of (hypothetical) jump detection sequences $q(t)$, now defining the following configurations. See Section \ref{ssec:an2mod} for the definition of signal components, and Figure \ref{fig:wave3MC} for a more visual explanation.
\begin{itemize}
\item $U$: any $q(t)$ ending with $\{ U \}$, followed by an arbitrary sequence of $\{k \, + \}$ or $\{ k\, L\}$ all with $k>1$;
\item $\{j_1+,\; j_2\,L\}$ for $j_2<j_1$:   any $q(t)$ ending with $\{ U \}$ , followed by an arbitrary sequence of $\{k \, + \}$ or $\{ k\, L\}$ containing two subsequences $s_{+,j_1} :=\{1 \, +\}, \{2 \, +\},... ,\{j_1 \, +\}$  and  $s_{L,j_2} := \{1 \, L\}, \{2 \, L\},... ,\{j_2 \, L\}$;  those sequences are interleaved such that $\{j \, L\}$ of subsequence $s_{L,j_2}$ comes after $\{(j+1) \, +\}$ of subsequence $s_{+,j_1}$, for all $j=1,2,...,j_2$. Furthermore, $q(t)$ contains no corresponding subsequences for $j_1' > j_1$ or $j_2' > j_2$.

In particular, we have $\{n\,+,\; (n-1)\,L\} =: G_{HZ}$.
\item $E$: any other $q(t)$, i.e. not containing $\{ U \}$ or ending with $\{ k \, E \}$ for some $k$ not followed by $\{ U \}$.
\end{itemize}
All signals $q(t)$ should take one of these forms. More precisely, an initial transient may yield arbitrary detections until $\{ U \}$ is detected once; the former are all covered by configuration $E$. After that, we cannot have $\{1 \, L\}$ before having seen either $\{1 \, + \}$ or $\{1 \, E \}$, i.e.~we are in configuration $U$ until switching either to $\{1+,\, 0L\}$ or back to $E$. The states $\{j_1+,\; j_2\,L\}$ then cover all possible combinations along the nominal chains of events. Note that after having detected a subsequence $\{ U \} \{1 \, +\}, \{2 \, +\},... ,\{j \, +\}$, it is impossible to re-detect any of the $\{k \, +\}$ with $k \leq j$, without re-encountering either $\{U \}$ or some error $\{k' \, E\}$ before. Therefore, the effect of the $\{k\, L\}$ will indeed be preserved, progressing towards $\q{GHZ_+}$, unless we re-encounter $\{U \}$ or some $\{k \, E\}$ and thus switch to configuration $U$ or $E$.

Like in Section \ref{app:aq3:a}, this Markov chain aggregates some output signals in a pessimistic way, e.g.~assuming that a full stabilization chain has to be re-applied after any $\{k \, E \}$ has occurred at the end of any detection sequence. However, at first order this approximation has no effect, and it greatly simplifies the analysis. Our goal is to compute and maximize the steady state population on $\{n\,+,\; (n-1)\,L\} =: G_{HZ}$, which we view as the sole configuration contributing to $\q{GHZ_+}$. We readily take $\kappa_c = \kappa_{st}$ at the maximal achievable reservoir rate.\\

\begin{figure}
\begin{center}
\includegraphics[width=150mm, trim=40 100 40 100, clip]{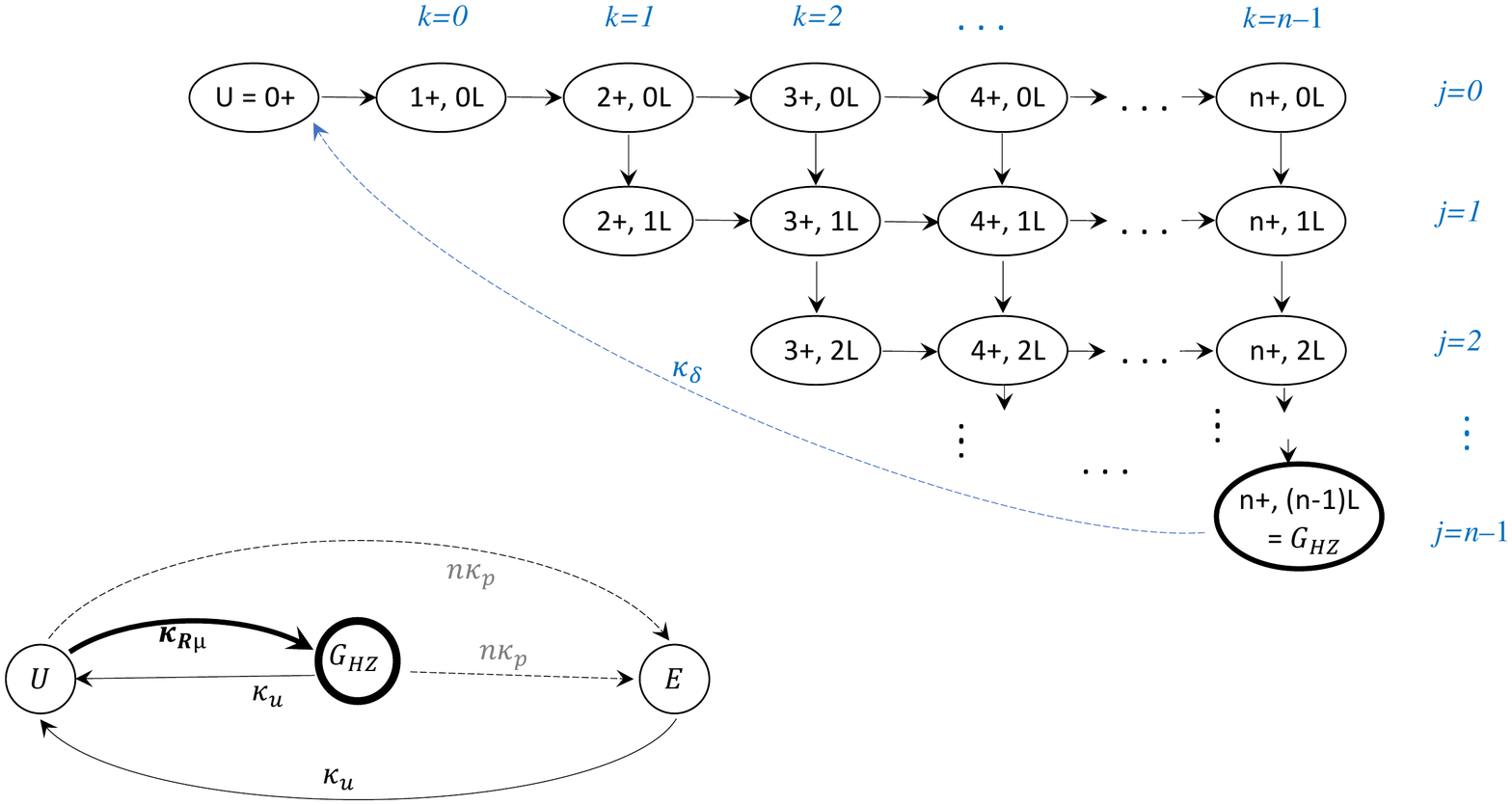}
\end{center}
\caption{Networks used for the simplified analysis of the Markov chain of Fig.~\ref{fig:wave3MC}. Bottom left: reduced Markov chain where the fast transitions of Fig.~\ref{fig:wave3MC}, i.e.~those involving $\kappa_c$ and $\kappa_{st}$, have all been aggregated into a single jump at effective rate $\kappa_{R\mu}$. Top right: The rate $\kappa_{R\mu}$ is computed such that $1/\kappa_{R\mu}$ is the expected time to cross this lattice from $\q{U}$ to $\q{G_{HZ}}$. All transitions here are at rate $\kappa_c = \kappa_{st}$. The coordinates labeled in blue, as well as the transition back from $\q{G_{HZ}}$ to $\q{U}$ at rate $\kappa_{\delta}$, are used in the analysis leading to the value of $\kappa_{R\mu}$. \label{fig:wave3MCb}}
\end{figure}

The Markov chain represented on Fig.~\ref{fig:wave3MC} is harder to analyze exactly. Instead, we directly resort to the technique of characteristic times mentioned at the end of Section \ref{app:aq3:a}. Namely, we compute the steady state of the Markov chain shown on the bottom left of Fig.~\ref{fig:wave3MCb}, which has summarized all the ``fast'' transitions (i.e.~those with $\kappa_c$ and $\kappa_st$) as a single jump with effective rate $\kappa_{R\mu}$, where $1/\kappa_{R\mu}$ is the expected time to cross the ``fast transition lattice'' represented on the top right of Fig.~\ref{fig:wave3MCb}. 

\begin{figure}
\begin{center}
\includegraphics[width=110mm]{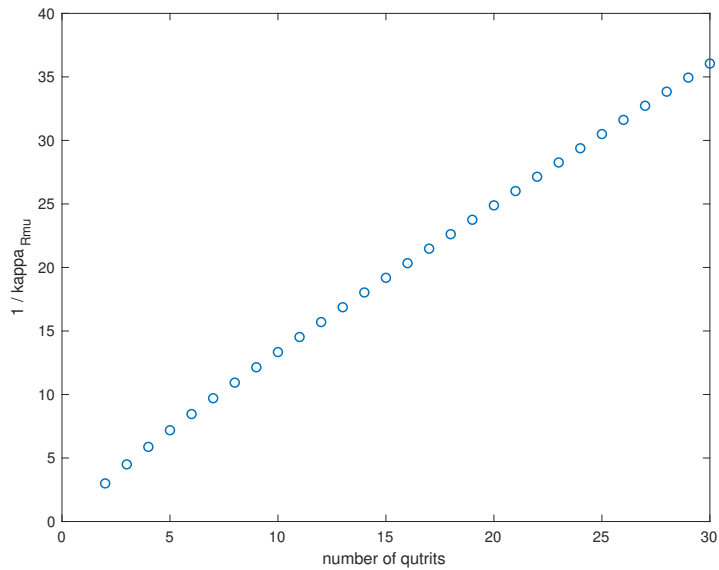}
\end{center}
\caption{Rate $\kappa_{R\mu}$ computed as the inverse of the expected time to cross the lattice of Figure \ref{fig:wave3MCb} from $\q{U}$ to $\q{n+,\; (n-1)L}$, as a function of the lattice size $n$. \label{fig:wave3MC2b}}
\end{figure}

The main analysis work is to properly estimate the transition rate $\kappa_{R\mu}$. Once this is fixed, a simple calculation gives
\begin{equation}\label{eq:ZDTYUI}
p_{G_{HZ}} = \frac{1}{(1+n\epsilon_p)(1+\tilde{\epsilon}+n\epsilon_p \tilde{\epsilon})} \simeq 1 - n\epsilon_p - \tilde{\epsilon} + n^2\epsilon_p^2 + \tilde{\epsilon}^2
\end{equation}
where $\epsilon_p = \kappa_p/\kappa_u$ and $\tilde{\epsilon} = \kappa_u / \kappa_{R\mu}$.

The expected time to cross the lattice on the top right of Fig.\ref{fig:wave3MCb} can be obtained as 
$$1/\kappa_{R\mu} = 1/\kappa_{\delta} \left( \frac{1}{\bar{p}_{G_{HZ}}} -1\right) \; ,$$
where $\bar{p}_{G_{HZ}}$ is the population on $\q{G_{HZ}}$ in the steady state of the Markov chain associated to the lattice with an additional transition at rate $\kappa_{\delta}$ back from $\q{G_{HZ}}$ to $\q{U}$ (blue dotted arrow on Fig.\ref{fig:wave3MCb}). The following properties are easy to show recursively.

\begin{Proposition}
Denote by $\bar{p}_{j,k}$ the steady state population on the node depicted at coordinates $j,k$ in the lattice of Figure Fig.\ref{fig:wave3MCb}. We denote $C^j_{j+k} = \tfrac{(j+k)!}{j! k!}$.
\begin{itemize}
\item (Steady-state relative population) For $j=0,1,...,n-2$ and $k=j,j+1,...,n-2$, we have
$$\bar{p}_{j,k} =  \bar{p}_{0,0} \;  C_{j+k}^j \; / \;   2^{k+j} \; .$$
For $k=n-1$, we can compute recursively with 
\begin{eqnarray*}
\bar{p}_{0,n-1} &=& \bar{p}_{0,n-1} \quad , \quad \bar{p}_{n-1,n-1} = \bar{p}_{0,0} \;  \tfrac{\kappa_{st}}{\kappa_{\delta}} \\
\bar{p}_{j,n-1} &=& \bar{p}_{j-1,n-1} + \bar{p}_{j,n-2} \quad \text{for } j=1,2,...,n-2 \; .
\end{eqnarray*}
One can double-check the recursion with the formula $\bar{p}_{n-2,n-1} = \bar{p}_{0,0}$.

Finally, $\bar{p}_{U} = \bar{p}_{0,0}$.
\item (Marginal and explicit population) The total population on row $j$ of the lattice is
\begin{eqnarray*}
\text{For } j=0: && \bar{p}_{U} + {\textstyle \sum_{k=0}^{n-1}} \bar{p}_{0,k} \; = \; 3\, \bar{p}_{0,0}   \\
\text{For } j=1,2,...,n-2: &&  {\textstyle \sum_{k=j}^{n-1}} \bar{p}_{0,k} \; = \; \bar{p}_{0,0} \, \left(1 + C^{j-1}_{2j-1} \; / \; 2^{(2j-1)}\right) \; . \\
\text{For } j=n-1: && \bar{p}_{n-1,n-1} = \bar{p}_{0,0} \; \tfrac{\kappa_{st}}{\kappa_{\delta}} \; .
\end{eqnarray*}
From this, one deduces
$$\bar{p}_{G_{HZ}} = \frac{\kappa_{st}}{\kappa_{\delta}}\, \bar{p}_{0,0} =  \frac{\tfrac{\kappa_{st}}{\kappa_{\delta}}}{(n-1) + 2 + \tfrac{\kappa_{st}}{\kappa_{\delta}} + {\textstyle \sum_{j=1}^{n-2}}   \tfrac{C^{j-1}_{2j-1}}{2^{(2j-1)}} } \; $$
and thus finally the rate
\begin{equation}\label{eq:kappard}
\kappa_{R\mu} = \kappa_{st} \; \frac{1}{n+1 + {\textstyle \sum_{j=1}^{n-2}}   \tfrac{C^{j-1}_{2j-1}}{2^{(2j-1)}}} 
\end{equation}
\end{itemize}
\end{Proposition}

The rate $\kappa_{R\mu}$ is graphically represented on Figure \ref{fig:wave3MC2b}. For low values of $n$, it takes the values:
\begin{eqnarray*}
&& n=2: \;\;  \kappa_{R\mu} = \tfrac{\kappa_{st}}{3}  \quad ; \quad n=3: \;\; \kappa_{R\mu} = \tfrac{\kappa_{st}}{4.5} \; ; \\
&& n=4: \;\;  \kappa_{R\mu} = \tfrac{\kappa_{st}}{5+7/8}  \quad ; \quad n=5: \;\; \kappa_{R\mu} = \tfrac{\kappa_{st}}{6+19/16} \; .
\end{eqnarray*}
For large $n$ it roughly scales as 
$$\kappa_{R\mu} \simeq \frac{\kappa_{st}}{n}  \; .$$
Plugging this into the formula \eqref{eq:ZDTYUI}, we obtain the result announced in Proposition \ref{prop:WaveFResult1}, part 2.

\end{document}